  \documentclass[journal]{IEEEtran}
  \usepackage{amsfonts}
\IEEEoverridecommandlockouts
\usepackage{amsfonts,amsmath,amssymb,yfonts}
\usepackage{cite}
\usepackage{graphicx}
\usepackage{url}
\usepackage{epstopdf}
\usepackage{bm}
\usepackage{xcolor}

\newtheorem{theorem}{Theorem}
 \newtheorem{lemma}{Lemma}
\newtheorem{sublemma}{Lemma}[lemma]
 \newtheorem{statement}{Statement}
 \newtheorem{proposition}{Proposition}
 \newtheorem{corollary}{Corollary}
 \newtheorem{definition}{Definition}
  \newtheorem{remark}{Remark}

 \newtheorem{subsidiary theorem}{Subsidiary Theorem}

\ifCLASSINFOpdf
\else
\fi
\usepackage{lineno}

\begin{document}
%

\title{Complex Linear Physical-Layer Network Coding}

\author{Long~Shi,~\IEEEmembership{Member,~IEEE} and Soung~Chang~Liew,~\IEEEmembership{Fellow,~IEEE}}
%
\maketitle

\begin{abstract}

This paper presents the results of a comprehensive investigation of complex linear physical-layer network (PNC) in two-way relay channels (TWRC). In this system, two nodes A and B communicate with each other via a relay   R. Nodes A and B send complex symbols,  $w_A$ and  $w_B$, simultaneously to relay R. Based on the simultaneously received signals, relay R computes a linear combination of the symbols, $w_N=\alpha w_A+\beta w_B$,  as a network-coded symbol and then broadcasts $w_N$ to nodes A and B. Node A then obtains $w_B$ from $w_N$ and its self-information $w_A$ by $w_B=\beta^{-1}(w_N-\alpha w_A)$. Node B obtains $w_B$ in a similar way. A critical question at relay R is as follows: ``Given channel gain ratio $\eta = h_A/h_B$, where $h_A$ and $h_B$ are the complex channel gains from nodes A and B to relay R, respectively, what is the optimal coefficients $(\alpha,\beta)$  that minimizes the symbol error rate (SER) of  $w_N=\alpha w_A+\beta w_B$  when we attempt to detect $w_N$  in the presence of noise?'' Our contributions with respect to this question are as follows: (1) We put forth a general Gaussian-integer formulation for complex linear PNC in which $\alpha,\beta,w_A, w_B$, and $w_N$  are elements of a finite field of Gaussian integers, that is, the field of $\mathbb{Z}[i]/q$ where $q$ is a Gaussian prime. Previous vector formulation, in which $w_A$, $w_B$, and $w_N$  were represented by $2$-dimensional vectors and $\alpha$ and $\beta$ were represented by  $2\times 2$ matrices, corresponds to a subcase of our Gaussian-integer formulation where $q$ is real prime only. Extension to Gaussian prime $q$, where $q$  can be complex, gives us a larger set of signal constellations to achieve different rates at different SNR. (2) We show how to divide the complex plane of $\eta$ into different Voronoi regions such that the $\eta$ within each Voronoi region share the same optimal PNC mapping $(\alpha_{opt},\beta_{opt})$. We uncover the structure of the Voronoi regions that allows us to compute a minimum-distance metric that characterizes the SER of $w_N$ under optimal PNC mapping $(\alpha_{opt},\beta_{opt})$. Overall, the contributions in (1) and (2) yield a toolset for a comprehensive understanding of complex linear PNC in $\mathbb{Z}[i]/q$. We believe investigation of linear PNC beyond $\mathbb{Z}[i]/q$ can follow the same approach.

\end{abstract}

\begin{IEEEkeywords}

Complex linear physical-layer network coding, Gaussian integer, minimum distance, Voronoi region

\end{IEEEkeywords}

\IEEEpeerreviewmaketitle

\section{Introduction}
Physical-layer network coding (PNC) can potentially boost the throughput of relay networks, such as a two-way relay channel (TWRC) \cite{pnc,liew}. \emph{Linear} PNC allows PNC decoding to be performed in a simpler manner than nonlinear PNC. In TWRC, data exchange between two isolated   nodes A and B is facilitated by a relay R. When PNC is employed in TWRC, the data exchange consists of two phases. In the uplink phase,    nodes A and B transmit $w_A$  and $w_B$ to   relay R simultaneously. For {linear} PNC,  the relay aims to decode a {linear} combination of  $w_A$  and $w_B$ as a network-coded (NC) symbol, $w_N=\alpha w_A+\beta w_B$, from the simultaneously received signals. We refer to the linear combination $w_N=\alpha w_A+\beta w_B$   as a linear PNC mapping. Equivalently, we also refer to the coefficient pair $(\alpha,\beta)$  as a PNC mapping, with the understanding that the coefficients are used in the linear combination  $w_N=\alpha w_A+\beta w_B$. In the downlink phase,   relay R broadcasts $w_N$ to the  nodes A and B. Node A then obtains  $w_B$ from $w_N$  and its self-information $w_A$   by   $w_B=\beta^{-1}(w_N-\alpha w_A)$.  Node B obtains $w_A$ by  $w_A=\alpha^{-1}(w_N-\beta w_B)$.

Linear PNC has been extensively studied because of its scalability in terms of the network coding operation for high-order modulations \cite{pnc,liew,pop,zhang,yang3,choi,chang,yanglet,yangtwc,long, lei}. The original version of linear PNC mapping was formulated as binary XOR mapping with BPSK \cite{pnc,liew,pop,zhang,yang3}. This was later extended to higher-order signal modulations \cite{lei,yanglet,yangtwc,long}.  Prior work in \cite{pnc,liew,pop,zhang,yang3,choi,chang} assumed ideal communication scenarios in which signals of the two end nodes received at the relay have balanced powers with perfect phase alignments. However, these ideal scenarios rarely occur in practice because of factors such as  imperfect power control, relative carrier frequency offset, and phase noise induced by the use of different oscillators at nodes A and B. In general, the powers will not be perfectly balanced and the phases will not be perfectly aligned.

The authors of \cite{yangtwc} formulated a PNC scheme to take into account   imbalanced received powers and relative phase offset, assuming the use of $q$-PAM and $q^2$-QAM  modulations by the nodes A and B, where $q$ is a prime integer.  Building on \cite{yangtwc}, we investigated the error performance of $q$-PAM linear PNC in \cite{long} via a systematic analysis of the effect of power imbalance on a signal-constellation minimum distance that characterizes the symbol error rate (SER) of decoding $w_N$  at the relay.  In particular, in \cite{long}, we  found that the performance of $q$-PAM linear PNC can be highly sensitive to small changes in the channel gains (i.e., small variations in channel gains of the two end nodes can cause significant performance changes).

This paper further extends the work in \cite{long}. Whereas \cite{long} assumed real channels for the two end nodes (i.e., the channel gains are real and there is no relative phase offset between them; there is only power imbalance), this paper assumes complex channels to take into account possible relative phase offset between the end nodes besides the power imbalance. We present a comprehensive investigation of optimal complex linear PNC. Our main contributions are as follows:

\begin{itemize}
  \item   	\emph{Gaussian-integer formulation}---We put forth a Gaussian-integer formulation for the complex linear PNC mapping in the finite field of Gaussian integer, where  $\alpha,\beta, w_A, w_B\in \mathbb{Z}[i]/q$, where $q$ is a Gaussian-integer prime. Compared with the vector formulation in [9], our Gaussian-integer formulation yields more choices of signal constellations for use in complex linear PNC. Specifically, the complex linear PNC in \cite{yangtwc} is a subset of the complex linear PNC here: specifically, the vector formulation in \cite{yangtwc} is equivalent to our Gaussian-integer formulation with $q$ being limited to a real prime; in general, $q$ can be a complex prime in our Gaussian-integer formulation, yielding additional signal constellations that can be used in  complex linear PNC mappings. In this paper, we also recast linear PNC using the coset theory to uncover the isomorphism among different linear PNC mappings. Beyond the mapping arithmetic in \cite{yanglet,yangtwc,long}, the coset theory offers us with a new angle to understand the principle of linear PNC mapping.
  \item   	\emph{Characteristic difference}---We put forth the concept of characteristic difference that is fundamental to the study of optimal PNC mapping and the minimum distance between constellation points that determines the SER performance of $w_N$. Specifically, a \emph{characteristic difference} is the difference between two distinct joint symbols, $(\delta^{char}_A, \delta^{char}_B)=(w_A, w_B)-(w'_A, w'_B)$, such that there is no common Gaussian-integer factor between $\delta^{char}_A$ and $ \delta^{char}_B$  (i.e.,  $\gcd(\delta^{char}_A, \delta^{char}_B)={\rm unit}$). Given a set of joint symbols, $\mathcal{W}_{(A,B)}=\{(w_A,w_B)|w_A,w_B\in \mathbb{Z}[i]/q\}$, there is a corresponding set of characteristic differences encompassing all possible characteristic differences under all possible joint symbols. For a given channel-gain ratio  $\eta=\frac{h_A}{h_B}$, where $h_A$ and $h_B$ are the complex channel gains from nodes A and B to relay R respectively, the minimum distance between any two constellation points in the received overlapped signals, $l_{\min}$, is given by  the  particular characteristic difference that yields the minimum  $|\eta\delta^{char}_A+\delta^{char}_B|$. The optimal PNC mapping $(\alpha_{opt},\beta_{opt})$ for that $\eta$ is the mapping that maps two pairs of symbols $(w_A,w_B)$  and  $(w'_A, w'_B)$ separated by that $(\delta^{char}_A,\delta^{char}_B)$  to the same NC symbol (i.e.,  $w_N=\alpha_{opt}w_A+\beta_{opt}w_B=\alpha_{opt}w'_A+\beta_{opt}w'_B$). Hence, there is no need to distinguish between  the constellations points corresponding to   $(w_A,w_B)$  and  $(w'_A, w'_B)$   as far as the decoding of $w_N$  is concerned. As a result, $l_{\min}$  is not a concern. What matters to SER performance is the minimum distance  $d^{(\alpha_{opt},\beta_{opt})}_{\min}$ between two pairs of symbols  $(w_A,w_B)$  and  $(w''_A, w''_B)$   mapped to different NC symbols under  $(\alpha_{opt},\beta_{opt})$.  For complex linear PNC, \emph{characteristic difference} is more convenient for the identification of the optimal PNC mapping and the study of  $d^{(\alpha_{opt},\beta_{opt})}_{\min}$ than the \emph{reference symbol} used in \cite{long}, which was devised for the study of real linear PNC.
  \item   	\emph{Voronoi-region characterization of optimal PNC mapping}---For a global understanding of $l_{\min}$ and $d^{(\alpha_{opt},\beta_{opt})}_{\min}$ for all $\eta$, we investigate how the complex plane of $\eta$ can be divided into different Voronoi regions. Associated with the $\eta$  within each Voronoi region is a characteristic difference $(\delta^{char}_A, \delta^{char}_B)$  that determines the $l_{\min}$  within that region, and an optimal PNC mapping that causes $l_{\min}$  to be not a performance concern, as explained in the previous paragraph.  We developed a systematic approach to identify the $d^{(\alpha_{opt},\beta_{opt})}_{\min}$ for all $\eta$ within a Voronoi region by considering the characteristic differences associated with the Voronoi regions adjacent to it.
\end{itemize}

The remainder of this paper is organized as follows. Section II overviews prior related work. Section III describes the general idea of complex linear PNC and raises the key outstanding problems. Section IV presents the advantages of the Gaussian-integer formulation over the vector formulation in complex linear PNC systems. Section  V characterizes the optimal PNC mappings for $\eta$  at which $l_{\min}=0$ and identifies the $d^{(\alpha_{opt},\beta_{opt})}_{\min}$ for these $\eta$  after the optimal PNC mappings.   Section VI considers the overall complex plane of $\eta$  and partitions it into different Voronoi regions. In particular, we show in Section VI that the continuum of $\eta$  within each Voronoi region has the same optimal PNC mapping. Importantly, we give a systematic approach to finding the $d^{(\alpha_{opt},\beta_{opt})}_{\min}$ for the $\eta$  within each Voronoi region.

\section{Related Work}
 In the previous few paragraphs, we have reviewed prior work on linear PNC that is most related to our work in this paper. Here, we review other related work.

\emph{Nonlinear PNC}: In nonlinear PNC systems, the NC mapping at the relay cannot be expressed as a linear weighted sum of the symbols transmitted from the end nodes. A representative work on nonlinear PNC is \cite{koi}. Based on an exclusive law to avoid ambiguity in the decoding of NC symbols at the relay, \cite{koi} made use of the closest-neighbor clustering principle (corresponding to mapping constellation points of two superimposed symbols separated by $l_{\min}$ to the same NC symbol in this paper) to map the superimposed symbols of two QPSK symbols of two users to NC symbols in $5$QAM constellation at the relay.

Nonlinear PNC mapping based on Latin square was proposed in \cite{latin}. Here, the row of the Latin square corresponds to the symbols of one node, and the column represents the symbols of the other node. Entry $(i,j)$ of the Latin square contains the NC symbol mapped to symbol $i$ and symbol $j$ of the two users. The exclusive law of PNC mapping is satisfied by the Latin square's constraint: an NC symbol appears once and only once in each row and in each column.  The study of Latin-square nonlinear PNC in \cite{latin} focused on low-order $M$-PSK (the end nodes transmit $M$-PSK signals), and the extension to high-order modulations requires high-order Latin squares. By contrast, as we will show, our Gaussian-integer formulation for linear PNC mapping is scalable with the NC operation with various high-order modulations such as $q$-PAM in \cite{yanglet,yangtwc,long} and complex modulations in this paper. In particular, for higher-order modulations, the Gaussian-integer formulation only requires selecting the optimal  coefficients $(\alpha, \beta)$   among a larger set of non-zero Gaussian integers.


As far as we know, how to analyze the minimum distances that characterize the decoding performance of $w_N$ (i.e., what is referred to as $d_{\min}$ in this paper) is still an open problem for Latin-square nonlinear PNC.  For complex linear PNC, on the other hand, as will be shown in this paper, we can explicitly formulate the optimal NC mapping for arbitrary channel gains and characterize the associated minimum distances. Specifically, our paper makes use a Voronoi-region analysis to characterize the optimal NC mapping, and in doing so, we find a systematic approach to identify the minimum distances that affect decoding performance of $w_N$.

\emph{Channel-Coded Linear PNC}: In channel-coded linear PNC systems, the two end nodes employ channel coding to encode the transmitted symbols to improve communication reliability. In general, channel-coded PNC can operate in two different ways: link-by-link or end-to-end. For end-to-end channel-coded PNC, the relay is oblivious of the channel coding employed by the two end nodes, and the PNC mapping at the relay is the same as that for nonchannel-coded PNC. Specifically, the relay applies PNC mapping on a symbol-by-symbol basis in both cases. It is at the end nodes after self-information is removed that channel decoding is performed.

For link-by-link channel-coded PNC, the relay is aware of the channel coding employed by the two end nodes (specifically, the relay knows the codebooks used by the two end nodes), and the relay can exploit the correlations among the symbols within each of the channel-coded packets to further improve the accuracy of PNC decoding/mapping.  The study of channel-coded PNC systems also originated from low-order modulations such as BPSK \cite{zhang,yang3}, and then evolved to high-order modulations in search of higher   throughput in the high SNR regime \cite{lei,erez,cf,hern,chen,naz1,nam}.

The linear PNC studied in this paper falls into class of nonchannel-coded PNC, and it can be naturally integrated into end-to-end channel-coded PNC. Compared with link-by-link channel-coded PNC, end-to-end channel-coded PNC is simpler to operate, at the expense of performance.

\section{Complex Linear PNC in $\mathbb{Z}[i]/q$}\label{sec:syn}

 \subsection{Choosing Representative Elements of $\mathbb{Z}[i]/q$ as Transmitted Symbols}

\begin{figure}[t]
  \centering
        \includegraphics[height=0.3\columnwidth]{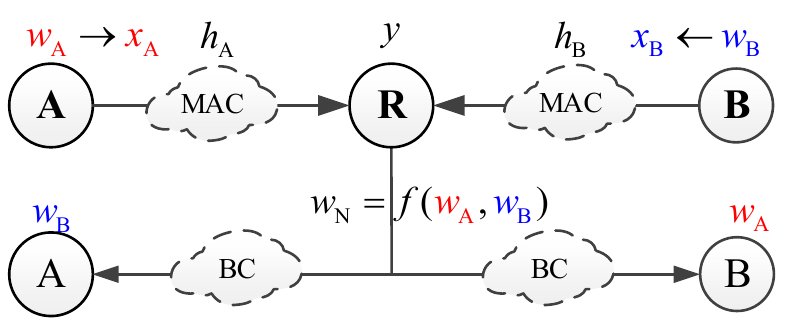}
       \caption{System model of a TWRN.}
        \label{fig:sym}
\end{figure}

Fig.\ref{fig:sym} shows a two-way relay network (TWRN) where nodes A and B communicate with each other via a relay R. In our system model, all nodes (A, B, and R) operate in the half-duplex mode, and each node has single antenna. We assume that there is no direct link between nodes A and B.

\begin{figure}[t]
  \centering
        \includegraphics[height=0.5\columnwidth]{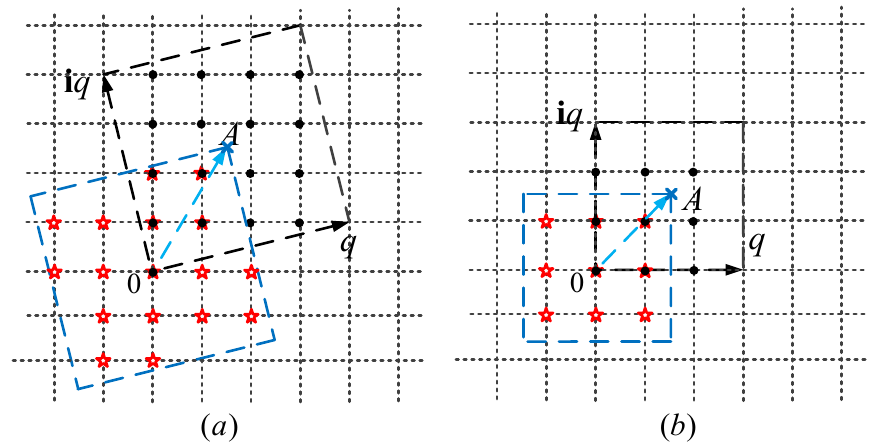}
       \caption{The congruence class of $\mathbb{Z}[i]/q$ with (a) $q=4+i$   (b) $q=3$, where red stars are the representative elements of $\mathbb{Z}[i]/q$ by \emph{Definition \ref{def:res}}, and black dots within the square formed by $q$ and $iq$ are a congruence class of $\mathbb{Z}[i]/q$.}
        \label{fig:suf}
\end{figure}

Nodes A and B send complex symbols, $w_A$ and $w_B$, simultaneously to relay R. We assume that $w_A$ and $w_B$  is selected from $\mathbb{Z}[i]/q$ (i.e., modulo $q$ in Gaussian integers), where $q$ is a Gaussian prime. Note that $q$ can be complex and that a real prime integer may not be a Gaussian prime \cite{JBF}. Given prime $q$, $\mathbb{Z}[i]/q$ is therefore a finite field of order $|q|^2$. If $q$ happens to be also a prime integer (i.e., $q= 3 ({\rm mod} ~4)$), then $\mathbb{Z}[i]/q=\big\{a+bi|a,b \in\{\frac{1-q}{2},\ldots,0,\ldots, \frac{q-1}{2}\}\big\}$. An example of a complex $q$ is $q=1+2i$, for which $\mathbb{Z}[i]/q\in\{-1,1,0,i,-i\}$. Formally, for arbitrary Gaussian prime $q$, we identify the elements in $\mathbb{Z}[i]/q$ as follows  (note: the physical meaning of the \emph{Definition \ref{def:res}} will be clearer if the reader refers to the  two illustrating examples in Fig. \ref{fig:suf} for $q=4+i$ and $q=3$ while reading the definition):

\begin{definition}[{Residue field of $\mathbb{Z}[i]/q, |q|\geq\sqrt{5}$}]\label{def:res}
  To identify a set of representative elements of the residue field of $\mathbb{Z}[i]/q$ when $|q|\geq\sqrt{5}$, we set up a coordinate system on the 2-D complex plane with the basis $(x,y)=(x^R+x^I i,y^R+y^I i )= (\frac{q}{|q|}, \frac{q}{|q|}i)$. Given a Gaussian integer $w=w^R+iw^I\in\mathbb{Z}[i]$ with basis $(1,i)$, we define a new coordinate system of $w$ with the  basis $(x,y)=(\frac{q}{|q|}, \frac{q}{|q|}i)$ as follows:
   \begin{align}\label{dre:1}
   \nonumber w^{x} &=  w^R x^R+w^I x^I, \\
    w^{y} &=  w^R y^R+w^I y^I = -w^R x^I +w^I x^R.
  \end{align}
  To be concise, we rewrite \eqref{dre:1} as
   \begin{align}\label{dre:2}
\nonumber \left[\begin{array}{c}
   w^{x}\\
     w^{y}
  \end{array}
  \right] &=\left[\begin{array}{cc}
   x^R & x^I\\
    -x^I & x^R
  \end{array}
  \right]\left[\begin{array}{cc}
   w^R \\
    w^I  \end{array}
  \right]\\&=\frac{1}{|q|}\left[\begin{array}{cc}
   q^R & q^I\\
    -q^I & q^R
  \end{array}
  \right]\left[\begin{array}{cc}
   w^R \\
    w^I  \end{array}
  \right].
    \end{align}
 We say that $w\in \mathbb{Z}[i]/q$ if and only if $|w^x|,|w^y|<|q|/2$ in the new coordinate system. In the context of our communication system model, such a $w$ is said to be a \emph{valid} symbol.

\end{definition}\rightline{$\blacksquare$}

\begin{remark}\label{rem:sqrt2}
For $|q|<\sqrt{5}$, the only possible Gaussian prime $q$ is $|q|=\sqrt{2}$,  for which  the only possible  $q$ are $\{1+i,1-i,-1-i,-1+i\}$. \emph{Definition \ref{def:res}} above applies to finding the representative elements in $\mathbb{Z}[i]/q$  with $|q|=\sqrt{2}$. For $|q|=\sqrt{2}$, we will let the representative elements be $\{0,1\}$. In general, the number of representative elements in $\mathbb{Z}[i]/q$ is $|q|^2$ for all $q$.

\end{remark}\rightline{$\blacksquare$}

Note that if $q$ is real  (e.g., in Fig. \ref{fig:suf}(b), $q=3$ corresponds to this case), then the basis $(x,y)$ remains the same as the original basis $(1,i)$, since $x^I=y^R=0$, and $w^x=w^R, w^y=w^I$. In other words, in the real $q$ case, the basis consists of the unit vector along the real line and the unit vector along the imaginary line. If $q$ is complex  (e.g., in Fig. \ref{fig:suf}(a), $q=4+i$ corresponds to this case), the new basis $(x,y)$  is a rotation of $(1,i)$ according to $q$. Whether $q$ is real or complex, we require the magnitude of $w^x$ and $w^y$ to be strictly less than $|q|/2$ (for $|q|\geq\sqrt{5}$). With reference to the two illustrating examples in Fig. \ref{fig:suf}, this means the representative elements must be within the prescribed squares centered around the origin (i.e., the red lattice points). Mathematically, this set of representative elements of $\mathbb{Z}[i]/q$  is not the only choice. Since congruence modulo $q$ is an equivalence relation, any  $|q|^2$ elements selected from the $|q|^2$ congruence classes can be used to represent $\mathbb{Z}[i]/q$ \cite{JBF}. In this paper, we choose the representative elements of $\mathbb{Z}[i]/q$ by \emph{Definition \ref{def:res}} to serve as the transmitted symbols in our communications systems, since each of such a representative element is the element with the smallest magnitude within its congruence class (i.e., the transmitted power of the corresponding symbol is the smallest). In particular, our definition requires an element of $\mathbb{Z}[i]/q$ to lie within the zero-centered square of side length $|q|$ with orientation aligned with the directions as indicated by the basis $(x,y)$.

\begin{proposition}\label{pro:scw}
 Consider a Gaussian prime $q$ that defines the valid symbols in $\mathbb{Z}[i]/q$ by \emph{Definition \ref{def:res}}, where $|q|\geq\sqrt{5}$. A sufficient condition for $w\in\mathbb{Z}[i]/q$  is $|w|<|q|/2$.
\end{proposition}

\begin{IEEEproof}[Proof of Proposition \ref{pro:scw}]
Consider the basis $(x,y)$ in \emph{Definition \ref{def:res}}, we have
 \begin{align}
|w|=\sqrt{(w^R)^2+(w^I)^2}=\sqrt{(w^x)^2+(w^y)^2}<|q|/2,
  \end{align}
 since norms are invariant under basis transformation. This implies $|w^x|, |w^y|<|q|/2$. Therefore, by \emph{Definition \ref{def:res}}, if $w\in\mathbb{Z}[i]/q$  if $|w|<|q|/2$.

\end{IEEEproof}

At node $m$, $m\in \{A,B\}$, a modulated symbol $x_m$ is given by $x_m=w_m/\mu$, where $\mu$ is a power normalization constant such that $E(|x_m|^2)=1$. If the Gaussian prime $q$ is a prime integer, the bijective mapping from $w_m$ to $x_m$ is equivalent to $q^2$-level quadrature amplitude modulation (QAM).

With respect to the TWRN system model shown in Fig. \ref{fig:sym}, in the MAC phase, nodes A and B transmit $x_A$ and $x_B$ simultaneously. At relay R, we assume that the arrival times of the symbols from nodes A and B are aligned, so that the received signal at the relay is given by
\begin{align}\label{eqn:yr}
\nonumber   y_R = h_A \sqrt{P} x_A  +  h_B \sqrt{P} x_B +z\\
    = \frac{\sqrt{P}}{\mu} (h_A w_A  +  h_B w_B) +z,
    \end{align}
where $h_m$  is the complex channel coefficient between node $m$, $m\in \{A,B\}$, and the relay; and $z$ is a complex additive white Gaussian noise with zero mean and variance of $\sigma^2=N_0$. We assume $h_A$ and $h_B$ are available at relay R, but not at nodes A and B.  In addition, nodes A and B transmit with equal power $P$.

\subsection{General Idea of Complex Linear PNC}\label{sec:GI}

Upon receiving ${y}$, relay R adopts a linear PNC strategy that tries to derive a network-coded symbol from $y$. To understand the details, let us first imagine that $(w_A,w_B)$ were perfectly known to relay R.  Relay R then encodes $(w_A,w_B)$ to a complex network-coded symbol. We refer to $(w_A,w_B)$  as a joint symbol. The overall set of joint symbols is  $\mathcal{W}_{(A,B)}=\{(w_A,w_B)|w_A,w_B\in \mathbb{Z}[i]/q\}$, and $|\mathcal{W}_{(A,B)}|=|q|^4$.

Under linear network coding, a joint symbol $(w_A,w_B)\in \mathcal{W}_{(A,B)}$ is mapped to an NC symbol:
\begin{align}\label{eqn:wn}
 w_N^{(\alpha,\beta)}\triangleq f_N^{(\alpha,\beta)}(w_A,w_B)\triangleq \alpha w_A+\beta w_B  ~({\rm mod}~ q),
\end{align}
where  $(w_A,w_B)\in \mathbb{Z}[i]/q$,  $\alpha,\beta\in \mathbb{Z}[i]/q\backslash\{0\}$. In this paper, we mark equations in which the multiplications and additions are finite-field operations in $\mathbb{Z}[i]/q$ by putting the notation $({\rm mod} ~q)$ right after the equations, such as in \eqref{eqn:wn}. For equations in which the multiplications and additions are not finite-field operations, there will be no $({\rm mod} ~q)$ after the equations. Since the NC mapping of \eqref{eqn:wn} is operated in $\mathbb{Z}[i]/q$, we refer to it as the \emph{Gaussian-integer formulation}. The advantage of Gaussian-integer formulation over the vector formulation in \cite{yangtwc} will be  elaborated in  {Section \ref{sec:adv}}.

In \eqref{eqn:wn}, since the field $\mathbb{Z}[i]/q$ is closed under addition and multiplication, $w_N^{(\alpha,\beta)}\in \mathbb{Z}[i]/q$. We refer to $\alpha,\beta$ as the NC mapping coefficients.
We denote the set of all possible NC symbols $w_N^{(\alpha,\beta)}$ by $\mathcal{W}_N^{(\alpha,\beta)}$. Given that $w_A,w_B\in \mathbb{Z}[i]/q$, it is easy to see from \eqref{eqn:wn} that $\mathcal{W}_N^{(\alpha,\beta)}=\mathbb{Z}[i]/q$, and that $|\mathcal{W}_N^{(\alpha,\beta)}|=|q|^2$.

Now, in the actual system, what is known to relay R is $y$ (which includes the noise $z$) and not the joint symbol $(w_A,w_B)$.  Conceptually, the decoding process at relay R can be thought of as a two-step process. The first step consists of finding the most likely joint symbol $(w_A,w_B)$ from $y$. The second step consists of the NC mapping as expressed in \eqref{eqn:wn}. Note that the decoded $w_N^{(\alpha,\beta)}$ can still be correct even if the decoding of $(w_A,w_B)$ is wrong. Specifically, let $(w_A,w_B)$  be the actual transmitted joint symbols by nodes A and B, and let  $({w}'_A, {w}'_B)$ be the decoded joint symbol. As long as $ f_N^{(\alpha,\beta)}(w_A,w_B )= f_N^{(\alpha,\beta)}({w}'_A, {w}'_B)$, the decoded NC symbol $w_N^{(\alpha,\beta)}$ is still correct. Thus, in general, the decoding error rate of $w_N^{(\alpha,\beta)}$ is smaller than that of  $(w_A,w_B)$. The goal of relay R is to find the NC coefficients  $(\alpha,\beta)$ that minimize the decoding error rate of $w_N^{(\alpha,\beta)}$.

Returning to Fig. \ref{fig:sym}, in the BC phase, relay R broadcasts the decoded $w_N^{(\alpha,\beta)}$ to nodes A and B. If the decoding of $w_N^{(\alpha,\beta)}$  at the relay is correct and the transmission of $w_N^{(\alpha,\beta)}$ in the broadcast phase is error-free, then node A can recover the message $w_B$ with the knowledge of $(\alpha,\beta)$, as follows:
\begin{align}\label{eqn:bro}
\beta^{-1} (w_N^{(\alpha,\beta)}-\alpha w_A )=\beta^{-1} \beta w_B=w_B  ~({\rm mod}~ q),
\end{align}
where $\beta^{-1}$ is the multiplicative inverse of $\beta$ in $\mathbb{Z}[i]/q$, i.e. $\beta^{-1} \beta=1({\rm mod} ~q)$. Note that the inverse $\beta^{-1}$ for the nonzero $\beta$ exists since $\mathbb{Z}[i]/q$ is a field. Similarly, node B can recover $w_A$ if $\alpha^{-1}$ exists in $\mathbb{Z}[i]/q$. The recovery of $w_m$ at each node is feasible if and only if both $\alpha$ and $\beta$ are nonzero in $\mathbb{Z}[i]/q$. As long as $q$ is a Gaussian prime (and therefore $\mathbb{Z}[i]/q$ is a field), the complex NC mapping under nonzero $\alpha$ and $\beta$ is \emph{valid}, in the sense that node A (B) can recover $w_B$ $(w_A)$ using $w_N^{(\alpha,\beta)}$ and $w_A$ $(w_B)$.

Consider two distinct joint symbols $(w_A,w_B),(w'_A,w'_B)\in\mathcal{W}_{(A,B)}$. The difference between these two distinct joint symbols is defined to be
\begin{align}\label{eqn:del}
(\delta_A, \delta_B)\triangleq (w_A,w_B)-(w'_A,w'_B).
\end{align}
We refer to such a $(\delta_A, \delta_B)$ as a \emph{difference pair}.  Note that \eqref{eqn:del} is not a finite-field equation: the regular integer subtraction is involved, not the finite-field subtraction. We define the set that collects all possible $(\delta_A, \delta_B)$   induced by  two distinct joint symbols in $\mathcal{W}_{(A,B)}$, as follows:
\begin{eqnarray}\label{eqn:dels}
\nonumber  \Delta  \triangleq  \big\{(\delta_A, \delta_B)\big|(\delta_A, \delta_B)=(w_A,w_B)-(w'_A, w'_B), \\ (w_A,w_B)\neq(w'_A,w'_B), (w_A,w_B ),(w'_A,w'_B) \in \mathcal{W}_{(A,B)}\big\}.
\end{eqnarray}
Note that unlike $w_A$ and $w_B$, $\delta_A, \delta_B$ may not be elements of $\mathbb{Z}[i]/q$ although their possible values depend on $\mathbb{Z}[i]/q$.

We also refer to $\delta_A$ or $\delta_B$ as a \emph{difference}. With respect to $\Delta$, we define the set that collects all possible  $\delta_A$ or $\delta_B$ as $\Lambda$. Note that the element in $\Lambda$ can be zero.

Given $(\delta_A, \delta_B)\in\Delta$, we define the associated mod-$q$ difference pair as follows:
\begin{align}\label{eqn:delq}
(\delta_A^{(q)},\delta_B^{(q)})\triangleq  \big(\delta_A ({\rm mod} ~q), \delta_B ({\rm mod} ~q)\big),
\end{align}
where $(\delta_A^{(q)},\delta_B^{(q)})\in \mathbb{Z}^2[i]/q$  (e.g., if $(\delta_A, \delta_B)=(8,  9i)$, then the corresponding $(\delta_A^{(7)}, \delta_B^{(7)}) = (1, 2i)$).

\begin{proposition}\label{pro:2}
 An NC mapping $f_N^{(\alpha, \beta)}(w_A,w_B)$ maps two distinct joint symbols $(w_A,w_B)$ and $(w'_A,w'_B)$ in $\mathcal{W}_{(A,B)}$ to the same NC symbol if and only if $\alpha\delta_A^{(q)} +\beta\delta_B^{(q)}=0 ~({\rm mod} ~q)$.
\end{proposition}
\begin{IEEEproof}[Proof of Proposition \ref{pro:2}]
An NC mapping under $(\alpha, \beta)$ maps $(w_A,w_B)$ and $(w'_A,w'_B)$ to the same NC symbol if and only if $f_N^{(\alpha, \beta)}(w_A,w_B)=f_N^{(\alpha, \beta)}(w'_A,w'_B)$. Thus,
\begin{align}\label{eqn:pro11}
\alpha w_A+\beta w_B=\alpha w'_A+\beta w'_B  ~({\rm mod} ~q).
\end{align}
Equivalently, we can rewrite \eqref{eqn:pro11} as
\begin{align}\label{eqn:pro12}
\alpha(w_A-w'_A )+ \beta(w_B-w'_B)=\alpha \delta_A^{(q)}+ \beta\delta_B^{(q)}=0 ~({\rm mod} ~q).
\end{align}
\end{IEEEproof}

In this paper, an NC mapping  $(\alpha,\beta)$   is said to \emph{cluster} $(\delta_A,\delta_B)$ if and only if $\alpha\delta_A^{(q)} +\beta\delta_B^{(q)}=0 ~({\rm mod} ~q)$, where $\delta_A^{(q)}=\delta_A ({\rm mod}~q)$ and $\delta_B^{(q)}=\delta_B ({\rm mod}~q)$. This is a quick way of saying $(\alpha,\beta)$  map any two joint symbols $(w_A,w_B)$ and $(w'_A,w'_B)$ satisfying  $(\delta_A^{(q)},\delta_B^{(q)})= (w_A,w_B)-(w'_A,w'_B) ~({\rm mod} ~q)$  to the same NC symbol.

\begin{proposition}\label{pro:3}
Consider two distinct joint symbols $(w_A,w_B),(w'_A,w'_B)\in \mathcal{W}_{(A,B)}$. Suppose that we want to find an NC mapping $f_N^{(\alpha, \beta)}$ such that $f_N^{(\alpha, \beta)}(w_A,w_B)=f_N^{(\alpha, \beta)}(w'_A,w'_B)$. Then, such an NC mapping exists if and only if $w_A\neq w'_A$ and $w_B\neq w'_B$.
\end{proposition}
\begin{IEEEproof}[Proof of Proposition \ref{pro:3}]
Since $(w_A,w_B)$ and $(w'_A,w'_B)$ are distinct, we cannot have $w_A=w'_A$ and $w_B=w'_B$ at the same time. Without loss of generality (w.l.o.g.), we will show that an NC mapping with  $f_N^{(\alpha, \beta)}(w_A,w_B)=f_N^{(\alpha, \beta)}(w'_A,w'_B)$ is not possible if $w_A=w'_A$ and $w_B\neq w'_B$, and is possible if $w_A\neq w'_A$ and $w_B\neq w'_B$.

If $w_A=w'_A$ and $w_B\neq w'_B$, we have $\delta_A=0$ and $\delta_B\neq0$. Suppose that  $f_N^{(\alpha, \beta)}(w_A,w_B)=f_N^{(\alpha, \beta)}(w'_A,w'_B)$  were possible. Then by \emph{Proposition \ref{pro:2}}, we have $\beta\delta_B^{(q)}=0 ~({\rm mod} ~q)$. Since $\beta\in \mathbb{Z}[i]/q\backslash\{0\}$, there exists $\beta^{-1}$ such that $\beta^{-1}\beta\delta_B^{(q)}=\delta_B^{(q)}=0 ~({\rm mod} ~q)$. That is, $\delta_B=uq$ for some $u\in \mathbb{Z}[i]$.
According to \emph{Definition \ref{def:res}}, we have $\delta_B=[\frac{w_A^x-w'^x_A}{|q|} +i \frac{w_A^y-w'^y_A}{|q|}]q$ and $|w_A^x|,|w'^x_A|,|w_A^y|,|w'^y_A|<|q|/2$. Thus, $\frac{w_A^x-w'^x_A}{|q|}$  and $\frac{w_A^y-w'^y_A}{|q|}$  cannot be non-zero integers and the only possibility for $u$ is $u=0$.  That is, $\delta_B=0$, contradicting the supposition that $w_B\neq w'_B$. Thus, an NC mapping is not possible when $(w_A,w_B)$ and $(w'_A,w'_B)$ are distinct and $w_A=w'_A$.

Next, we prove that we can find $(\alpha,\beta)$ such that  $f_N^{(\alpha, \beta)}(w_A,w_B)=f_N^{(\alpha, \beta)}(w'_A,w'_B)$ if $w_A\neq w'_A$ and $w_B\neq w'_B$.  In this case, $\delta_A^{(q)},\delta_B^{(q)}\neq0 ({\rm mod} ~q)$. This means the inverses $(\delta_A^{(q)})^{-1}$ and $(\delta_B^{(q)})^{-1}$ exist if $q$ is a Gaussian prime. We need to find a pair, $\alpha,\beta\neq 0 ({\rm mod} ~q)$, such that $\alpha\delta_A^{(q)}+\beta\delta_B^{(q)}=0 ({\rm mod} ~q)$.  A possible pair is $(\alpha,\beta)=(-(\delta_A^{(q)})^{-1}\delta_B^{(q)}, 1) ~({\rm mod} ~q)$.

\end{IEEEproof}

Motivated by \emph{Proposition \ref{pro:3}}, the definition below specifies a difference pair of two joint symbols that can be mapped to the same NC symbol by NC mapping:

\begin{definition}[{NC-validity of $(\delta_A,\delta_B)$}]\label{def:ncd}
 A difference pair $(\delta_A,\delta_B)$ is said to be an \emph{NC-valid difference pair} if and only if (1) $(\delta_A,\delta_B)\in \Delta$ and (2) $\delta_A\neq0, \delta_B\neq0$. That is, $(\delta_A,\delta_B)$ is NC-valid only if there is an NC mapping $(\alpha,\beta)$ that maps two distinct joint symbols $(w_A, w_B)$ and $(w'_A, w'_B)$ separated by $(\delta_A,\delta_B)$  to the same NC symbol.
\end{definition}\rightline{$\blacksquare$}

\begin{definition}[{Isomorphism of NC mappings}]\label{def:iso}
 Two NC mappings $f_N^{(\alpha, \beta)}: \mathcal{W}_{(A,B)}\rightarrow \mathcal{W}_N^{(\alpha,\beta)}$ and $f_N^{(\alpha', \beta')}: \mathcal{W}_{(A,B)}\rightarrow \mathcal{W}_N^{(\alpha',\beta')}$ are said to be \emph{isomorphic} if for any two distinct joint symbols $(w_A,w_B),(w'_A,w'_B)\in \mathcal{W}_{(A,B)}$, $f_N^{(\alpha, \beta)}(w_A,w_B)=f_N^{(\alpha, \beta)}(w'_A,w'_B)$ if and only if $f_N^{(\alpha', \beta')}(w_A,w_B)=f_N^{(\alpha',\beta')}(w'_A,w'_B)$.
\end{definition}\rightline{$\blacksquare$}

Given an NC mapping $f_N^{(\alpha, \beta)}$, we can always find another NC mapping isomorphic to $f_N^{(\alpha, \beta)}$ with a simpler expression, as stated in \emph{Proposition \ref{pro:4}} below.

\begin{proposition}\label{pro:4}
 Given any NC mapping $f_N^{(\alpha, \beta)}:\mathcal{W}_{(A,B)}\rightarrow \mathcal{W}_N^{(\alpha,\beta)}$ of the form $f_N^{(\alpha, \beta)}(w_A,w_B)=\alpha w_A+\beta w_B ~({\rm mod} ~q)$ with $\beta\in \mathbb{Z}[i]/q\backslash\{0\}$, the NC mapping $f_N^{(\alpha', 1)}:\mathcal{W}_{(A,B)}\rightarrow \mathcal{W}_N^{(\alpha',1)}$, where $\alpha'=\beta^{-1} \alpha ~({\rm mod} ~q)$, is an   isomorphic NC mapping.
\end{proposition}
\begin{IEEEproof}[Proof of Proposition \ref{pro:4}]
Since $\beta\neq 0$, $\beta^{-1}$ exists. We have
\begin{align}\label{eqn:pro41}
\nonumber &\beta^{-1}f_N^{(\alpha, \beta)}(w_A,w_B)=\beta^{-1} (\alpha w_A+\beta w_B)\\
&=\beta^{-1} \alpha w_A+w_B= f_N^{(\alpha',1)}(w_A,w_B)  ~({\rm mod} ~q).
\end{align}

From \eqref{eqn:pro41}, since $\beta^{-1}$ is non-zero, we can see that for two distinct $(w_A,w_B),(w'_A,w'_B)\in \mathcal{W}_{(A,B)}$, $f_N^{(\alpha, \beta)}(w_A,w_B)=f_N^{(\alpha, \beta)}(w'_A,w'_B)$ if and only if $f_N^{(\alpha',1)}(w_A,w_B)=f_N^{(\alpha',1)}(w'_A,w'_B)$.

\end{IEEEproof}

\begin{definition}[{Clustered difference pairs}]\label{def:cd}
 We  refer to the set of NC-valid  $(\delta_A,\delta_B)$ clustered by $(\alpha,\beta)$ as its \emph{clustered-difference set}:
\begin{align}\label{eqn:cds}
\nonumber&\Delta_{(\alpha,\beta)}=\big\{(\delta_A,\delta_B)\in\Delta\big| \\ & \alpha(\delta_A ({\rm mod} ~q)) +\beta(\delta_B ({\rm mod} ~q))=0~({\rm mod} ~q)\big\}.
\end{align}
We refer to the elements in $\Delta_{(\alpha,\beta)}$ as the \emph{clustered difference pairs}.
\end{definition}\rightline{$\blacksquare$}

The significance of studying $\Delta_{(\alpha,\beta)}$  lies in that two joint symbols, $(w_A,w_B )$ and $(w'_A,w'_B)$, separated by the clustered difference pair $(\delta_A,\delta_B)\in\Delta_{(\alpha,\beta)}$ will be mapped to the same NC symbol under $(\alpha,\beta)$.

 In \emph{Appendix I}, we use coset theory to interpret the linear PNC mapping in $\mathbb{Z}[i]/q$, uncovering the structure of the isomorphism among different possible PNC mappings $(\alpha,\beta)$. The isomorphism substantially reduces the search space of $(\alpha,\beta)$  when we look for the optimal PNC mapping.

\emph{Appendix I} further deduces that the complex NC mapping $f_N^{(\alpha, \beta)}:\mathcal{W}_{(A,B)}\rightarrow \mathcal{W}_N^{(\alpha,\beta)}$ is a $|q|^2$-to-$1$ mapping. This NC mapping partitions $\mathcal{W}_{(A,B)}$ into $|q|^2$ subsets (i.e., $|q|^2$ cosets), each corresponding to a unique NC symbol, as follows:
\begin{align}\label{eqn:par1}
\nonumber  \mathcal{W}_{(A,B)} (w_N^{(\alpha,\beta)})\triangleq&\big\{(w_A,w_B)\in\mathbb{Z}^2[i]/q\big|\\
& w_N^{(\alpha,\beta)}=f_N^{(\alpha,\beta)} (w_A,w_B)\big\}.
\end{align}
We refer to the  partitioning of $\mathcal{W}_{(A,B)}$ into $|q|^2$ subsets, each with $|q|^2$ elements, as the \emph{NC partitioning}  under $(\alpha,\beta)$.

\subsection{Distance Metrics of Superimposed Constellation at Relay}

Given a pair of $h_A$ and $h_B$, we define a superimposed symbol as
\begin{align}\label{eqn:si}
w_S\triangleq f_S(w_A, w_B)\triangleq h_Aw_A+h_Bw_B.
\end{align}
Furthermore, we refer to the set of all possible $w_S$ as $\mathcal{W}_S$. Since $h_A$ and $h_B$ are selected from the set of all complex numbers, $\mathcal{W}_S\subset \mathbb{C}$.

Each joint symbol $(w_A,w_B)\in\mathcal{W}_{(A,B)}$ corresponds to a superimposed symbol $w_S$. Therefore, an NC mapping $f_N^{(\alpha,\beta)}$ also partitions $\mathcal{W}_S$ into $|q|^2$ subsets, each subset being labeled by a specific $w_N^{(\alpha,\beta)}$ (i.e., elements in a subset are mapped to the same NC symbol). The subset of $\mathcal{W}_S$ associated with a particular NC symbol $w_N^{(\alpha,\beta)}$ can be written as
\begin{align}\label{eqn:ws}
\nonumber & \mathcal{W}_S{(w_N^{(\alpha,\beta)})} \triangleq\big\{w_S\in \mathcal{W}_S\big|\exists(w_A, w_B)\in \mathcal{W}_{(A,B)}:\\
  & w_N^{(\alpha,\beta)} =f_N^{(\alpha,\beta)} (w_A,w_B) ~{\rm and}~ w_S=f_S(w_A, w_B)\big\}.
\end{align}

In the constellation of $\mathcal{W}_S$, the Euclidean distance between any two superimposed symbols $w_S$ and $w'_S$ associated with two distinct joint symbols $(w_A,w_B)$ and $(w'_A,w'_B)$ is given by
\begin{align}\label{eqn:l}
l\triangleq |w_S-w'_S| =|h_A\delta_A+h_B\delta_B|.
\end{align}
We remark that for a particular $(\delta_A,\delta_B)$, $(\epsilon\delta_A, \epsilon\delta_B)$ with $\epsilon\in\{-1, \pm i\}$ has the same distance as $(\delta_A,\delta_B)$ (i.e., multiplying both $\delta_A$ and $\delta_B$ by a unit does not change the distance).

\begin{definition}[{Validity of $\delta_A$ or $\delta_B$}] \label{def:dvds}
$\delta_A$ or $\delta_B$ is said to be a \emph{valid difference} if and only if $\delta_A$ or $\delta_B$ $\in \Lambda$ .
\end{definition}

\begin{definition}[{Distance validity of $(\delta_A,\delta_B)$}] \label{def:dvd}
$(\delta_A,\delta_B)$ is said to be a \emph{distance-valid difference pair} if and only if $(\delta_A,\delta_B) \in \Delta$ .
\end{definition}

\begin{remark}\label{rem:ncd}
 Note that for $(\delta_A,\delta_B)$ to be NC-valid, according to \emph{Definition \ref{def:ncd}}, we need both  $\delta_A\neq0$ and $\delta_B\neq0$  (i.e., $(\delta_A,\delta_B)$ is the difference of two distinct joint symbols that can be mapped to the same NC symbol). On the other hand, for $(\delta_A,\delta_B)$ to be distance-valid, we only need   $\delta_A\neq0$ or $\delta_B\neq0$ (i.e., $(\delta_A,\delta_B)$ corresponds to the difference of two distinct joint symbols, and it makes sense to talk about the distance between the corresponding two superimposed symbols given by \eqref{eqn:l}). Thus, the set of distance-valid difference pairs is a strict superset of the set of NC-valid difference pairs. The elements that are in the former but not in the latter are those in the former with either  $\delta_A=$ or $\delta_B=0$.
\end{remark}

Two distance metrics relevant to decoding errors are defined as follows \cite{yangtwc,long}:
\begin{itemize}
  \item  \emph{Minimum symbol distance} $l_{\min}$
\begin{align}\label{l:lmin}
l_{\min}\triangleq
\mathop{\arg\min}\limits_{\substack{ (w_{A}, w_B) \neq (w'_{A}, w'_B), \\ (w_A, w_B), (w'_A, w'_B) \in \mathcal{W}_{(A,B)}, \\ w_{S}=f_S(w_A, w_B), w'_{S}=f_S(w'_A, w'_B) }} |w_{S}-w'_{S}|.
\end{align}

\item  \emph{Minimum NC-symbol distance} $d^{(\alpha, \beta)}_{\min}$
\begin{eqnarray}\label{l:dmin}
d^{(\alpha, \beta)}_{\min}\triangleq
\mathop{\arg\min}\limits_{\substack{(w_{A}, w_B) \neq (w'_{A}, w'_B), \\  (w_A, w_B), (w'_A, w'_B) \in \mathcal{W}_{(A,B)}, \\ w_{S}=f_S(w_A, w_B), w'_{S}=f_S(w'_A, w'_B),\\ f^{(\alpha, \beta)}_N(w_A, w_B) \neq f^{(\alpha, \beta)}_N(w'_A, w'_B)}}  |w_S-w'_S|.
\end{eqnarray}
\end{itemize}
In other words, $l_{\min}$ is the minimum distance among all pairs of superimposed symbols $w_S$ and $w'_S$ in the superimposed constellation, and it depends on $h_A$ and $h_B$ only. On the other hand, $d_{\min}^{(\alpha,\beta)}$ is the minimum distance among all pairs of superimposed symbols $w_S$ and $w'_S$ in the superimposed constellation that belong to different partitions in \eqref{eqn:ws} (i.e., as far as $d_{\min}^{(\alpha,\beta)}$ is concerned, $w_S$ and $w'_S$ must be associated with different NC symbols). We see that $d_{\min}^{(\alpha,\beta)}\geq l_{\min}$ in general and, unlike $l_{\min}$, $d_{\min}^{(\alpha,\beta)}$ depends on the NC coefficients $(\alpha,\beta)$ as well as $h_A$ and $h_B$.

This paper focuses on the use of a minimum NC-symbol distance  mapping rule at the relay. In the high SNR regime, the SER of decoding NC symbols at the relay is dominated by $d_{\min}^{(\alpha,\beta)}$ \cite{yangtwc,long}, which in turn depends on the NC coefficients $(\alpha,\beta)$. The minimum NC-symbol distance mapping rule, given below, finds the $(\alpha,\beta)$ that maximizes $d_{\min}^{(\alpha,\beta)}$ to minimize SER:
 \begin{align}\label{eqn:optab}
(\alpha_{opt}, \beta_{opt})=\mathop{\arg\max}\limits_{\alpha, \beta \in \mathbb{Z}[i]/q\backslash \{0\}} d^{(\alpha,\beta)}_{\min}.
\end{align}


W.l.o.g., we consider a normalized version of \eqref{eqn:yr} as follows:
\begin{align}\label{eqn:yrn}
\frac{y_R}{h_B} = \eta \sqrt{P}x_A  +  \sqrt{P} x_B +\frac{z}{h_B},
\end{align}
where $\eta=\frac{h_A}{h_B}\in \mathbb{C}$. For simplicity, and w.l.o.g., we assume that $h_B=1$ and thereby $\eta=h_A$. Accordingly, the superimposed symbol in \eqref{eqn:si} is scaled as $w_S=\eta w_A+w_B$ and the Euclidean distance in \eqref{eqn:l} becomes $l=|\eta\delta_A+\delta_B|$.

\begin{figure}[t]
  \centering
        \includegraphics[height=0.43\columnwidth]{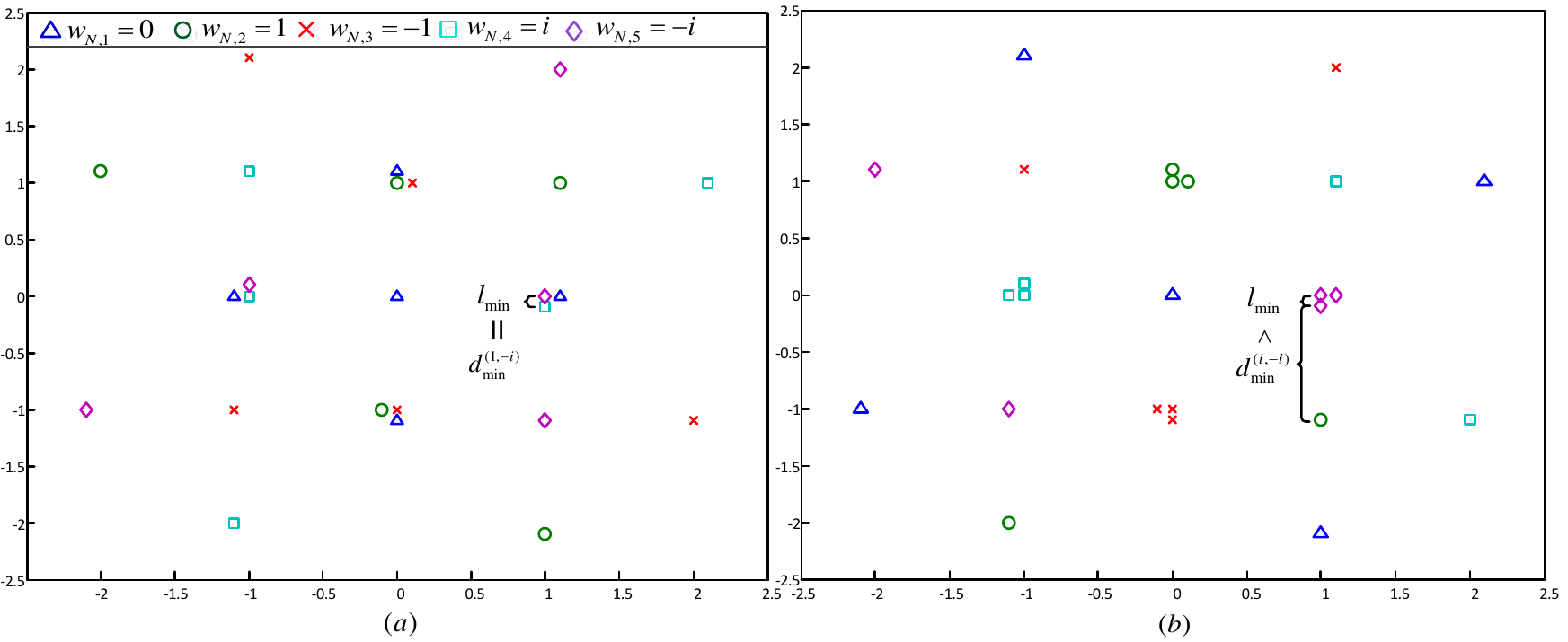}
       \caption{Constellations of superimposed symbols for complex linear PNC in $\mathbb{Z}[i]/(2+i)$ when $(h_A,h_B)=(1.1+i,1)$ and (a) $(\alpha,\beta)=(1,-i)$; (b) $(\alpha,\beta)=(i,-i)$.}
        \label{fig:eg}
\end{figure}

As an illustrating example, let us consider the case of $q=2+i$ and $\eta=1.1+i$. When $q=2+i$, according to \emph{Definition \ref{def:res}}, $w_A,w_B\in \mathbb{Z}[i]/(2+i)=\{0,\pm1,\pm i\}$.  To see the effect of $(\alpha, \beta)$ on $l_{\min}$ and $d_{\min}^{(\alpha, \beta)}$, we plot the constellations of the superimposed symbols based on different $(\alpha, \beta)$ in Fig. \ref{fig:eg}. In Fig. \ref{fig:eg}, we use different shapes to label the superimposed symbols mapped to distinct NC symbols; the superimposed symbols with the same shape are mapped to the same NC symbol under the particular $(\alpha, \beta)$. In Fig. \ref{fig:eg}(a), $(\alpha,\beta)=(1,-i)$; in Fig. \ref{fig:eg}(b), $(\alpha,\beta)=(i,-i)$. We observe that $d_{\min}^{(\alpha, \beta)}$ varies with different $(\alpha, \beta)$  while $l_{\min}$ is constant. In particular, $d_{\min}^{(i,-i)}$ in Fig. \ref{fig:eg}(b) is larger than $d_{\min}^{(1,-i)}$ in Fig. \ref{fig:eg}(a), thus, the NC mapping under $(\alpha,\beta)=(i,-i)$ should have a better SER performance than under $(\alpha,\beta)=(1,-i)$ for the decoding of $w_N^{(\alpha, \beta)}$ in the high SNR regime.

The above example illustrates how $(\alpha, \beta)$ affects $d_{\min}^{(\alpha, \beta)}$. This example also brings out two key problems we aim to attack in this paper.

\underline{Key Problems:}

\begin{enumerate}
  \item[(1)] What is the optimal complex linear PNC mapping $(\alpha_{opt},\beta_{opt})$ given $\eta=h_A/ h_B$?
  \item[(2)] How to characterize $l_{\min}$ and $d_{\min}^{(\alpha, \beta)}$ given $\eta=h_A/ h_B$?
\end{enumerate}

To address problem (1) in a systematic manner,  {Section \ref{sec:adv}} will elaborate the advantage of the Gaussian-integer formulation of the complex linear PNC mapping in $\mathbb{Z}[i]/q$. In  {Section \ref{sec:l0}}, we will identify the optimal PNC mapping $(\alpha_{opt},\beta_{opt})$ for the special $\eta$  at which $l_{\min}=0$ (these $\eta$  are defined as the zero-$l_{\min}$ channel gains). {Section \ref{sec:vor}} then considers the general  $\eta$. In particular,   {Section \ref{sec:vor}}  shows how to divide the complex plane of $\eta$ into different Voronoi regions, with an optimal linear PNC mapping $(\alpha_{opt},\beta_{opt})$ being associated with each Voronoi region (i.e., $(\alpha_{opt},\beta_{opt})$ is optimal for  $\eta$ in the Voronoi region).

To address problem (2),  {Sections \ref{sec:l0}} and  {\ref{sec:vor}} will give systematic approaches to identify $d_{\min}^{(\alpha_{opt},\beta_{opt})}$ at each zero-$l_{\min}$ channel gain and its associated Voronoi region. In particular, in  {Section \ref{sec:vor}}, $d_{\min}^{(\alpha_{opt},\beta_{opt})}$ for a given  $\eta$ can be derived through a ``Voronoi-region analysis''.

\section{The Advantage of Gaussian-Integer Formulation over Vector Formulation}\label{sec:adv}

In this section, we elaborate the relationship between the Gaussian-integer formulation and the vector formulation, and show that the Gaussian-integer formulation gives us more choices of linear PNC mappings with a larger set of signal constellations than the vector formulation in \cite{yangtwc}.

When $q$ is an prime integer, we can also formulate linear PNC mappings in a vector space  in $GF(q)$, as in \cite{yangtwc}. However, as we will see, this formulation has a limitation when such a prime integer $q$ is not a Gaussian prime  (e.g., $q=2=(1+i)(1-i)$ is a prime integer, but not a Gaussian prime since it can be factorized)--this is the reason we only consider Gaussian prime $q$ in this paper. Specifically, in this case, it may not be able to map all the joint symbols separated by the minimum distance $l_{\min}$ to the same NC symbol when $l_{\min}$ is very small (including the case where $\eta$ is such that $l_{\min}=0$), resulting in small $d_{\min}=l_{\min}$.

Overall, this section establishes the following:

\begin{itemize}
  \item  	The \emph{equivalence} between the vector formulation in \cite{yangtwc} under a ``dual mapping'' that minimizes $d_{\min}$ and the Gaussian-integer formulation in this paper when $q$ is both an integer prime and a Gaussian prime;
  \item 	The \emph{limitation} of the vector formulation when $q$ is an integer prime but not a Gaussian prime.
  \item 	The \emph{broader scope} of the Gaussian-integer formulation over the vector formulation when $q$ is a Gaussian prime but not an integer prime.
\end{itemize}

The vector mapping scheme of \cite{yangtwc} that corresponds to \eqref{eqn:wn} is as follows:
 \begin{align}\label{eqn:vec}
\mathbf{w}_N^{(\bm{\alpha},\bm{\beta})}\triangleq f_N^{(\bm{\alpha},\bm{\beta})}(\mathbf{w}_A,\mathbf{w}_B)
\triangleq \bm{\alpha}\mathbf{w}_A+\bm{\beta}\mathbf{w}_B  ~({\rm mod} ~q)
\end{align}
In the above, $\bm{\alpha}=[\alpha_{ij}]_{(2\times2)}$ and $\bm{\beta}=[\beta_{ij}]_{(2\times2)},{i,j\in\{1,2\}}$, are two $2\times 2$ NC mapping matrices. The joint symbol and the NC symbol are expressed in vector form as $(\mathbf{w}_A,\mathbf{w}_B) \triangleq\big((w_A^R,w_A^I)^T, (w_B^R,w_B^I)^T \big)$ and $w_N^{(\bm{\alpha},\bm{\beta})}=(w_N^{(\bm{\alpha},\bm{\beta}),R},w_N^{(\bm{\alpha},\bm{\beta}),I})^T$ respectively, where ${}^R$ and ${}^I$ denote the real and imaginary parts of a complex number respectively. In \cite{yangtwc}, the linear NC mapping in \eqref{eqn:vec} is said to be valid if and only if $\alpha$ and $\beta$ are invertible matrices.

Consider two distinct joint symbols $(\mathbf{w}_A,\mathbf{w}_B)$ and $(\mathbf{w}'_A,\mathbf{w}'_B)$. The difference between these two different joints symbols in the vector formulation is $(\bm{\delta}_A ,\bm{\delta}_B)\triangleq(\mathbf{w}_A,\mathbf{w}_B)-(\mathbf{w}'_A,\mathbf{w}'_B)$. Given a prime integer $q$, $\delta_A^R,\delta_A^I,\delta_B^R,\delta_B^I\in\{-(q-1),\ldots,0,\ldots,(q-1)\}$. The mod-$q$ difference pair is defined as $(\bm{\delta}_A^{(q)},\bm{\delta}_B^{(q)})\triangleq(\bm{\delta}_A ({\rm mod}\  q),\bm{\delta}_B ({\rm mod}\ q))$. Given a valid $(\delta_A,\delta_B)$ in the vector formulation, the associated Gaussian-integer formulation is $(\delta_A,\delta_B)=(\delta_A^R+i\delta_A^I,\delta_B^R+i\delta_B^I)$.

$\bullet$ {Equivalence under dual mapping when $q$ is both an integer prime and a Gaussian prime}

Consider all $(\bm{\delta}_A ,\bm{\delta}_B)$ (or $(\delta_A,\delta_B)$ in Gaussian integer form) that yields $l_{\min}$. To cluster these $(\bm{\delta}_A ,\bm{\delta}_B)$, we show that the NC mapping in the vector formulation under a dual mapping   is equivalent to the Gaussian-integer formulation when $q$, an integer prime, also happens to be a Gaussian prime.


\emph{Dual Mapping of Vector Formulation:}

For the vector formulation, we define the dual of a particular $(\bm{\delta}_A,\bm{\delta}_B)$ as $(\bar{\bm{\delta}}_A,\bar{\bm{\delta}}_B)\triangleq(-\delta_A^I,\delta_A^R ,-\delta_B^I, \delta_B^R )$\cite{yangtwc}. Note that both $(\bm{\delta}_A,\bm{\delta}_B)$ and its dual $(\bar{\bm{\delta}}_A,\bar{\bm{\delta}}_B)$ yield a same distance $l$, since $|h_A (\delta_A^R+i\delta_A^I)+h_B (\delta_B^R+i\delta_B^I)|=|h_A (-\delta_A^I+i\delta_A^R)+h_B (-\delta_B^I+i\delta_B^R)|$. In other words, under a specific $\eta$, if $(\bm{\delta}_A, \bm{\delta}_B)$ yields $l_{\min}$, then so does its dual $(\bar{\bm{\delta}}_A,\bar{\bm{\delta}}_B)$. To maximize  $d_{\min}$, the NC mapping needs to cluster both $(\bm{\delta}_A,\bm{\delta}_B)$ and its dual $(\bar{\bm{\delta}}_A,\bar{\bm{\delta}}_B)$. Suppose that this NC mapping is $(\bm{\alpha},\bm{\beta})=\bigg(\left(
                   \begin{array}{cc}
                     \alpha_1 & \alpha_2 \\
                     \alpha_3 & \alpha_4 \\
                   \end{array}
                 \right)
,\mathbf{I}\bigg)$ such that
 \begin{align}\label{eqn:dual}
  \bigg(\left(
                   \begin{array}{cc}
                     \alpha_1 & \alpha_2 \\
                     \alpha_3 & \alpha_4 \\
                   \end{array}
                 \right),\mathbf{I}\bigg)
                 \left(
                   \begin{array}{cc}
                     \delta_A^{R(q)} & -\delta_A^{I(q)} \\
                     \delta_A^{I(q)} & \delta_A^{R(q)} \\
                     \delta_B^{R(q)} & -\delta_B^{I(q)} \\
                     \delta_B^{I(q)} & \delta_B^{R(q)}
                   \end{array}
                 \right)=\mathbf{0}
                   ~({\rm mod} ~q)
\end{align}
where $\mathbf{I}$ is an identity matrix. For \emph{Proposition \ref{pro:4}} in {Section \ref{sec:syn}} concerning the Gaussian integer formulation, we can find a corresponding \emph{Proposition \ref{pro:4}} for the vector formulation. Specifically, for a valid NC mapping where both $\bm{\alpha}$ and $\bm{\beta}$ are invertible, by isomorphism, we can set $\bm{\beta}=\mathbf{I}$ and so that we only need to  look for an appropriate $\bm{\alpha}$.

The solution of $(\bm{\alpha},\bm{\beta})$  in  \eqref{eqn:dual} is given by
 \begin{align}\label{eqn:duso}
 (\bm{\alpha},\bm{\beta}) =\bigg(\left(
                   \begin{array}{cc}
                     \alpha^R & -\alpha^I \\
                     \alpha^I & \alpha^R \\
                   \end{array}
                 \right),\mathbf{I}\bigg)
\end{align}
where
 \begin{align}\label{eqn:val}
\nonumber \left(\begin{array}{c}
\alpha^R \\
\alpha^I \\
\end{array}
\right)
 &=\big((\delta^{R(q)}_A)^2+(\delta^{I(q)}_A)^2\big)^{-1} \\
  &                 \left(\begin{array}{cc}
 \delta^{R(q)}_A \delta^{R(q)}_B+\delta^{I(q)}_A\delta^{I(q)}_B \\
 \delta^{R(q)}_A \delta^{I(q)}_B-\delta^{I(q)}_A\delta^{R(q)}_B  \\
                   \end{array}\right) ~({\rm mod} ~q)
\end{align}
Note that $(\delta^{R(q)}_A)^2+(\delta^{I(q)}_A)^2({\rm mod} ~q)$ is invertible, as explained below. First, the case of $\delta^{R(q)}_A=\delta^{I(q)}_A=0$ is eliminated because for an NC-valid $({\bm\delta}^{(q)}_A,{\bm\delta}^{(q)}_B)$,   both ${\bm\delta}^{(q)}_A$ and ${\bm\delta}^{(q)}_B$ cannot be zero (see \emph{Definition \ref{def:ncd}}). W.l.o.g., suppose that $\delta^{I(q)}_A\neq 0$ and that $(\delta^{R(q)}_A)^2+(\delta^{I(q)}_A)^2$ is not invertible (i.e., $(\delta^{R(q)}_A)^2+(\delta^{I(q)}_A)^2 = 0 ({\rm mod} ~q)$). We can write $((\delta^{R(q)}_A)(\delta^{I(q)}_A)^{-1})^2=-1({\rm mod} ~q)$. However, this contradicts \emph{Lemma 1} below, which states that $((\delta^{R(q)}_A)(\delta^{I(q)}_A)^{-1})^2=-1({\rm mod} ~q)$ has a solution if and only if $q=1 ({\rm mod} ~4)$; but the $q$ being considered here is a prime integer  as well as a Gaussian prime, which requires  $q=3({\rm mod} ~4)$.

\begin{lemma}[Law of Quadratic Reciprocity]
The congruence $x^2=-1 ({\rm mod} ~q)$ is solvable if and only if $q=1({\rm mod} ~4)$ \cite{qua}.
\end{lemma}

Furthermore, we can verify that $(\bm{\alpha},\bm{\beta})$ in \eqref{eqn:duso} is equivalent to $(\alpha,\beta)=(\alpha^R+i\alpha^I,1)=(-(\delta_A^{(q)})^{-1} \delta_B^{(q)},1)$ in the Gaussian-integer formulation, where $(\delta_A^{(q)})^{-1}=((\delta_A^{R(q)})^2+(\delta_A^{I(q)})^2)^{-1} (\delta_A^{R (q)}+i\delta_A^{I(q)})$. Therefore, the vector formulation in \cite{yangtwc} under the dual mapping is equivalent to the Gaussian-integer formulation in \eqref{eqn:wn} when $q$ is a prime integer that is also a Gaussian prime.

$\bullet$ {Limitation of the vector formulation when $q$ is an integer prime but not a Gaussian prime.}

From our previous discussions in \eqref{eqn:duso}, dual mapping under the vector formulation is  desired. Otherwise, $d_{\min}=l_{\min}$ for all $\eta$ and the system performance will be poor (any arbitrary NC mapping can achieve $d_{\min}=l_{\min}$ and a system adopting the vector formulation without insisting on dual mapping is an unoptimized systems).

Furthermore, we also know that dual mapping under the vector formulation is always feasible when the integer prime $q$ also happens to be  a Gaussian prime. However, when $q$ is an integer prime, but not a Gaussian prime, dual mapping may not be possible  (specifically, this occurs when $(\delta_A^{R(q)})^2+(\delta_A^{I(q)})^2=0 ({\rm mod} ~q)$  in \eqref{eqn:val} so that a nonzero $\bm{\alpha}$ is not possible). For the vector formulation, when the dual mapping is not satisfied, we   have $d_{\min}=l_{\min}$.  In the following, we give an example showing that when $q$ is an integer prime but not a Gaussian prime, and when $\eta$ is such that $l_{\min}$ is small, we cannot find an NC mapping in the vector formulation to cluster all $(\bm{\delta}_A,\bm{\delta}_B)$ that yield the same $l_{\min}$.

Let us consider $q=2$ (i.e., $4$QAM), which is an integer prime but not Gaussian prime. Let the four representative elements in $\mathbb{Z}[i]/2$   be $\{0, 1 ,i, 1+i\}$  (note that \emph{Definition \ref{def:res}} does not apply to $q=2$). Therefore, for $q=2$, we have $\delta_A,\delta_B\in \{0, \pm 1 , \pm i, \pm(1+i), \pm(1-i)\}$.
At $\eta=\frac{1+i}{2}$, we   find that $(\bm{\delta}_A,\bm{\delta}_B)=(1,-1,-1,0)$ and its dual $(1,1,0,-1)$ yield $l_{\min}=0$. Their corresponding mod-$q$ difference pairs are   $(\bm{\delta}_A^{(q)},\bm{\delta}_B^{(q)})=(1,1,1,0)$ and $(1,1,0,1)$. At this $\eta$, we cannot find a valid dual mapping, since $(\delta_A^{R(q)})^2+(\delta_A^{I(q)})^2=0 ({\rm mod} ~2)$ is not invertible in $GF(2)$. We can choose to cluster either $(1,1,1,0)$ or $(1,1,0,1)$ but not both  (i.e., with respect to \eqref{eqn:dual}, we could design $(\alpha_1, \alpha_2)$ to cluster the former, or design $(\alpha_3, \alpha_4)$ to cluster the latter, but not both at the same time). As a consequence, $d_{\min}=l_{\min}=0$. When $\eta$ deviates from $\frac{1+i}{2}$ a little bit so that $(\bm{\delta}_A,\bm{\delta}_B)=(1,-1,-1,0)$ and its dual $(1,1,0,-1)$  still yield $l_{\min}$, but $l_{\min}$ is slightly larger than $0$, the dual mapping cannot be satisfied either. Thus, $d_{\min}=l_{\min}\approx0$. For $q>2$, the same problem arises when $q$ is an integer prime such that  $q= 1({\rm mod} ~4)$, i.e., when $q$ is not a Gaussian prime.

$\bullet$ Broader scope of Gaussian-integer formulation

\begin{figure}[t]
  \centering
        \includegraphics[height=0.3\columnwidth]{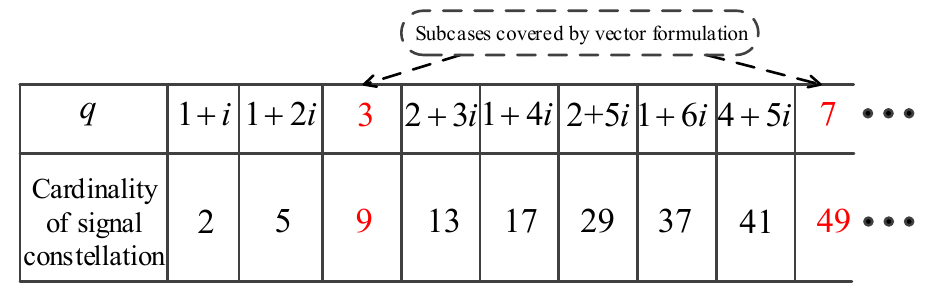}
       \caption{Possible values of $q$ and the cardinalities of the corresponding signal constellations possible with the Gaussian integer formulation, and the subcases possible with the vector formulation.}
        \label{table}
\end{figure}

For good performance, adopting dual mapping under the vector formulation limits us to $q$ that are  prime integers as well as Gaussian primes ($q=3,7,11,19,\ldots$). The Gaussian-integer formulation, on the other hand, can also solve the same dual mapping problem of the vector formulation in more concise way. Going beyond that, the Gaussian-integer formulation allows us to adopt complex $q$ (not just real $q$) that are Gaussian primes. There are many such complex Gaussian primes (e.g., $q=1+i,1+2i,\ldots$ as listed in Fig. \ref{table}).

To deal with these Gaussian primes, Gaussian-integer formulation uses the residues of the associated Gaussian prime field $\mathbb{Z}[i]/q$ as the signal constellation (modulation) used by nodes A and B. The cardinality of such a signal constellation is $|q|^2$. Therefore, with Gaussian integer formulation, we have more flexibility than with vector formulation in terms of the choices for signal constellations.

Returning to the example of $q=2$. Both the vector formulation and the Gaussian-integer formulation cannot satisfy the dual mapping requirement at some $\eta$ when $l_{\min}$ is very small. The case of $q=2$ corresponds to nodes A and B adopting 4-QAM as their signal constellation, for which the number of points on the constellation (the cardinality of the modulation) is $4$. If we insist on using an integer $q$, the next available constellation is that of $q=3$, with cardinality $9$; and after that, that of $q=7$, with cardinality $49$. We cannot find a constellation close to the 4-QAM for our purpose.

Complex $q$ in the Gaussian-integer formulation fills in this gap. Let us consider $q=1+2i$ as an example. Under the Gaussian integer formulation, the constellation points (residues of $({\rm mod} ~q)$, i.e., $\mathbb{Z}[i]/(1+2i)$) in this case are $\{0,1,-1,+i,-i\}$. The constellation cardinality is $5$, closer to the cardinality of $q=2$, which is $4$. To be a linear NC mapping in $\mathbb{Z}[i]/q$, we require $\alpha,\beta\in\mathbb{Z}[i]/q\backslash\{0\}$. With the same channel gain as the $q=2$ example above where $\eta=\frac{1+i}{2}$, we find that $(\delta_A,\delta_B)=(1-i,-1)$ and $(1+i,-i)$ yield $l_{\min}=0$. Recall that when $q=2$, the dual mapping cannot be satisfied under both the vector formulation and Gaussian-integer formulation at this $\eta$. However, when $q=1+2i$, $(\delta_A^{R(q)})^2+(\delta_A^{I(q)})^2$ is invertible in $\mathbb{Z}[i]/(1+2i)$. By \emph{Proposition \ref{pro:2}}, we can easily verify that the NC mapping with $\alpha=-1$ and $\beta=1$ can cluster the NC-valid $(\delta_A,\delta_B)=(1-i,-1)$ and $(1+i,-i)$ together.

Finally, as shown in Fig. \ref{table}, if we order the cardinality from small to large, between two real $q$ that can be used for our purpose, there are many complex $q$ offering cardinalities in between the cardinalities of the two real $q$. In other words, the Gaussian-integer formulation offers us more choices in terms of constellation cardinality than the vector formulation.

\section{$d_{\min}^{(\alpha_{opt},\beta_{opt})}$ Analysis at Zero-$l_{\min}$ Channel Gains}\label{sec:l0}

This section analyzes $l_{\min}$ as a function of $\eta$, and  focuses on those special $\eta$ at which $l_{\min}=0$ for the study of optimal NC mapping $(\alpha_{opt},\beta_{opt})$   and $d_{\min}^{(\alpha_{opt},\beta_{opt})}$. Building on the foundation established in this section,  {Section \ref{sec:vor}} will consider the optimal NC mapping  $(\alpha_{opt},\beta_{opt})$  and $d_{\min}^{(\alpha_{opt},\beta_{opt})}$ for general $\eta$.


Consider a particular valid $(\delta_A,\delta_B)$. When $\eta=-\delta_B/\delta_A$, we have $l_{\min}=0$, i.e., $|\eta\delta_A+\delta_B |=0$. We refer to   such an $\eta$ as \emph{a zero-$l_{\min}$ channel gain}. Each distance-valid $(\delta_A,\delta_B)\in\Delta$  induces a zero-$l_{\min}$  channel gain. Two superimposed symbols separated by $(\delta_A,\delta_B)$ overlap with each other at the zero-$l_{\min}$ channel gain $\eta=-\delta_B/\delta_A$.  Note that there could be multiple $(\delta_A,\delta_B)$ associated with the same $\eta$, since $\eta$ is a ratio of  $-\delta_B$ and $\delta_A$.  If there is a common factor between $\delta_A$ and $\delta_B$, we could factor out the common factor to find another $(\delta'_A,\delta'_B)$ and still retain the same $\eta=-\delta_B/\delta_A = -\delta'_B/\delta'_A$. In this paper, we refer to the $(\delta_A,\delta_B)$ with no common factor between $\delta_A$ and $\delta_B$ as a \emph{characteristic difference pair}  and denote such difference pair by $(\delta^{char}_A,\delta^{char}_B)$. Note that, strictly speaking, a characteristic difference is actually a difference pair rather than a difference. We opt to use the term ``characteristic difference'' for simplicity. As will be seen, $l_{\min}$ and $d_{\min}^{(\alpha_{opt},\beta_{opt})}$ are determined by characteristic differences; non-characteristic differences are not fundamental to the study of $l_{\min}$ and $d_{\min}^{(\alpha_{opt},\beta_{opt})}$.

We  remark that $\eta=0$  and $\eta=\epsilon\cdot\infty$  are also zero-$l_{\min}$ channel gains ($\epsilon=1,-1,i$ or $-i$ is the unit). The former corresponds to the case where $\delta_B=0$ and $\delta_A\neq0$, and the latter corresponds to the case where $\delta_A=0$ and $\delta_B\neq0$. The characteristic difference for $\eta=0$  is $(\delta_A, \delta_B)= (\epsilon, 0)$ and the characteristic difference for $\eta=\epsilon\cdot\infty$  is $(0, \epsilon)$ (this is the outcome of all Gaussian integers being a factor of $0$, and we will see later that it also makes sense for our problem of identifying the minimum distance in this paper). We will refer to $\eta=0$  and  $\eta=\epsilon\cdot\infty$  as the trivial zero-$l_{\min}$ channel gains because communication basically breaks down at this $\eta$ (e.g., at $\eta=0$ , $h_A=0$). Also, by \emph{Definition \ref{def:dvd}}, the distance-valid difference pairs that induce trivial zero-$l_{\min}$ channel gains are not NC-valid and they cannot be clustered by an NC mapping $(\alpha, \beta)$ (the implication is that, as with $l_{\min}$,  $d_{\min}$ is also $0$ at such $\eta$---note that this is not unreasonable from an intuitive viewpoint, because we should not expect good communication performance anyway since the channel gain of one node is $0$). On the other hands, all non-trivial distance-valid difference pairs are also NC-valid and they can be clustered by an NC mapping (the implication is that at non-trivial zero-$l_{\min}$ channel gains, $d_{\min}>0$ at such $\eta$). The notion will be made clear later in this paper.

Before delving into the details, let us review some fundamental definitions of  Gaussian integers.

\begin{definition}
The \emph{norm} of a Gaussian integer $a=a^R+ia^I\in\mathbb{Z}[i]$  is $|a|^2=(a^R)^2+(a^I)^2$.
\end{definition}

\begin{definition}
The \emph{units} of $\mathbb{Z}[i]$ are those elements with norm $1$, i.e., the units are $1,-1,i,-i$.
\end{definition}

\begin{definition}
Consider $a,b\in\mathbb{Z}[i]$ where at least one of $a$ or $b$ is non-zero. A \emph{greatest common divisor} (gcd) of $a$ and $b$, $\gcd(a,b)$, is a common divisor with maximal norm. Note that $\gcd(0,a)=a$, where $a$ is non-zero.
\end{definition}

\begin{definition}
 The \emph{associates} of a Gaussian integer $a$ are $a,-a,ia$, and $-ia$.
\end{definition}\rightline{$\blacksquare$}

Note that the $\gcd(a,b)$ is not unique. If $c$ is a $\gcd$ of $a$ and $b$, then so are the associates of $c$. This is because the factorization of a Gaussian integers is not unique: a factor and all its associates are all valid factors (e.g., if $a=cd$ where $c$ and $d$ are the factors, then $a$ can also be written as $a=(-c)(-d),(ic)(-id)$, or$(-ic)(id)$.)

In this paper, when we say $\gcd(a,b)=1$, we mean the unit associates are the $\gcd$ of $a$ and $b$.

\begin{definition}
Consider $a,b\in\mathbb{Z}[i]$ where at least one of $a$ or $b$ is non-zero. Then $a$ and $b$ are said to be \emph{coprime} if they only have unit factors in common (i.e., $\gcd(a,b)=1$).
\end{definition}\rightline{$\blacksquare$}

\subsection{$l_{\min}$ Versus $\eta$ Analysis and Characteristic Difference}

Before analyzing zero-$l_{\min}$ channel gains in detail, let use first quickly show how  $l_{\min}$  varies as a function of $\eta$. As defined in \eqref{eqn:l}, the distance $l$ induced by a distance-valid $(\delta_A,\delta_B)$ at a particular $\eta$ (i.e., this is the distance between two superimposed symbols separated by $(\delta_A,\delta_B)$ is given by
\begin{align}\label{eqn:l1}
l_{(\delta_A,\delta_B)}(\eta)\triangleq|\eta \delta_A+\delta_B|.
\end{align}

Let us write the real and imaginary parts of the following variables explicitly: $\delta_A=\delta^R_A+i\delta^I_A$, $\delta_B=\delta^R_B+i\delta^I_B$, and $\eta=\eta^R+i\eta^I$. We can then rewrite \eqref{eqn:l1} as
\begin{align}\label{eqn:l11}
\nonumber l^2_{(\delta_A,\delta_B)}(\eta)&=(\eta^R\delta^R_A-\eta^I\delta^I_A+\delta^R_B)^2
+(\eta^R\delta^I_A+\eta^I\delta^R_A+\delta^I_B)^2\\
\nonumber &=(\eta^R)^2|\delta_A|^2+2\eta^R(\delta^R_A\delta^R_B+\delta^I_A\delta^I_B)+
(\eta^I)^2|\delta_A|^2\\
&+2\eta^I(\delta^R_A\delta^I_B-\delta^I_A\delta^R_B)
+|\delta_B|^2,
\end{align}
\begin{align}\label{eqn:ld}
\frac{l^2_{(\delta_A,\delta_B)}(\eta)}{|\delta_A|^2}=
(\eta^R+\frac{\delta^R_A\delta^R_B+\delta^I_A\delta^I_B}{|\delta_A|^2})^2
+(\eta^I+\frac{\delta^R_A\delta^I_B-\delta^I_A\delta^R_B}{|\delta_A|^2})^2.
\end{align}
or equivalently,
\begin{align}\label{eqn:ld1}
\frac{l_{(\delta_A,\delta_B)}(\eta)}{|\delta_A|}=|\eta+\frac{\delta_B}{\delta_A}|.
\end{align}
From \eqref{eqn:ld} (or \eqref{eqn:ld1}), we can see that $l_{(\delta_A,\delta_B)}(\eta)$ as a function of $\eta$  is a cone with vertex at $\eta^o\triangleq-\delta_B/\delta_A$.
Following the definition of $l_{\min}$  in \eqref{l:lmin}, which is the minimum distance among the distances of $l_{(\delta_A,\delta_B)}(\eta)$ for all distance-valid $(\delta_A,\delta_B)\in\Delta$, where $\Delta$ is defined in \eqref{eqn:dels}, we can write
\begin{align}\label{l:lmin1}
l_{\min}(\eta) = \min_{(\delta_A,\delta_B)\in\Delta}l_{(\delta_A,\delta_B)}(\eta).
\end{align}

\begin{figure}[t]
  \centering
        \includegraphics[height=0.42\columnwidth]{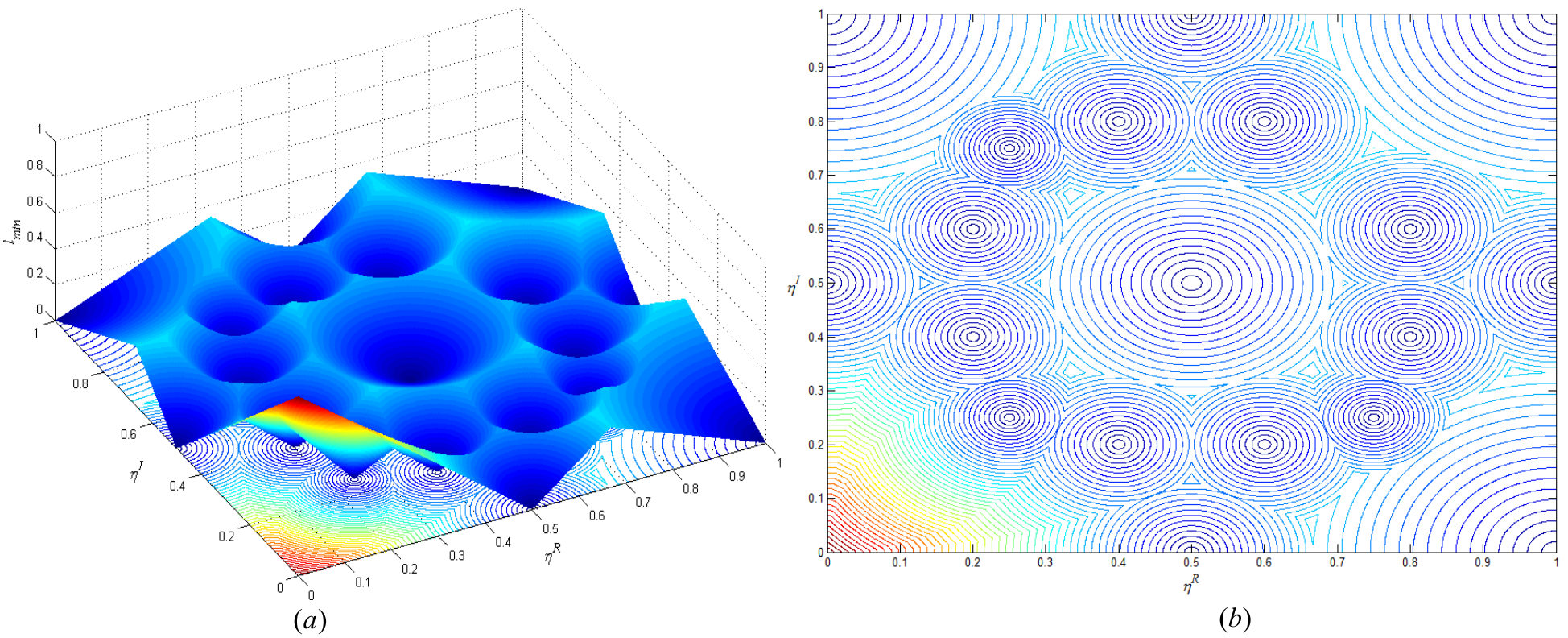}
       \caption{(a) $l_{\min}(\eta)$  surface; (b) contour graph of $l_{\min}(\eta)$.}
        \label{fig:lmin}
\end{figure}

To see how $l_{\min}(\eta)$ varies with $\eta$, we plot a three-dimensional graph of $l_{\min}(\eta)$   surface in Fig. \ref{fig:lmin}(a) and its contour graph in Fig. \ref{fig:lmin}(b) when $q=3$. We observe that
\begin{itemize}
  \item $l_{\min}(\eta)$ reaches a minimum value of zero value at the vertices of the cones as defined in \eqref{eqn:ld} for all distance-valid  $(\delta_A,\delta_B)\in\Delta$ (i.e., at zero-$l_{\min}$ channel gains);
  \item $l_{\min}(\eta)$ reaches a local maximum point at the intersections between three or more adjacent cones, and a ``local maximum edge'' at the intersections between two adjacent cones (this will be verified in Section \ref{sec:vor}).
\end{itemize}

Now, consider a particular cone induced by a particular distance-valid $(\delta_A,\delta_B)$. At the vertex of the cone, $\eta^o=-\delta_B/\delta_A$, there could be other distance-valid $(\delta'_A,\delta'_B)\in\Delta$ that also yield  $l_{\min}(\eta^o)=0$. In particular, this happens if $\frac{\delta'_B}{\delta'_A}=\frac{\delta_B}{\delta_A}$. At this $\eta^o$, we define the set that collects all $(\delta_A,\delta_B)\in\Delta$ that yield $l_{\min}(\eta^o)=0$ as follows:
\begin{align}\label{eqn:e0s}
\Delta_{\eta^o}\triangleq\{(\delta_A,\delta_B)\in\Delta|\eta^o\delta_A+\delta_B=0\}.
\end{align}

\begin{definition}[{Characteristic difference}]\label{def:char}
For a given zero-$l_{\min}$ channel gain $\eta^o$ and its associated $\Delta_{\eta^o}$, we define the characteristic difference $(\delta_A^{char},\delta_B^{char})$ as $(\delta_A,\delta_B)$: $\gcd(\delta_A,\delta_B)=1$, $(\delta_A,\delta_B)\in\Delta_{\eta^o}$ .
\end{definition}\rightline{$\blacksquare$}

\begin{remark}
At a particular $\eta^o$, the characteristic difference is unique except for the collection of associates $(\delta_A,\delta_B)= (\epsilon\delta_A^{char},\epsilon\delta_B^{char})$ for $\epsilon\in\{\pm 1,\pm i\}$.
\end{remark}\rightline{$\blacksquare$}

Each of $(\delta_A,\delta_B)\in\Delta_{\eta^o}$ induces a cone centered at ${\eta^o}$ given by $l_{(\delta_A,\delta_B)}(\eta)=|\eta \delta_A+\delta_B|$. Among all the cones, $(\delta_A^{char},\delta_B^{char})$ gives the smallest $l$ for all $\eta$, since $|\eta\delta_A^{char}+\delta_B^{char}|<|\gcd(\delta_A,\delta_B)||\eta\delta_A^{char}+\delta_B^{char} |=|\eta\delta_A+\delta_B|$. Thus, as shown in Fig. \ref{fig:lmin}(a), $l_{\min}$ at $\eta$  in the neighborhood of $\eta^o$  is given by the cone of  $l_{(\delta_A^{char},\delta_B^{char})}(\eta)$. In particular, $l_{\min}(\eta)$ can never be given by the cones of non-characteristic  $(\delta_A,\delta_B)$ except at ${\eta^o}$. In studying $l_{\min}(\eta)$  as a function of ${\eta}^o$, it suffices to restrict our attention to $(\delta_A^{char},\delta_B^{char})$.

Furthermore, consider a cone with vertex $\eta^o$. For a fixed value $l_{(\delta_A,\delta_B)}(\eta)$ of the cone, note from \eqref{eqn:ld1} that the contour of $\eta$  that achieves this fixed $l_{(\delta_A,\delta_B)}(\eta)$, is a circle of radius $\frac{l_{(\delta_A,\delta_B)}}{|\delta_A|}$ centered at
$(-\frac{\delta^R_A\delta^R_B+\delta^I_A\delta^I_B}{|\delta_A|^2},-\frac{\delta^R_A\delta^I_B-\delta^I_A\delta^R_B}{|\delta_A|^2})$. Thus, as shown in Fig. \ref{fig:lmin}(b), the contour lines of $l_{(\delta_A^{char},\delta_B^{char})}$  for a particular $(\delta_A^{char},\delta_B^{char})$ are concentric circles centered at
$(-\frac{\delta^{char,R}_A\delta^{char,R}_B+\delta^{char,I}_A\delta^{char,I}_B}{|\delta^{char}_A|^2},
-\frac{\delta^{char,R}_A\delta^{char,I}_B-\delta^{char,I}_A\delta^{char,R}_B}{|\delta^{char}_A|^2}). $

For a particular $q$, the aforementioned minima and local maxima for $l_{\min}$ characterize the performance of NC mapping at various $\eta$, since post-NC mapping $d_{\min}^{(\alpha,\beta)}$ is related to $l_{\min}$. In {Part B} below, we first study minima at the zero-$l_{\min}$ channel gains. In Section \ref{sec:vor}, we will consider the local maxima.

\subsection{Identifying Zero-$l_{\min}$ Channel Gains and Characteristic Differences}

By the definition of zero-$l_{\min}$ channel gain, we can identify all $\eta$ at which $l_{\min}=0$ in the complex plane of $\eta$ and their associated characteristic differences. To be specific, given a $q$, we can go through all $(\delta_A,\delta_B)\in\Delta$ of \eqref{eqn:dels}  to find all $\eta$ such that $\eta\delta_A+\delta_B=0$. Then, we have a set that collects all distinct zero-$l_{\min}$ channel gains as $\mathcal{H}^o=\{\eta^o|\eta^o=-\delta_B/\delta_A ,(\delta_A,\delta_B )\in\Delta\}$. Then, for each $\eta^o \in \mathcal{H}^o$, we select the characteristic difference $(\delta_A^{char},\delta_B^{char})\in \Delta_{\eta^o}$ in \eqref{eqn:e0s} by \emph{Definition \ref{def:char}}.

\begin{remark}\label{rem:nont}
 The distance-valid difference pairs $(\delta_A,\delta_B)\in\Delta_{\eta^o}$ where $\delta_A\neq 0$ and $\delta_B=0$ correspond to a \emph{trivial} zero-$l_{\min}$ channel gain $\eta^o=0$. Similarly, we also have a trivial zero-$l_{\min}$ channel gain at $\eta^o=\infty$, which is induced by $\delta_A=0,\delta_B\neq0$. From \emph{Remark \ref{rem:ncd}}, the distance-valid difference pair associated with this trivial zero-$l_{\min}$ channel gain cannot be clustered by any NC mapping. By \emph{Definition \ref{def:char}}, the characteristic difference at the  trivial zero-$l_{\min}$ channel $\eta^o=0$ is $(\delta_A^{char},\delta_B^{char}) =(1,0)$ and the characteristic difference at the trivial zero-$l_{\min}$ channel $\eta^o=\infty$ is $(\delta_A^{char},\delta_B^{char})=(0, 1)$. The distance-valid difference pairs where both $\delta_A\neq0$ and $\delta_B\neq0$, i.e., NC-valid difference pairs, correspond to the \emph{non-trivial} zero-$l_{\min}$ channel gains $\eta^o\neq0$.
\end{remark}

The proposition below simplifies our search for the zero-$l_{\min}$ channel gains in the complex plane of $\eta$, by exploiting a symmetry property.

\begin{proposition}[{Zero-$l_{\min}$ symmetry}]\label{pro:zeros}
 Consider a zero-$l_{\min}$  channel gain $\eta^o=|\eta^o|e^{i\theta^o}$ within $0<\theta^o<\pi/4$   in the complex plane of $\eta$. Suppose that the characteristic difference at this $\eta^o$ is $(\delta_A^{char},\delta_B^{char})$. Given this $\eta^o$, we can find seven other ``symmetric'' zero-$l_{\min}$ channel gains in the complex plane of $\eta$ as follows:
 \begin{align}
\nonumber &|\eta^o|e^{i(\frac{\pi}{2}-\theta^o)}, |\eta^o|e^{i(\frac{\pi}{2}+\theta^o)},
|\eta^o|e^{i({\pi}-\theta^o)},\\
\nonumber &|\eta^o|e^{i({\pi}+\theta^o)},
|\eta^o|e^{i(\frac{3\pi}{2}-\theta^o)},\\
&|\eta^o|e^{i(\frac{3\pi}{2}+\theta^o)},
 {\rm~and~} |\eta^o|e^{i({2\pi}-\theta^o)}.
 \end{align}
The corresponding characteristic differences are
 \begin{align}
\nonumber&(\delta_A^{char *},i\delta_B^{char*}),(\delta_A^{char},i\delta_B^{char})    ,(\delta_A^{char *},-\delta_B^{char*}),\\
\nonumber&(\delta_A^{char},-\delta_B^{char} ),(\delta_A^{char *},-i\delta_B^{char*} ),\\&(\delta_A^{char},-i\delta_B^{char}),  {\rm~and~} (\delta_A^{char *},\delta_B^{char*}),
 \end{align}
 respectively, where $*$ denote the complex conjugate.
\end{proposition}

\begin{IEEEproof}[Proof of {Proposition \ref{pro:zeros}}]
It can be verified that the seven vertices and the corresponding characteristic differences identified below correspond to those listed in the proposition.

For $\eta^o=-\delta_B/\delta_A$, its reflection on the real axis is $\eta^{o*}=-\delta^*_B/\delta^*_A$. The reflections of these two vertices on the imaginary axis are $-\eta^{o*}=\delta^*_B/\delta^*_A$  and   $-\eta^{o}=-\delta_B/\delta_A$ respectively. Furthermore, if $\gcd(\delta_A,\delta_B)=1$, we have $\gcd(\delta_A^*,\delta_B^* )=1$ as well. We now have four vertices and their characteristic differences.

A $90$-degree rotation of gives $i\eta^o=-i\delta_B/\delta_A$. Going through the reflection process as above gives us $-i\eta^{o*}=i\delta^*_B/\delta^*_A$, $i\eta^{o*}=-i\delta^*_B/\delta^*_A$, and $-i\eta^{o}=i\delta_B/\delta_A$. We now have the other four vertices and their characteristic differences.

\end{IEEEproof}

\begin{remark}
Note that the eight zero-$l_{\min}$ channel gains in \emph{Proposition \ref{pro:zeros}} are symmetric points with respect to a circle in the complex plane of $\eta$. We refer to this as the symmetry of zero-$l_{\min}$ channel gains.
\end{remark}\rightline{$\blacksquare$}

Therefore, to identify all zero-$l_{\min}$ channel gains in the complex plane of $\eta$, we only need to consider the zero-$l_{\min}$ channel gains within $0\leq\theta\leq\pi/4$.

\begin{figure}[t]
  \centering
        \includegraphics[height=0.8\columnwidth]{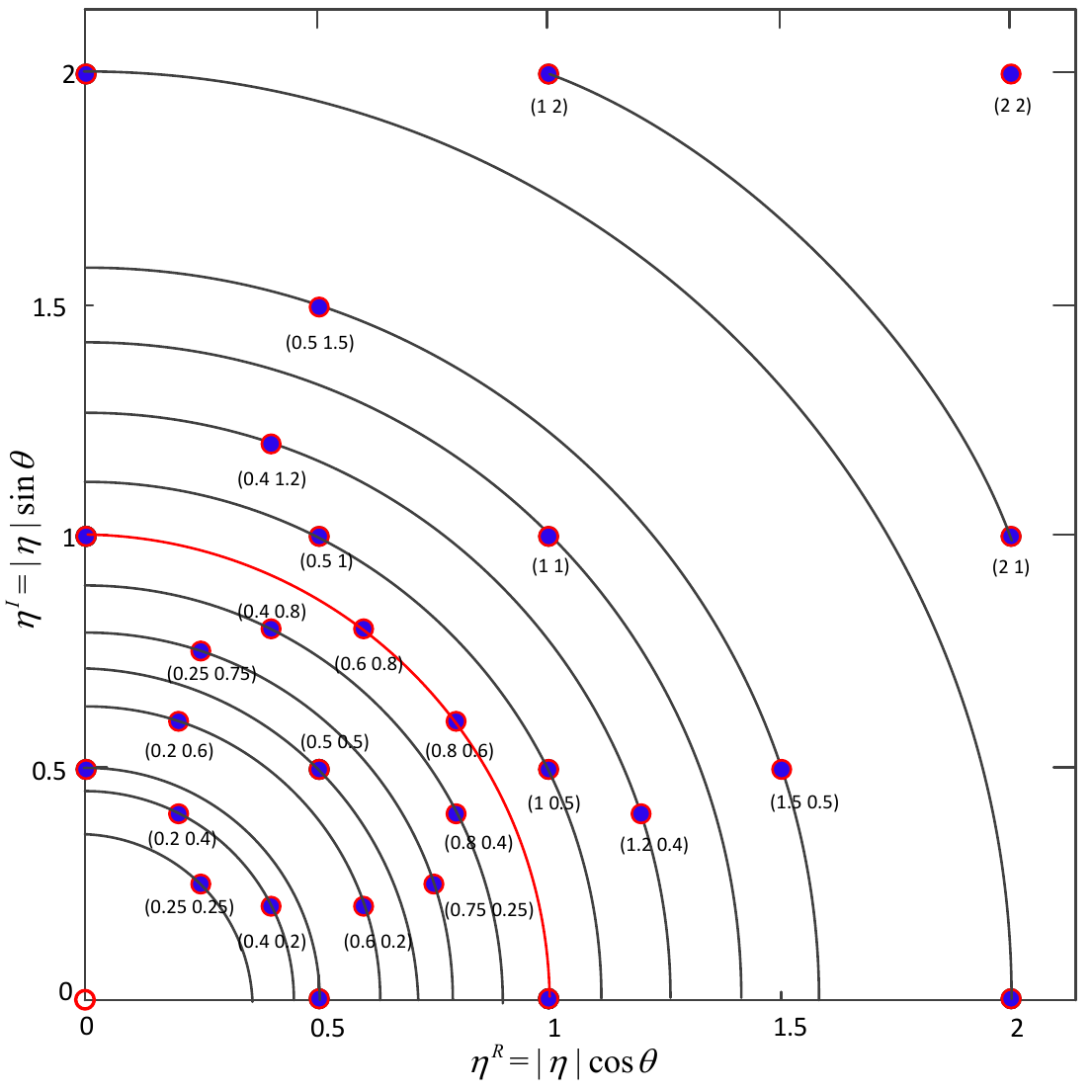}
       \caption{The zero-$l_{\min}$ channel gains within $0\leq\theta\leq\pi/2$ in the complex plane of $\eta$ when $q=3$. The blue dots with red outline are the non-trivial zero-$l_{\min}$ channel gains induced by the NC-valid difference pairs, and the red circle is the trivial zero-$l_{\min}$ channel gain induced by the distance-valid difference pairs with $\delta_A\neq0$ and $\delta_B=0$. }
        \label{fig:lzero}
\end{figure}

As an example, we  plot the zero-$l_{\min}$ channel gains (marked with blue dots) within $0\leq\theta\leq\pi/2$ when $q=3$ in Fig. \ref{fig:lzero}. Through an arc centered at the origin, we can find two symmetric zero-$l_{\min}$ channel gains with respect to $\theta=\pi/4$ (the zero-$l_{\min}$ channel gains with $\theta=\pi/4$ do not have the symmetric points within $0\leq\theta\leq\pi/2$). Consider a zero-$l_{\min}$ channel gain $\eta=1+i$. At this $\eta$, we find $(\delta_A,\delta_B)=(2,-2-2i),(1,-1-i),(1+i,-2i),(i,1-i),(2i,2-2i)$ that yield $l_{\min}=0$. Among these $(\delta_A,\delta_B)$, we can select $(\delta^{char}_A,\delta^{char}_B)=(1,-1-i)$ or $(i,1-i)$ as the characteristic difference of $\eta=1+1i$, since $\gcd(1,-1-i)=1$.

 {Nonzero-$l_{\min}$ Symmetry:}

The above considered the symmetry of zero-$l_{\min}$  channel gains on the complex plane of $\eta$. Similar symmetry apply to non-zero $l_{\min}$  channel gains and that it suffices to consider $\eta$  with angle between $0$ and $\pi/4$. To see this,  let us consider a particular cone with the vertex $\eta^o=|\eta|e^{i\theta^o}$ and $\theta^o\in[0,\pi/4]$. Suppose that the characteristic difference at this $\eta^o$ is  $(\delta_A^{char},\delta_B^{char})$. Now, consider a channel gain $\eta'=|\eta'|e^{i\theta'}$ in the neighborhood of $\eta^o$ where $l_{\min}(\eta')$  is still determined by the cone with the vertex $\eta^o$. i.e., $l_{\min}(\eta')=l_{(\delta_A^{char},\delta_B^{char})}(\eta')$. By \emph{Proposition \ref{pro:zeros}}, given this $(\delta_A^{char},\delta_B^{char})$ of $\eta^o$, we can identify seven other symmetric zero-$l_{\min}$ channel gains in the complex plane of $\eta$ and the associated characteristic differences. Correspondingly, we can generalize this symmetry of  characteristic differences to the $l_{\min}(\eta')$. That is, given $\eta'$ and  $l_{\min}(\eta')$, we can identify seven other channel gain  $\eta$ in the seven other octants by the same transformations stated in \emph{Proposition \ref{pro:zeros}} (replacing $\eta^o$  by $\eta'$  and  $\theta^o$  by $\theta'$  in the statement of the transformations to get the new  $\eta$) such that $l_{\min}(\eta)=l_{\min}(\eta')$. For example, the reflection of $\eta'$ on the real axis is $\eta'^*=|\eta'|e^{i(2\pi-\theta)}$. At $\eta=\eta'^*$, $l_{\min}(\eta)$ is determined with respect to the transformed characteristic difference $(\delta_A^{char*},\delta_B^{char*})$  and we can verify that $l_{\min}(\eta)=l_{\min}(\eta')$. Therefore, we only need to focus on the  $l_{\min}(\eta)$ induced by the zero-$l_{\min}$ channel gains within $0\leq\theta\leq\pi/4$.

Since the distance-valid difference pairs with $\delta_A=0$ or $\delta_B=0$ cannot be clustered by any NC-valid difference pair ($d_{\min}$ will also be $0$ at the non-trivial zero-$l_{\min}$ channel gain), we focus on the optimal NC mapping for the NC-valid difference pair at the non-trivial zero-$l_{\min}$ channel gains in the following parts.

\subsection{Optimal NC Mapping at  Nontrivial  Zero-$l_{\min}$ Channel Gains}

\begin{theorem}\label{th:op0}
For a particular  nontrivial  zero-$l_{\min}$ channel gain $\eta^o$ and its associated characteristic difference $(\delta_A^{char},\delta_B^{char})$, an optimal NC mapping is
\begin{align}\label{eqn:thop0}
(\alpha_{opt},\beta_{opt})=\big(-(\delta_A^{char(q)})^{-1}\delta_B^{char(q)}, 1\big)
\end{align}
\end{theorem}
where $\delta_A^{char(q)}=\delta_A^{char} ({\rm mod}~q) \in \mathbb{Z}[i]/q$ and $\delta_B^{char(q)}=\delta_B^{char} ({\rm mod}~q) \in \mathbb{Z}[i]/q$.

\begin{IEEEproof}[Proof of Theorem \ref{th:op0}]
To ensure $d_{\min}^{(\alpha,\beta)}>0$ at the non-trivial  zero-$l_{\min}$ channel gain $\eta^o$, we need to cluster $(\delta_A^{char},\delta_B^{char})$ and all the other  $(\delta_A,\delta_B)\in \Delta_{\eta^o}$ (defined in \eqref{eqn:e0s}) yielding $l_{\min}=0$. If we did not cluster all $(\delta_A,\delta_B)\in \Delta_{\eta^o}$, then $d_{\min}^{(\alpha,\beta)}=l_{\min}=0$ for sure. The solution for this clustering is $(\alpha,\beta)=(-(\delta_A^{char(q)})^{-1} \delta_B^{char (q)},1)$. \emph{Remark \ref{rem:iso}} in \emph{Appendix I} indicates that the NC partitioning under this $(\alpha,\beta)$ is unique. In particular, once $(\delta_A^{char},\delta_B^{char})$ is clustered by $(\alpha,\beta)$, the NC partitioning is fixed, and there is no further freedom to cluster another NC-valid $(\delta_A,\delta_B)$ that does not belong to the clustered-difference set of $(\alpha,\beta)$.  Therefore, the NC mapping in \eqref{eqn:thop0} is optimal for the nontrivial  zero-$l_{\min}$ channel gain.

\end{IEEEproof}

\subsection{Identifying $d_{\min}^{(\alpha_{opt},\beta_{opt})}$ at  Zero-$l_{\min}$ Channel Gains}

In  {Part B}, we have identified all zero-$l_{\min}$ channel gains in the complex plane of $\eta$ and the associated characteristic differences. In particular, each characteristic difference $(\delta_A^{char},\delta_B^{char})$ yields the $l_{\min}$ at $\eta$ in the neighborhood of the associated channel gain $\eta^o$ through the relationship $l_{\min}(\eta)=\eta \delta^{char}_A+\delta^{char}_B$. Specifically, this $l_{\min}$  corresponds to a cone centered at $\eta^o$, as illustrated in Fig. \ref{fig:lmin}. In this part, we aim to identify $d_{\min}^{(\alpha_{opt},\beta_{opt})}$ at zero-$l_{\min}$ channel gains.

\emph{Trivial Theorem:} At the trivial zero-$l_{\min}$ channel gain $\eta^o=0$ and $\eta^o=\infty$, $d_{\min}^{(\alpha_{opt},\beta_{opt})}=l_{\min}=0$.
\rightline{$\blacksquare$}

By \emph{Remark \ref{rem:ncd}}, the characteristic difference $(\delta_A^{char},\delta_B^{char})=(1,0)$ and $(0, 1)$ associated with the trivial zero-$l_{\min}$ channel gains  $\eta^o=0$ and $\eta^o=\infty$   cannot be clustered by any NC mapping. Therefore, at the trivial zero-$l_{\min}$ channel gains, $d_{\min}^{(\alpha_{opt},\beta_{opt})}=l_{\min}=0$.

\begin{theorem}\label{thm:eta0}
 Consider a particular nontrivial zero-$l_{\min}$ channel gain $\eta^o$ associated with the characteristic difference $(\delta_A^{char},\delta_B^{char})$ that can be clustered by the NC mapping $(\alpha_{opt},\beta_{opt})$. There exists a characteristic difference $({{\delta}^{char}_A}',{{\delta}^{char}_B}')$  that determines the $d_{\min}^{(\alpha_{opt},\beta_{opt})}$ at this $\eta^o$ such that $|\delta^{char}_B {\delta^{char}_A}'-\delta_A^{char}{\delta^{char}_B}'|=1$ and that  $d_{\min}^{(\alpha_{opt},\beta_{opt})}=|\eta^o{\delta_A^{char}}'+{\delta_B^{char}}'|=\frac{1}{|\delta_A^{char}|}$.
\end{theorem}

\begin{IEEEproof}[Proof of Theorem \ref{thm:eta0}]
Let us assume the validity of the statements of (T2-1) and (T2-2) below. They will be proved separately.

\begin{enumerate}
  \item [(T2-1)] There exists a characteristic difference  $({{\delta}^{char}_A}',{{\delta}^{char}_B}')$ yielding
       $|\delta^{char}_B {\delta^{char}_A}'-\delta_A^{char}{\delta^{char}_B}'|=1$
       (in \emph{Lemma \ref{lem:dmin0}}), and
  \item [(T2-2)]  For a $({{\delta}^{char}_A}',{{\delta}^{char}_B}')$ that satisfies $|\delta^{char}_B {\delta^{char}_A}'-\delta_A^{char}{\delta^{char}_B}'|=1$,  $({{\delta}^{char}_A},{{\delta}^{char}_B})$ and  $({{\delta}^{char}_A}',{{\delta}^{char}_B}')$ cannot be clustered by the same NC mapping $(\alpha,\beta)$.
\end{enumerate}

(T2-1) together with (T2-2) imply $d_{\min}^{(\alpha_{opt},\beta_{opt})}=|\eta^o{\delta_A^{char}}'+{\delta_B^{char}}'|=\frac{1}{|\delta_A^{char}|}$  for the following reason. Given that $({{\delta}^{char}_A}',{{\delta}^{char}_B}')$ and $({{\delta}^{char}_A},{{\delta}^{char}_B})$ cannot be clustered by the same NC mapping, then under the optimal NC mapping that clusters $({{\delta}^{char}_A},{{\delta}^{char}_B})$ at  $\eta^o=-\delta_B/\delta_A$,  $(\alpha_{opt},\beta_{opt})=(-(\delta_A^{char(q)})^{-1}\delta_B^{char(q)}, 1)$,  the distance $l=|\eta^o{\delta_A^{char}}'+{\delta_B^{char}}'|$
is a potential candidate for $d_{\min}^{(\alpha_{opt},\beta_{opt})}$. Now, $|\eta^o{\delta_A^{char}}''+{\delta_B^{char}}''|>0$  for all characteristic differences  $({\delta^{char}_A}'',{\delta^{char}_B}'')$ that cannot be clustered by $(\alpha_{opt},\beta_{opt})$. This means $|\delta_A^{char}||\eta^o{\delta_A^{char}}''+{\delta_B^{char}}''|=|\delta^{char}_B{\delta^{char}_A}''-\delta_A^{char}{\delta^{char}_B}''|>0$. But then $|\delta^{char}_B{\delta^{char}_A}''-\delta_A^{char}{\delta^{char}_B}''|$  must be an integer, meaning $|\delta^{char}_B {\delta^{char}_A}''-\delta_A^{char}{\delta^{char}_B}''|\geq 1$.
Thus, $|\eta^o{\delta_A^{char}}''+{\delta_B^{char}}''|\geq \frac{1}{|\delta^{char}_A|}$. Among all $({\delta^{char}_A}'',{{\delta}^{char}_B}'')$  that cannot be clustered by $(\alpha_{opt},\beta_{opt})$,  $({\delta^{char}_A}',{\delta^{char}_B}')$
can meet the lower bound of the inequality: i.e., $|\eta^o{\delta_A^{char}}'+{\delta_B^{char}}'|=\frac{1}{|\delta^{char}_A|}$  according to (T2-1) and (T2-2).
Thus,  $d_{\min}^{(\alpha_{opt},\beta_{opt})}=|\eta^o{\delta_A^{char}}'+{\delta_B^{char}}'|=\frac{1}{|\delta^{char}_A|}$.

\end{IEEEproof}


\begin{remark}\label{rem:cri}
Note that the $d_{\min}^{(\alpha_{opt},\beta_{opt})}$-determining differences for a given zero-$l_{\min}$ channel gain $\eta^o=-\delta^{char}_B/\delta^{char}_A$  may not be unique, since the characteristic differences that satisfy \emph{Theorem \ref{thm:eta0}} may not be unique. For example, $\eta^o=\frac{1+i}{2}$ in Fig. \ref{fig:lmin} is associated with the characteristic difference $(\delta_A^{char},\delta_B^{char})=(1+i,-i)$. The multiple  $d_{\min}^{(\alpha_{opt},\beta_{opt})}$-determining differences $({{\delta}^{char}_A}',{{\delta}^{char}_B}')$ for this $\eta^o$ within $\theta^o\in[0,\pi/4]$ are $(1+2i,-i),(2+i,-1-i),(2+2i,-1-2i),(1+2i,-2i)$, and $(2+i,-1-2i)$---they all yield  $|\delta^{char}_B {\delta^{char}_A}'-\delta_A^{char}{\delta^{char}_B}'|=1$.
\end{remark}\rightline{$\blacksquare$}

The proof of (T2-2) is straightforward and is as follows: A solution for the NC mapping that can cluster $({\delta}^{char}_A, {\delta}^{char}_B)$  is $(\alpha_{opt},\beta_{opt})=(-{\delta}^{char (q)}_B, {\delta}^{char (q)}_A)$   (other solutions are isomorphic). Suppose that $(\alpha_{opt},\beta_{opt})=(-{\delta}^{char (q)}_B, {\delta}^{char (q)}_A)$  can also cluster $({\delta_A^{char}}', {\delta_B^{char}}')$.  Then, $-\delta_B^{char(q)}{\delta_A^{char(q)}}'+ \delta_A^{char(q)}{\delta_B^{char(q)}}'=0({\rm mod}~q)$. Therefore, $\delta_B^{char(q)}{\delta_A^{char(q)}}'- \delta_A^{char(q)}{\delta_B^{char(q)}}'=uq$ for some Gaussian integer $u$. Given that $q$ is not a unit, we must have that $|\delta_B^{char(q)}{\delta_A^{char(q)}}'- \delta_A^{char(q)}{\delta_B^{char(q)}}'|=|u||q|\neq 1$, leading to a contradiction. Thus, $(\alpha_{opt},\beta_{opt})$  cannot cluster $({\delta_A^{char}}', {\delta_B^{char}}')$  at the same time.

The proof of (T2-1) is much more involved, and is given  through a series of lemmas (\emph{Lemmas \ref{lem:suf}} to \emph{\ref{lem:dmin0}}) in the following. Let us first define a few distance measures to clarify the issue.

Consider two distinct zero-$l_{\min}$ channel gains $\eta_i^o$ and $\eta_j^o$ ($\eta_i^o$ and $\eta_j^o$ can be trivial or nontrivial zero-$l_{\min}$ channel gains). Suppose that $(\delta_{A,i}^{char},\delta_{B,i}^{char})$ and $(\delta_{A,j}^{char},\delta_{B,j}^{char})$ are the two characteristic differences associated with $\eta_i^o$ and $\eta_j^o$ respectively. We define the \emph{Euclidean distance}, \emph{normalized distance}, and \emph{weighted distance} between $\eta_i^o$ and $\eta_j^o$ as follows:
\begin{itemize}
  \item The \emph{Euclidean distance} between  $\eta_i^o$ and $\eta_j^o$
  \begin{eqnarray}\label{l:ed}
\nonumber &d_{ij}\triangleq |\eta^o_i-\eta^o_j|\\
 & = \big|  \frac{\delta_{B,i}^{char}}{\delta_{A,i}^{char}}
-\frac{\delta_{B,j}^{char}}{\delta_{A,j}^{char}}\big|
=\frac{|\delta_{B,i}^{char}\delta_{A,j}^{char}-\delta_{A,i}^{char}\delta_{B,j}^{char}|}
{|\delta_{A,i}^{char}||\delta_{A,j}^{char}|}
\end{eqnarray}
  \item 	The \emph{normalized distance} between $\eta_i^o$ and $\eta_j^o$
  \begin{align}\label{l:nd}
d^*_{ij}&\triangleq |\delta_{B,i}^{char}\delta_{A,j}^{char}-\delta_{A,i}^{char}\delta_{B,j}^{char}|
\end{align}

  \item 	The \emph{weighted distance} from $\eta_j^o$ to $\eta_i^o$
  \begin{align}\label{l:wd}
\nonumber d_{j\rightarrow i}&\triangleq |\eta^o_i\delta_{A,j}^{char}+\delta_{B,j}^{char}| \\ &=|-\frac{\delta_{B,i}^{char}}{\delta_{A,i}^{char}}\delta_{A,j}^{char}+\delta_{B,j}^{char}|
\end{align}
\end{itemize}

From the definitions above, we have the following remarks:
\begin{itemize}
  \item The weighted distance is a distance induced by $(\delta_{A,j}^{char},\delta_{B,j}^{char})$ at $\eta_i^o$, i.e., $l_{(\delta_{A,j}^{char},\delta_{B,j}^{char})}(\eta^o_i)=|\eta^o_i\delta_{A,j}^{char}
      +\delta_{B,j}^{char}|$.  Note that $d_{j\rightarrow i}\neq d_{i\rightarrow j}$ if $|\delta_{A,i}^{char}|\neq |\delta_{A,j}^{char}|$. 	 That is, the distance induced by $(\delta_{A,j}^{char},\delta_{B,j}^{char})$ at  $\eta_i^o$ is not the same as the distance induced by $(\delta_{A,j}^{char},\delta_{B,j}^{char})$ at    $\eta_j^o$ in general.
  \item  The relationships among Euclidean distance, normalized distance, and weighted distance are
        \begin{align}\label{l:rel}
d_{j\rightarrow i}=|\delta_{A,j}^{char}|d_{ij}=\frac{d^*_{ij}}{\delta_{A,i}^{char}}
\end{align}

  \item  In \emph{Lemmas \ref{lem:suf}} and \emph{\ref{lem:nec}} below, we give a necessary condition and a sufficient condition  for the distance-validity of $(\delta_A, \delta_B)$. In \emph{Lemma \ref{lem:dmin0}} below, we identify the minimum normalized distance associated with any zero-$l_{\min}$ channel gain using \emph{Lemmas} \emph{\ref{lem:suf}} to \emph{\ref{lem:be1i}}.

\end{itemize}

\emph{Symmetry of $\mathbb{Z}[i]/q$ Under Rotations and Reflections}: In the following (including the Appendices), we assume w.l.o.g that $q^R>q^I\geq 1$ when the proofs are given under a complex Gaussian prime $q=q^R+iq^I$, $|q|\geq \sqrt{5}$, and $q^R,q^I\neq0$. Due to the symmetry property of Gaussian primes, there is no loss of generality in assuming positive $q^R$ and $q^I$ within $0<\theta<\pi/4$ in the complex plane. Specifically, for a Gaussian prime $q$, rotations by multiples of $\pi/2$ and reflections on the real and imaginary axes give other symmetric Gaussian primes. Similar symmetry applies to elements in $\mathbb{Z}[i]$. Therefore, elements in $\mathbb{Z}[i]/q$ also undergo similar symmetric transformations.

\begin{lemma}[Necessary condition of validity]\label{lem:suf}
Given a Gaussian prime $q$  and $q^R\neq0, q^I\neq0$, if $(\delta_A, \delta_B)\in \mathbb{Z}^2[i]$ is a distance-valid difference pair, then $|\delta_A|,|\delta_B|\leq \sqrt{2|q|^2-4q^R+2}$.
\end{lemma}\rightline{$\blacksquare$}

The proof of \emph{Lemma \ref{lem:suf}} is given in  {Appendix II}.

%

\begin{lemma}[Sufficient condition of validity]\label{lem:nec}
 Given a pair of Gaussian integers  $(\delta_A, \delta_B)\in \mathbb{Z}^2[i]$, and a Gaussian prime $q$ that defines valid symbols in $\mathbb{Z}[i]$ according to \emph{Definition \ref{def:res}}, a sufficient condition for $(\delta_A, \delta_B)$ to be a distance-valid difference pair is $|\delta_A|\neq 0$ or $|\delta_B|\neq0$ and $|\delta_A|, |\delta_B|<|q|$.
\end{lemma}\rightline{$\blacksquare$}

The proof of \emph{Lemma \ref{lem:nec}} is given in  {Appendix III}.

\emph{Lemma \ref{lem:be}} below is the well-known \emph{B\'{e}zout Identity} in $\mathbb{Z}[i]$ \cite{gi,bez}.

\begin{lemma}{[\emph{B\'{e}zout Identity in $\mathbb{Z}[i]$}]}\label{lem:be}
Consider two Gaussian integers $a$ and $b$, not both zero. There exist two Gaussian integers $x$ and $y$ such that $ax-by=\gcd(a,b)$. Furthermore, among the Gaussian integers that can be written in the form of $ax'-by'$  where $x'$ and $y'$ are Gaussian integers, the $x$ and $y$ that satisfy  $ax-by=\gcd(a,b)$ yield the smallest possible norm for  $ax'-by'$  (i.e., $|\gcd(a,b)|^2$  is the smallest possible norm for $ax'-by'$).
\end{lemma} \rightline{$\blacksquare$}

\begin{lemma}\label{lem:be1i}
 Given a Gaussian prime $q$,  for a distance-valid difference pair $(\delta_{A,i},\delta_{B,i})$ where $\delta_{A,i}\in\{\pm 1,\pm i\}$ or $\delta_{B,i}\in\{\pm 1,\pm i\}$, there exists a distance-valid difference pair $(\delta_{A,j},\delta_{B,j})$ such that $\delta_{A,j}\delta_{B,i}-\delta_{B,j}\delta_{A,i}=1$.
\end{lemma}
 \begin{IEEEproof}[Proof of Lemma \ref{lem:be1i}]
 W.l.o.g., consider $\delta_{B,i}\in \{\pm 1, \pm i\}$. For $\delta_{A,j}\delta_{B,i}-\delta_{B,j}\delta_{A,i}=1$, we choose a distance-valid $(\delta_{A,j},\delta_{B,j})=(\delta_{B,i},0)$ if $\delta_{B,i}\in \{1,-1\}$ and $(\delta_{A,j},\delta_{B,j})=(\delta^*_{B,i},0)$ if $\delta_{B,i}\in \{i,-i\}$, where $*$ denotes complex conjugate.

 \end{IEEEproof}

%
%
%
%
%
%
%

\emph{Lemma \ref{lem:be1i}} is similar to {B\'{e}zout Identity} for $\mathbb{Z}[i]$ in \emph{Lemma \ref{lem:be}} except that $a$ and $b$ are units, the RHS is a unit (i.e., $\gcd(a,b)=\{\pm 1,\pm i\}$), and that $x$ and $y$ are restricted to be components in a distance-valid difference pair $(\delta_A, \delta_B)$.

%

\begin{lemma}\label{lem:dmin0}
Given a Gaussian prime $q$, for a nontrivial zero-$l_{\min}$ channel gain $\eta_i^o$ associated with a characteristic difference $(\delta_{A,i}^{char},\delta_{B,i}^{char})$, there exists a characteristic difference $(\delta_{A,j}^{char},\delta_{B,j}^{char})$ yielding  $|\delta^{char}_{A,j}\delta^{char}_{B,i}-\delta^{char}_{B,j}\delta^{char}_{A,i}|=1$
\end{lemma}

\begin{IEEEproof}[Proof of Lemma \ref{lem:dmin0}]

For $|q|=\sqrt{2}$, the validity of the lemma can be easily verified. Specifically, for $|q|=\sqrt{2}$, the representative elements in $\mathbb{Z}[i]/q$ are limited to the two values in the set $\{0, 1\}$. The nontrivial zero-$l_{\min}$ channel gains can only be $\eta^o=-\delta^{char}_{B,i}/\delta^{char}_{A,i}=1$ or $-1$. To satisfy $|\delta^{char}_{A,j}\delta^{char}_{B,i}-\delta^{char}_{B,j}\delta^{char}_{A,i}|=1$, we can simply let $\delta^{char}_{A,j}=\delta^{char}_{B,i}$  and $\delta^{char}_{B,j}=0$.

We now consider $|q|\geq\sqrt{5}$.
The proof consists of two parts P1) and P2). In P1), we prove that given a Gaussian prime $q$, there exists a distance-valid difference pair $(\delta_{A,j},\delta_{B,j})$ such that
\begin{align}\label{eqn:dminb1}
\delta_{A,j} \delta_{B,i}^{char}-\delta_{B,j}\delta^{char}_{A,i}=1.
\end{align}
In P2), we prove that $\gcd(\delta_{A,j},\delta_{B,j})=1$, i.e., $(\delta_{A,j},\delta_{B,j})$ is a characteristic difference.


\textbf{P1)} Given a Gaussian prime $q$, the case where $\delta_{A,i}\in\{\pm 1,\pm i\}$ or $\delta_{B,i}\in\{\pm 1,\pm i\}$ has been covered by \emph{Lemma \ref{lem:be1i}}. Our proof here focuses on the case where $\delta_{A,i}\notin\{\pm 1,\pm i\}$ and $\delta_{B,i}\notin\{\pm 1,\pm i\}$.

By \emph{Lemma \ref{lem:be}}, given a Gaussian prime $q$, there exist Gaussian integers $x$ and $y$ such that
\begin{align}\label{eqn:dm1}
x \delta^{char}_{B,i}-y \delta^{char}_{A,i}=1.
\end{align}

Now, $(x,y)$ may or may not be a distance-valid difference pair. However, given that $(x,y)$ is a solution to \eqref{eqn:dm1}, the following are also solutions:
\begin{align}\label{eqn:dm2}
\nonumber &\delta_{A,j}=x + k \delta^{char}_{A,i}\\
&\delta_{B,j}=y + k \delta^{char}_{B,i},
\end{align}
for all Gaussian integers $k$. Our goal is to show that there exists a $k\in\mathbb{Z}[i]$ such that $(\delta_{A,j},\delta_{B,j})$ is distance-valid.

This paragraph shows that there exist $\delta_{A, j}$ for some $k \in \mathbb{Z}[i]$ in \eqref{eqn:dm2} such that $0 < |\delta_{A, j}| < |q|$. By  \cite[\emph{Theorem 3.1}]{gi}, for $\delta^{char}_{A}$ where $ |\delta^{char}_{A}|^2>1$ (i.e., $\delta^{char}_{A}\notin\{\pm 1,\pm i\}$), there exists a $k\in \mathbb{Z}[i]$ such that
\begin{align}\label{eqn:dx2}
|\delta_{A,j}|^2\leq \frac{1}{2}|\delta^{char}_{A,i}|^2
\end{align}
Thus, $|\delta_{A,j}|^2\leq \frac{1}{2}|\delta^{char}_{A,i}|^2\leq \frac{1}{2}(2|q|^2-4q^R+2)<|q|^2$, where the second inequality is due to the necessary condition in \emph{Lemma \ref{lem:suf}}. Furthermore, $\delta_{A,j}\neq 0$ (otherwise, we would have $\delta_{B,j} \delta^{char}_{A,i}=-1$ in \eqref{eqn:dminb1}, implying $|\delta^{char}_{A,i}|^2=1$; but the proof here assumes $|\delta^{char}_{A,i}|^2>1$ since the case $|\delta^{char}_{A,i}|^2=1$ has been covered by \emph{Lemma \ref{lem:be1i}}).

The next few paragraphs show that given the $\delta_{A,j}$ found in the previous paragraph, the corresponding $\delta_{B, j}$ that satisfies \eqref{eqn:dm2} is such that
$0 <|\delta_{B, j}| < |q|$. Thus, the pair $(\delta_{A, j}, \delta_{B, j})$ is distance-valid according to \emph{Lemma \ref{lem:nec}}. Recall that the pair $(\delta_{A,j}, \delta_{B,j})$ must satisfy \eqref{eqn:dminb1}. With respect to \eqref{eqn:dminb1}, Fig. \ref{fig:tri} below draws the relationship between the vectors $\delta_{A,j}\delta^{char}_{B,i}$, $-\delta_{B,j} \delta^{char}_{A,i}$ and $1$ in the complex plane:
\begin{figure}[h]
  \centering
        \includegraphics[height=0.24\columnwidth]{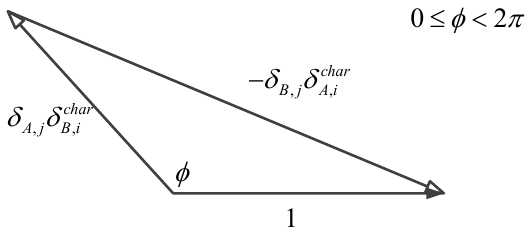}
       \caption{The triangle formed by the vectors $\delta_{A,j}\delta^{char}_{B,i}$, $-\delta_{B,j} \delta^{char}_{A,i}$, and $1$.}
        \label{fig:tri}
\end{figure}

By cosine rule,
\begin{align}
\nonumber |\delta_{B,j} \delta^{char}_{A,i}|^2&=|\delta_{A,j}\delta^{char}_{B,i}|^2+1
-2|\delta_{A,j}\delta^{char}_{B,i}|cos\phi\\
&\leq |\delta_{A,j}\delta^{char}_{B,i}|^2+1+2|\delta_{A,j}\delta^{char}_{B,i}|.
\end{align}

Then, we have
\begin{align}\label{eqn:cosr}
\nonumber |\delta_{B,j}|^2 &\leq \frac{ |\delta_{A,j}\delta^{char}_{B,i}|^2+1+2|\delta_{A,j}\delta^{char}_{B,i}|}{|\delta^{char}_{A,i}|^2}\\
&\leq \frac{1}{2}|\delta^{char}_{B,i}|^2+\frac{1}{2}+\sqrt{2}\frac{|\delta^{char}_{B,i}|}{|\delta^{char}_{A,i}|},
\end{align}
where the second inequality is due to \eqref{eqn:dx2} and $|\delta^{char}_{A,i}|^2>1$, because the case where $|\delta^{char}_{A,i}|=1$ has been covered by \emph{Lemma \ref{lem:be1i}}.

W.l.o.g., we assume  $\frac{|\delta^{char}_{B,i}|}{|\delta^{char}_{A,i}|}\leq 1$ (alternatively, if  $\frac{|\delta^{char}_{B,i}|}{|\delta^{char}_{A,i}|}> 1$, we switch the roles of A and B in \eqref{eqn:dx2} and a similar argument follows after that).

Given $|q|\geq \sqrt{5}$ there are two possibilities: $q$ is real or $q$ is complex. For real $q$, we have $q^R=|q|\geq 2$. For complex $q$, w.l.o.g., we assume $q^R>q^I\geq 1$.   (note: for a given complex $q$ with $|q|\geq \sqrt{5}$, we can always find one $q$ such that  $q^R>q^I\geq 1$; the proofs are similar for other cases if we apply the symmetry of $\mathbb{Z}[i]$ and $\mathbb{Z}[i]/q$ to our previous lemmas and the proof here). Thus, overall, whether $q$ is real or complex, we have $q^R\geq 2$. Continuing from \eqref{eqn:cosr}, we have
\begin{align}
\nonumber |\delta_{B,j}|^2 &\leq \frac{1}{2}|\delta^{char}_{B,i}|^2+\frac{1}{2}+\sqrt{2}\\
&\leq \frac{1}{2}(2|q|^2-6)+\frac{1}{2}+\sqrt{2}<|q|^2,
\end{align}
where the second inequality holds since $|\delta^{char}_{B,i}|^2\leq2|q|^2-6$  by substituting $q^R\geq2$ in $\sqrt{2|q|^2-4q^R+2}$ in the statement of \emph{Lemma \ref{lem:suf}}.

Furthermore, $\delta_{B,j}\neq 0$ since if $\delta_{B,j}=0$, \eqref{eqn:dminb1} would become $\delta_{A,j} \delta_{B,i}^{char}=1$, but we are not considering the case where $|\delta_{B,i}^{char}|^2=1$ here since it has been covered by \emph{Lemma \ref{lem:be1i}}.

\textbf{P2)}  Suppose that the distance-valid $(\delta_{A,j},\delta_{B,j})$ found in P1) are not coprime in $\mathbb{Z}[i]$, i.e., $\gcd(\delta_{A,j},\delta_{B,j})=d$ and $|d|^2>1$. By \emph{Lemma 3}, there exist some $(x',y')\in\mathbb{Z}[i]$ such that $x'\delta_{A,j}+y'\delta_{B,j}=d$, and $|d|^2$ is the smallest norm of $x'\delta_{A,j}+y'\delta_{B,j}$. However, the smallest norm of $x'\delta_{A,j}+y'\delta_{B,j}$ is $1$ from \eqref{eqn:dminb1}, contradicting $|d|^2>1$.

 \end{IEEEproof}

\section{Weighted Voronoi Region of Zero-$l_{\min}$ Channel Gains} \label{sec:vor}


 Section \ref{sec:l0} analyzed the distance properties at zero-$l_{\min}$ channel gains. This section moves on to the study of general channel gains where $l_{\min}$  is not necessary zero. Intuitively, for channel gains $\eta$  in the near neighborhood of a zero-$l_{\min}$ channel gain   $\eta^o_i$, the linear NC mapping  $(\alpha_{opt},\beta_{opt})=(-(\delta_{A,i}^{char (q)})^{-1} \delta_{B,i}^{char (q)},1)$ for $\eta^o_i$ is still optimal in that it will still yield the largest possible  $d^{(\alpha_{opt},\beta_{opt})}_{\min}$. In general, as we will see, the complex plane of $\eta$  can be partitioned into multiple Voronoi regions, with each region containing exactly one  at zero-$l_{\min}$  channel gain  and that the optimal NC mapping  $(\alpha_{opt},\beta_{opt})$ for that $\eta^o$  applies to all $\eta$ within the Voronoi region of  $\eta^o$.{\footnote{As far as $l_{\min}$ analysis is concerned, the Voronoi-region analysis in this section applies to both linear and nonlinear PNC mappings; it is the $d^{(\alpha_{opt},\beta_{opt})}_{\min}$  analysis in Part F that applies specifically  to linear PNC mapping. In particular, our $l_{\min}$ analysis for complex modulations also applies to nonlinear PNC, should someone wants to further study nonlinear PNC.}}

With reference to Fig. \ref{fig:lmin}, the set of channel gains that should adopt the NC mapping  $(\alpha_{opt},\beta_{opt})=(-(\delta_{A,i}^{char (q)})^{-1} \delta_{B,i}^{char (q)},1)$ is given by
\begin{align}\label{eqn:scg}
\nonumber &\mathcal{V}(\eta^o_i)\triangleq \{\eta\in\mathbb{C}\big|d_{\eta^o_i\rightarrow \eta}\leq d_{\eta^o_j\rightarrow \eta}, \forall j\neq i \}\\
=&\{\eta\!\in\!\mathbb{C}\big| |\delta_{A,i}^{char}\eta\!+\!\delta_{B,i}^{char}|\!\leq\!|\delta_{A,j}^{char}\eta\!+\!\delta_{B,j}^{char}|, \forall j\neq i\}.
\end{align}
In other words, for an  $\eta\in \mathcal{V}(\eta^o_i)$, $l_{\min}(\eta)$  is given by $|\delta_{A,i}^{char}\eta+\delta_{B,i}^{char}|$  and not by   $|\delta_{A,j}^{char}\eta+\delta_{B,j}^{char}|$,  $j \neq i$. Note that in \eqref{eqn:scg}, we have generalized the definition of the weighted distance from one zero-$l_{\min}$ channel gain to another zero-$l_{\min}$ channel gain in \eqref{l:wd} to a weighted distance from a zero-$l_{\min}$  channel gain $\eta^o_i$  to a general channel gain $\eta$  to as follows:
\begin{align}\label{eqn:ged}
d_{\eta^o_i\rightarrow \eta}=|\delta_{A,i}^{char}\eta+\delta_{B,i}^{char}|.
\end{align}
We refer to $\mathcal{V}(\eta^o_i)$ as the weighted Voronoi region of $\eta^o_i$.

Section \ref{sec:l0} showed that for   $\eta=\eta^o_i$, $d^{(\alpha_{opt},\beta_{opt})}_{\min}=\frac{1}{|\delta_{A,i}^{char}|}$  and that there is always another characteristic difference $(\delta_{A,j}^{char},\delta_{B,j}^{char})$  whose normalized distance with respect to $(\delta_{A,i}^{char},\delta_{B,i}^{char})$   is one---i.e., $|\delta_{A,j} \delta_{B,i}-\delta_{B,j}\delta_{A,i} |=1$. For a general $\eta$  within the Voronoi region of  $\eta^o_i$, however, the situation is more complicated.  In general, $\eta^o_i$  may have several neighbors whose Voronoi regions share a boundary with $\eta^o_i$   and the  $d^{(\alpha_{opt},\beta_{opt})}_{\min}$ at a particular  $\eta\in \mathcal{V}(\eta^o_i)$  is the weighted distance of one of these neighbors to  $\eta$; however,  $d^{(\alpha_{opt},\beta_{opt})}_{\min}$  at different   $\eta\in \mathcal{V}(\eta^o_i)$  may be determined by the weighted distances of different neighbors. The normalized distance between some of these neighbors and  $(\delta_{A,j}^{char},\delta_{B,j}^{char})$  may be larger than one.

\subsection{Preliminaries on the Weighted Voronoi Region}

\begin{figure}[t]
  \centering
        \includegraphics[height=0.5\columnwidth]{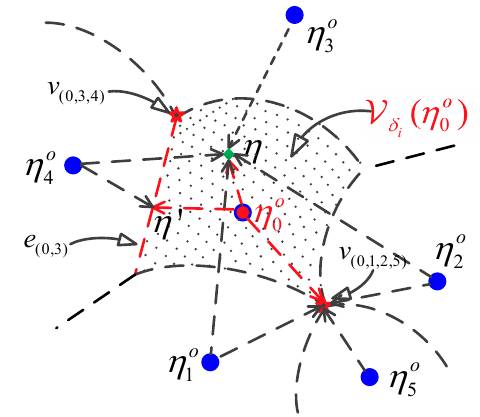}
       \caption{The weighted Voronoi of $\eta^o_0$ and its adjacent regions.}
        \label{fig:vorpre}
\end{figure}

Fig. \ref{fig:vorpre} shows the weighted Voronoi region of $\eta^o_0$ in the complex plane of $\eta$. Illustrated by this figure, we introduce some definitions and essential properties of the weighted Voronoi region \cite{vor1,vor2,vor3}.

In the complex plane of $\eta$, we consider a set of distinct generators $\{\eta^o_0, \eta^o_1, \ldots, \eta^o_I\}$ (generators are simply zero-$l_{\min}$ channel gains in our problem) and assign a weight $\delta_i$ to each $\eta^o_i$. With this weight, we define a distance from the generator $\eta^o_i$ to any other point as a weighted distance $d_{\eta^o_i\rightarrow \eta}$ from $\eta^o_i$ to $\eta$:
\begin{align}\label{eqn:vd}
d_{\eta^o_i\rightarrow \eta}\triangleq |\delta_i(\eta^o_i-\eta)|.
\end{align}

\emph{Weighted Voronoi Region}: The weighted Voronoi region of $\eta^o_i$ is defined as (see the shaded region in Fig. \ref{fig:vorpre})
\begin{align}\label{eqn:vdr}
\mathcal{V}(\eta^o_i)\triangleq \{\eta\in\mathbb{C}|d_{\eta^o_i\rightarrow \eta}\leq d_{\eta^o_j\rightarrow \eta}, \forall j\neq i \}.
\end{align}
Given any point $\eta\in \mathcal{V}(\eta^o_i)$, the weighted distance from $\eta^o_i$ to $\eta$ is no more than that from $\eta^o_j$ to $\eta$, $\forall j\neq i$. We call $\eta^o_i$ of $\mathcal{V}(\eta^o_i)$  the \emph{generator} of the weighted Voronoi region.


\emph{Voronoi Edge}: A \emph{closed} weighted Voronoi region of $\eta^o_i$  contains its boundary that consists of straight lines and circular arcs, which we call weighted Voronoi edges. Mathematically, if $\mathcal{V}(\eta^o_i)\cap \mathcal{V}(\eta^o_j) \neq \emptyset$, the set $\mathcal{V}(\eta^o_i)\cap \mathcal{V}(\eta^o_j)$ gives a Voronoi edge (which may degenerate into a point). We denote a Voronoi edge shared by $\mathcal{V}(\eta^o_i)$ and $\mathcal{V}(\eta^o_j)$ as
\begin{align}\label{eqn:ve}
e_{(i,j)}\triangleq \{\eta\in\mathbb{C}|d_{\eta^o_i\rightarrow \eta}= d_{\eta^o_j\rightarrow \eta}, \forall j\neq i \}.
\end{align}
That is, given any $\eta$ on the edge $e_{(i,j)}$, the weighted distance from  $\eta^o_i$ to $\eta$ is the same as the weighted distance from $\eta^o_j$ to $\eta$. In particular, in the complex plane with the coordinates  $(\eta^R, \eta^I)$, an edge is a circular arc if and only if the weights of the weighted Voronoi regions sharing the edge are different, i.e.,
\begin{eqnarray} \label{eqn:ve1}
\nonumber   & e_{(i,j)}: |\delta^{char}_{A,i}(\eta^o_i-\eta)|=|\delta^{char}_{A,j}(\eta^o_j-\eta)|\Rightarrow\\
 \nonumber  & \big(\eta^R-\frac{(\eta^o_i)^R|\delta^{char}_{A,i}|^2-(\eta^o_j)^R|\delta^{char}_{A,j}|^2}{|\delta^{char}_{A,i}|^2-|\delta^{char}_{A,j}|^2} \big)^2\\ &
  \nonumber   +\big(\eta^I-\frac{(\eta^o_i)^I|\delta^{char}_{A,i}|^2-(\eta^o_j)^I|\delta^{char}_{A,j}|^2}{|\delta^{char}_{A,i}|^2-|\delta^{char}_{A,j}|^2} \big)^2  \\ \nonumber &
 =\frac{|\delta^{char}_{A,i}|^2|\delta^{char}_{A,j}|^2}{(|\delta^{char}_{A,i}|^2-|\delta^{char}_{A,j}|^2)^2}[\big((\eta^o_i)^R-(\eta^o_j)^R\big)^2+\big((\eta^o_i)^I-(\eta^o_j)^I\big)^2],
\end{eqnarray}
and an edge is a straight line if and only if the weights of the weighted Voronoi regions sharing the edge are the same, i.e.,
\begin{align}\label{eqn:ve2}
\nonumber & e_{(i,j)}: |\eta^o_i-\eta|=|\eta^o_j-\eta|\Rightarrow\\
&\eta^I=\frac{(\eta^o_i)^R-(\eta^o_j)^R}{(\eta^o_j)^I-(\eta^o_i)^I}\eta^R+\frac{|\eta^o_j|^2-|\eta^o_i|^2}{2(\eta^o_j)^I-(\eta^o_i)^I}.
\end{align}

Note that $e_{(i,j)}$ is empty if $\mathcal{V}(\eta^o_i)\cap \mathcal{V}(\eta^o_j)= \emptyset$. In Fig. \ref{fig:vorpre}, the red dashed line $e_{(0,3)}$ is a Voronoi edge shared by $\mathcal{V}(\eta^o_0)$ and $\mathcal{V}(\eta^o_3)$.

\emph{Voronoi Vertex}: An end point of a Voronoi edge is called a Voronoi vertex. We denote a Voronoi vertex shared by three or more Voronoi regions  $\mathcal{V}(\eta^o_i), \mathcal{V}(\eta^o_j), \mathcal{V}(\eta^o_k), \ldots$ as
\begin{align}\label{eqn:vv}
 \nonumber e_{(i,j, k,\ldots)}\triangleq & \big\{\eta\in \mathbb{C}|d_{\eta^o_i\rightarrow \eta}=d_{\eta^o_j\rightarrow \eta}=d_{\eta^o_k\rightarrow \eta}=\ldots, \\ & \forall j\neq k \neq i\neq \ldots \big\}.
\end{align}
 In Fig. \ref{fig:vorpre},  $\mathcal{V}(\eta^o_0)$, $\mathcal{V}(\eta^o_3)$, and $\mathcal{V}(\eta^o_4)$ meet at a Voronoi vertex $v_{(0,3,4)}$.

\emph{Adjacent Voronoi Regions}: Two Voronoi regions are said to be adjacent if the Voronoi regions share a Voronoi edge or a Voronoi vertex. We also say that two characteristic differences,  $(\delta_{A,i}^{char},\delta_{B,i}^{char})$ and $(\delta_{A,j}^{char},\delta_{B,j}^{char})$, are adjacent if their associated Voronoi regions, $\mathcal{V}(-\frac{\delta_{B,i}^{char}}{\delta_{A,i}^{char}})$ and $\mathcal{V}(-\frac{\delta_{B,j}^{char}}{\delta_{A,j}^{char}})$, are adjacent.

In Fig. \ref{fig:vorpre},  $\mathcal{V}(\eta^o_0)$ and $\mathcal{V}(\eta^o_4)$ are adjacent since they share the same edge $e_{(0,4)}$, and $\mathcal{V}(\eta^o_0)$, $\mathcal{V}(\eta^o_1)$, $\mathcal{V}(\eta^o_2)$, and $\mathcal{V}(\eta^o_5)$ are adjacent since they meet at a point $v_{(0,1,2,5)}$.

\subsection{Optimal NC mapping of Voronoi Regions}

In this part, we show that the optimal NC mapping for an arbitrary $\eta\in \mathcal{V}(\eta^o_0)$ is $(-(\delta_{A}^{char (q)})^{-1} \delta_{B}^{char (q)},1)$, the same as that for  $\eta=\eta^o$, when $\eta^o$  is a nontrivial zero-$l_{\min}$  channel gain.

\begin{remark}\label{rem:optge}
 For  the two trivial zero-$l_{\min}$  channel gains $\eta^o=0$ and  $\eta^o=\infty$, we cannot find an NC mapping to cluster $(\delta^{char}_A, \delta^{char}_B)=(1,0)$  which induces a zero  $l_{\min}$  at  $\eta^o=0$, and to cluster  $(\delta^{char}_A, \delta^{char}_B)=(1,0)$ which induces a zero $l_{\min}$  at $\eta^o=\infty$. Both   $(\delta^{char}_A, \delta^{char}_B)=(1,0)$ and   $(\delta^{char}_A, \delta^{char}_B)=(0,1)$  are distance-valid, but not NC-valid, characteristic differences. See \emph{Trivial Theorem} in  {Section \ref{sec:l0}} for details. Extrapolating this observation to the Voronoi region of $\eta^o=0$ and  $\eta^o=\infty$, we conclude that $d_{\min}(\eta)=l_{\min}(\eta)$  within their Voronoi regions.
\end{remark}\rightline{$\blacksquare$}

\begin{theorem}\label{thm:optgd}
Consider a nontrivial zero-$l_{\min}$ channel gain $\eta^o$  associated with the characteristic difference $(\delta^{char}_A, \delta^{char}_B)$. For all $\eta\in \mathcal{V}(\eta^o)$, the optimal NC mapping is the same as that for the zero-$l_{\min}$  channel gain $\eta^o$, i.e., $(\alpha_{opt},\beta_{opt})=(-(\delta_{A}^{char (q)})^{-1} \delta_{B}^{char (q)},1)$.
\end{theorem}
\begin{IEEEproof}[Proof of Theorem \ref{thm:optgd}]
 $l_{\min}$ at an arbitrary $\eta\in \mathcal{V}(\eta^o)$ is the weighted distance from the generator $\eta^o=-\frac{\delta^{char}_B}{\delta^{char}_A}$ to $\eta$, i.e., $l_{\min}(\eta)=l_{(\delta^{char}_A, \delta^{char}_B)}(\eta)=|\eta \delta^{char}_A+\delta^{char}_B|$ for $\eta\in \mathcal{V}(\eta^o)$. If $\eta=\eta^o$, then the optimal NC mapping follows from \emph{Theorem \ref{thm:eta0}} for the zero-$l_{\min}$ channel gain. If  $\eta\neq\eta^o$ and $\eta\in \mathcal{V}(\eta^o)$, we have $l_{\min}>0$. In this case, to ensure $d^{(\alpha_{opt},\beta_{opt})}_{\min}\geq l_{\min}>0$, we need to cluster $(\delta^{char}_A, \delta^{char}_B)$. Otherwise, $d^{(\alpha_{opt},\beta_{opt})}_{\min}=l_{\min}$. Similar to the proof of \emph{Theorem  \ref{thm:eta0}}, the solution for this clustering is $(\alpha_{opt},\beta_{opt})=(-(\delta_{A}^{char (q)})^{-1} \delta_{B}^{char (q)},1)$. Once $(\delta^{char}_A, \delta^{char}_B)$ is clustered, the NC partitioning is fixed, and there is no further freedom to cluster another NC-valid difference pair that does not belong to the clustered-difference set of $(\alpha_{opt},\beta_{opt})$.
\end{IEEEproof}

\subsection{Identifying  $d_{\min}^{(\alpha_{opt},\beta_{opt})}$ in Voronoi Regions}

In Section \ref{sec:l0}, we have identified $d_{\min}^{(\alpha_{opt},\beta_{opt})}$  at zero-$l_{\min}$  channel gains.   This part aims to identify $d_{\min}^{(\alpha_{opt},\beta_{opt})}$    for general channel gains. Let us consider $\eta$ in a particular Voronoi region generated by zero-$l_{\min}$ channel gain $\eta^o$.   To explicitly identify the  $d_{\min}^{(\alpha_{opt},\beta_{opt})}$-determining difference, we first consider an exhaustively search method as follows,  before putting forth an efficient method for doing so in \emph{Theorem \ref{thm:gdmin}}.
\begin{figure}[t]
  \centering
        \includegraphics[height=0.41\columnwidth]{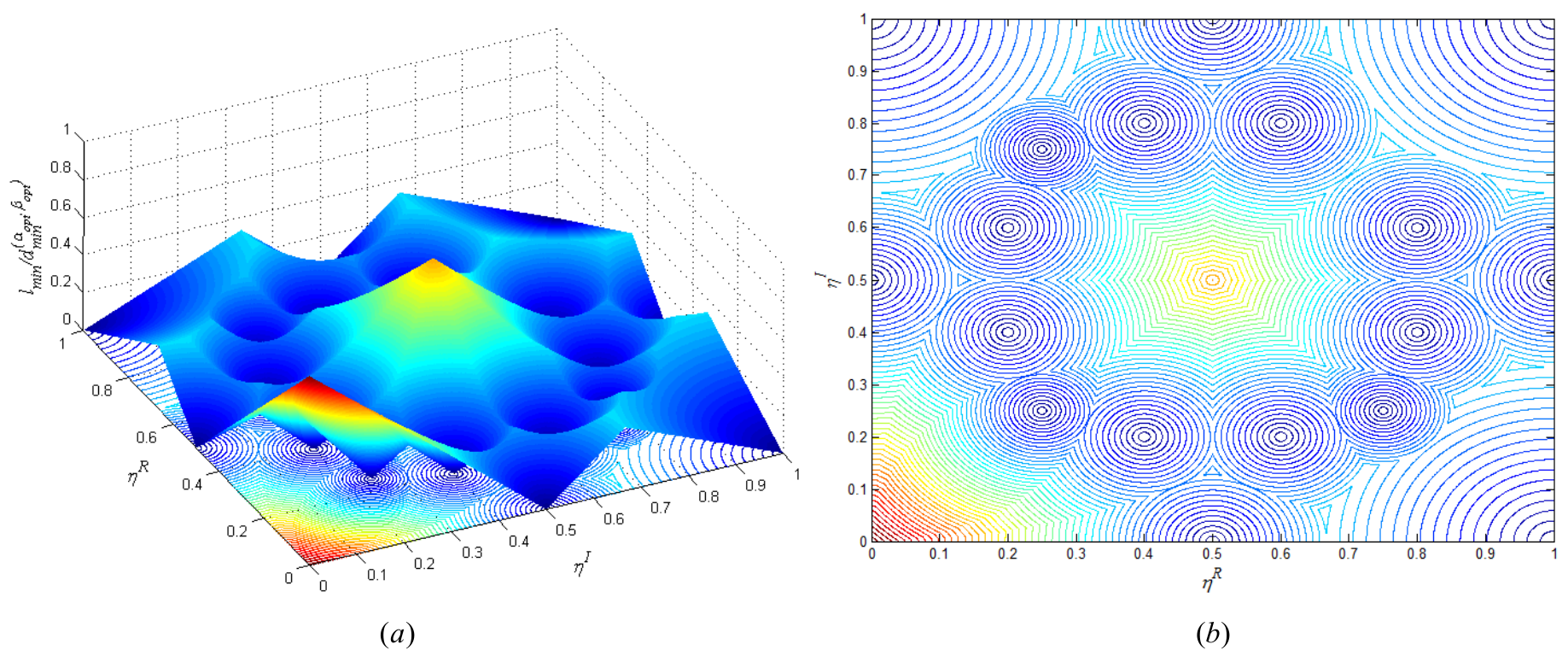}
       \caption{(a) $l_{\min}(\eta)$ for $\eta\notin \mathcal{V}(-\frac{\delta^{char}_B}{\delta^{char}_A})$ and $d_{\min}^{(\alpha_{opt},\beta_{opt})}(\eta)$ versus $\eta\in \mathcal{V}(-\frac{\delta^{char}_B}{\delta^{char}_A})$, where $(\delta^{char}_A,\delta^{char}_B)=(1+i,-i)$ and $q=3$; (b) the corresponding contour graphs of $l_{\min}(\eta)$ and $d_{\min}^{(\alpha_{opt},\beta_{opt})}(\eta)$.}
        \label{fig:rocd1}
\end{figure}

As shown in Fig. \ref{fig:rocd1}, one way to find the $d_{\min}^{(\alpha_{opt},\beta_{opt})}$-determining difference at an $\eta\in \mathcal{V}(-\frac{\delta^{char}_B}{\delta^{char}_A})$ is to first remove the characteristic difference  $(\delta^{char}_A, \delta^{char}_B)$ and the other clustered differences of  $(\alpha_{opt},\beta_{opt})$  (i.e., remove difference pairs in the clustered-difference set $\Delta_{(\alpha_{opt}, \beta_{opt})}$), and then numerically search for the new characteristic difference $({\delta^{char}_A}', {\delta^{char}_B}')$ that determines the ``new''  $l_{\min}$ for $\eta\in \mathcal{V}(-\frac{\delta^{char}_B}{\delta^{char}_A})$ in the absence of difference pairs in $\Delta_{(\alpha_{opt}, \beta_{opt})}$. The $l_{\min}$ for an $\eta\in \mathcal{V}(-\frac{\delta^{char}_B}{\delta^{char}_A})$  in the absence of $\Delta_{(\alpha_{opt}, \beta_{opt})}$  is the  $d_{\min}^{(\alpha_{opt},\beta_{opt})}$ at this $\eta$, and  $({\delta^{char}_A}', {\delta^{char}_B}')$ is the  $d_{\min}^{(\alpha_{opt},\beta_{opt})}$-determining difference at this $\eta$.

We formally define and describe ``removal of the characteristics differences'' as follows (an illustrating example is given in Fig. \ref{fig:rocd1}):

\emph{\textbf{Removal of Characteristic Differences Induced by Optimal NC mapping (ROCD)}}: The optimal NC mapping $(\alpha_{opt},\beta_{opt})$  \emph{removes} the difference pairs in $\Delta_{(\alpha_{opt}, \beta_{opt})}$  from consideration in the process of finding the  $d_{\min}^{(\alpha_{opt},\beta_{opt})}$-determining differences at an arbitrary  $\eta\in \mathcal{V}(-\frac{\delta^{char}_B}{\delta^{char}_A})$. Specifically, after the removal of such   difference pairs, we redraw the Voronoi regions of the remaining characteristic differences. The other characteristic differences ``close to'' $(\delta^{char}_A, \delta^{char}_B)$  may divide the region  among them so that their ``new'' Voronoi regions may include part of  the old $\mathcal{V}(-\frac{\delta^{char}_B}{\delta^{char}_A})$. The ``new'' $l_{\min}$   associated with an  $\eta\in \mathcal{V}(-\frac{\delta^{char}_B}{\delta^{char}_A})$   is the $d_{\min}^{(\alpha_{opt},\beta_{opt})}$  for that $\eta$.

 {Theorem \ref{thm:gdmin}}  below  puts forth an efficient approach to identify the $d_{\min}^{(\alpha_{opt},\beta_{opt})}$-determining differences at an arbitrary $\eta\in\mathcal{V}(\eta^o)$  by stating that only characteristic differences adjacent to $(\delta_A^{char}, \delta_B^{char})$ can be $d_{\min}^{(\alpha_{opt},\beta_{opt})}$-determining differences .

\begin{theorem}\label{thm:gdmin}
Consider an arbitrary channel gain $\eta\in\mathcal{V}(\eta^o)$, where $\eta^o=-\frac{\delta^{char}_B}{\delta^{char}_A}$ is a nontrivial zero-$l_{\min}$ channel gain associated with the characteristic difference $(\delta^{char}_A, \delta^{char}_B)$. With the optimal NC mapping $(\alpha_{opt},\beta_{opt})$, the $d_{\min}^{(\alpha_{opt},\beta_{opt})}$-determining difference at this $\eta$ is a characteristic difference $({\delta^{char}_A}', {\delta^{char}_B}')$   that is adjacent to  $(\delta^{char}_A, \delta^{char}_B)$.
\end{theorem}

\begin{IEEEproof}[Proof of Theorem \ref{thm:gdmin}]

Our proof depends on two results that will be proved later in Part F:
\begin{itemize}
  \item[(T4-1)] $(\alpha_{opt},\beta_{opt})$ cannot cluster any characteristic difference   $({\delta^{char}_A}', {\delta^{char}_B}')$ that is adjacent to $(\delta^{char}_A, \delta^{char}_B)$.

  \item[(T4-2)] For any characteristic difference $({\delta^{char}_A}'', {\delta^{char}_B}'')$ that is not adjacent to $(\delta^{char}_A, \delta^{char}_B)$, given any  $\eta\in\mathcal{V}(\eta^o)$, there exists a characteristic difference $({\delta^{char}_A}', {\delta^{char}_B}')$ adjacent to $(\delta^{char}_A, \delta^{char}_B)$ such that  $|\eta{\delta^{char}_B}''+{\delta^{char}_A}''|>|\eta{\delta^{char}_B}'+{\delta^{char}_A}'|$ $\forall \eta\in \mathcal{V}(\eta^o)\backslash\{\eta^o\}$, and $|\eta{\delta^{char}_B}''+{\delta^{char}_A}''|\geq|\eta{\delta^{char}_B}'+{\delta^{char}_A}'|$ at $\eta=\eta^o$.
\end{itemize}

 (T4-2) means that given any  $\eta\in\mathcal{V}(\eta^o)$ and a non-adjacent characteristic difference, there is always  an adjacent characteristic difference that is closer to  $\eta$  than any given non-adjacent characteristic difference. (T3-1) says that this adjacent characteristic difference cannot be clustered by the optimal NC mapping $(\alpha_{opt},\beta_{opt})$ applied within   $\mathcal{V}(\eta^o)$. Thus, under ROCD, the $d_{\min}^{(\alpha_{opt},\beta_{opt})}$  within $\mathcal{V}(\eta^o)$  must be determined by characteristic differences adjacent to  $({\delta^{char}_A}, {\delta^{char}_B})$, and not by non-adjacent characteristic differences.

\end{IEEEproof}

\begin{remark}\label{rem:adj}
Note that for different $\eta$  within  $\mathcal{V}(\eta^o)$, the  $d_{\min}^{(\alpha_{opt},\beta_{opt})}$  may be determined by different characteristic differences, but they must all be adjacent to  $(\delta^{char}_A, \delta^{char}_B)$.
\end{remark}\rightline{$\blacksquare$}


The proof of (T4-1) requires some background to be established regarding the properties of adjacent characteristic differences (specifically, the normalized distances between a characteristic difference and its adjacent characteristic differences). Parts D and E below will first establish this background. Part F will then provide the proofs for (T4-1) and (T4-2).

\subsection{Notations and Definitions}

 In  Part E, we will put forth an efficient way to identify characteristic differences that are adjacent to $(\delta^{char}_A, \delta^{char}_B)$ through their normalized distances to $(\delta^{char}_A, \delta^{char}_B)$  and other properties. We will draw heavily on the formalism in \cite{mah} when deriving our results. For easy cross-reference by the reader,  we redefine some notations in this part  for consistency with the notations used in \cite{mah}. In addition, we also put forth some new definitions in preparation for the discussion in Part E.

\emph{Notation Modifications}: We express a zero-$l_{\min}$ channel gain $\eta^o$ as a ratio of two Gaussian integers, e.g., $(\kappa, \tau)$. Then, we have $\eta^o=\frac{\kappa}{\tau}$ where $\kappa=-\delta^{char}_B$ and $\tau=\delta^{char}_A$ (note: we have switched the position of $(\delta_A, \delta_B)$ in $(\kappa, \tau)$ since this is notation used in \cite{mah}). Thus, $\gcd(\kappa, \tau)=1$. Furthermore, we denote the weighted Voronoi region of $(\kappa, \tau)$ as  $\mathcal{V}(\frac{\kappa}{\tau})$.

With respect to \eqref{eqn:dels}, we have defined $\Delta$ as the set of distance-valid difference pairs induced by elements of $\mathbb{Z}[i]/q$  for some Gaussian primes $q$. Here, we define the subset of $\Delta$ that collects all characteristic differences in $\Delta$ as the $\mathcal{Q}^{char}_q$-set:
\begin{align}\label{set:Q}
\mathcal{Q}^{char}_q\triangleq\big\{(\kappa, \tau)  \big|   {(-\tau, \kappa)\in\Delta}, \gcd(\kappa, \tau)=1\big\}.
\end{align}
  Note that by definition of  $\Delta$,   $\kappa$ and $\tau$ cannot both be zero at the same time.


With respect to \eqref{set:Q}, the elements in $\mathcal{Q}^{char}_q$-set  form a generalized \emph{Farey Sequence} in $\mathbb{Z}[i]$ \cite{mah}. Each element in the Farey Sequence is an irreducible fraction $\frac{\kappa}{\tau}, (\kappa, \tau)\in\Delta$. There is a bijective mapping between the elements in the $\mathcal{Q}^{char}_q$-set and the elements in the associated Farey Sequence.

We further define a dual set of $\mathcal{Q}^{char}_q$-set as follows:
\begin{align}\label{set:Qt}
\nonumber\mathcal{Q}_q&\triangleq\big\{(\kappa,\tau)\in\mathbb{Z}^2[i] \big|\\
&\exists\upsilon\in\{\pm 1, \pm i\}, (\frac{\upsilon\kappa}{\gcd(\kappa,\tau)}, \frac{\upsilon\tau}{\gcd(\kappa,\tau)})\in \mathcal{Q}^{char}_q\big\}.
\end{align}
Let us elaborate the definition in \eqref{set:Qt}. First, let us write    $(\kappa', \tau')=(\frac{\upsilon\kappa}{\gcd(\kappa,\tau)}, \frac{\upsilon\tau}{\gcd(\kappa,\tau)})$. Obviously,  $\frac{\kappa'}{\tau'}$  is an \emph{irreducible} fraction. According to the above definition, if   $(\kappa', \tau')\in\mathcal{Q}^{char}_q$, then  $(\kappa, \tau)\in\mathcal{Q}_q$. Furthermore, we note that $\gcd$ of two Gaussian integers is not unique. If $x$ is a particular gcd of $\kappa$  and  $\tau$, so are  $-x,ix,-ix$. In general, there are four possible ways to reduce  $(\kappa, \tau)$ to  $(\kappa', \tau')$. In \eqref{set:Qt}, we write  $(\kappa', \tau')=(\frac{\upsilon\kappa}{\gcd(\kappa,\tau)}, \frac{\upsilon\tau}{\gcd(\kappa,\tau)})$, where $\gcd(\kappa,\tau)$  refers specifically to one of the four possible $\gcd$'s. Now, in general, it is possible for some of the $(\kappa', \tau')$  to belong to the set  $\mathcal{Q}^{char}_q$ and some not. According to our definition in \eqref{set:Qt}, we require only at least one of the four $(\kappa', \tau')$  to belong to the set $\mathcal{Q}^{char}_q$  in order that  $(\kappa, \tau)\in \mathcal{Q}_q$. That is, $(\kappa, \tau)\in \mathcal{Q}_q$  if there is a $\upsilon \in \{\pm1, \pm i\}$ such that $(\kappa', \tau')\in \mathcal{Q}^{char}_q$.

Two elements $({\kappa}',  {\tau}')\in \mathcal{Q}^{char}_q$ and $({\kappa},{\tau})\in\mathcal{Q}_q$ are said to be \emph{equivalent} if $\frac{\kappa'}{\tau'}=\frac{{\kappa}}{{\tau}}$. These two ratios are the same and they correspond to the same zero-$l_{\min}$ channel gain in the communication problem of this paper.  The reason for defining  $\mathcal{Q}_q$ is for the convenience of the statements of some lemmas and proofs later (specifically, \emph{Q-criteria 1-3}). In the proofs, when we identify a pair $(\kappa,\tau)\in\mathcal{Q}_q$, that means we have also identified a pair  $(\kappa',  \tau')\in \mathcal{Q}^{char}_q$.


Now, our problem can be reformulated as how to characterize the weighted Voronoi regions of elements in $\mathcal{Q}^{char}_q$-set. Clearly, the weighted Voronoi region of each element in $\mathcal{Q}^{char}_q$-set is bounded by a finite set of lines and arcs, and these boundaries  are generated by the adjacent Voronoi regions. From the definition of adjacent Voronoi regions, two pairs $({\kappa}, {\tau}),({\gamma}, {\delta})\in \mathcal{Q}^{char}_q$ are said to be  adjacent if the weighted Voronoi regions $\mathcal{V}(\frac{\kappa}{\tau})$ and $\mathcal{V}(\frac{\gamma}{\delta})$ have a point $z'$ in common  (note: $z'$ here is $\eta$ in precious sections), i.e.,
\begin{align}\label{eqn:com}
\nonumber &d_{\frac{\kappa}{\tau}\rightarrow z'}=d_{\frac{\gamma}{\delta}\rightarrow z'}=|\tau(\frac{\kappa}{\tau}-z')|=|\delta(\frac{\gamma}{\delta}-z')|\\
&=\min_{\forall(\zeta,\vartheta)\in\mathcal{Q}^{char}_q}|\vartheta(\frac{\zeta}{\vartheta}-z')|
=\min_{\forall({\tilde{\zeta}},{\tilde{\vartheta}})\in\mathcal{Q}_q}|{\tilde{\vartheta}}(\frac{{{\tilde{\zeta}}}}{{\tilde{\vartheta}}}-z')|.
\end{align}

Before we detail our approach to identify the adjacent regions, let us introduce some relevant results from   \cite{mah} that provides useful insights to solve our problem. Specifically, \cite{mah} developed a systematic way to study  the approximation of complex numbers by numbers of the quadratic field $\mathbb{Q}(\sqrt{-1})$ (i.e., approximating complex numbers by  Gaussian rationals formed by ratios of two Gaussian integers $\mathbb{Z}[i]$, in a way that is analogous to approximating real numbers by rational numbers). The approximation problem can also be characterized by identifying the weighted Voronoi regions of a set of generators drawn from $\mathbb{Q}(\sqrt{-1})$, i.e., all complex numbers in a Voronoi region is approximated by its generator. However, \cite{mah} considered a different set of irreducible fractions $\frac{\kappa}{\tau}$ such that $|\kappa|^2, |\tau|^2\leq N$, where $N$ is a real integer. In this paper, we define the set Gaussian integers  used in \cite{mah} as the $\mathcal{N}^{char}_N$-set and the ${\mathcal{N}_N}$-set:{\footnote {$\mathcal{N}^{char}_N$ and $\mathcal{N}_N$ correspond to $\textfrak{F}_N$ and $\textfrak{G}_N$ in \cite{mah} respectively.}}
\begin{align}\label{set:Qt1}
\nonumber \mathcal{N}^{char}_N\triangleq\big\{(\kappa, \tau)\in \mathbb{Z}^2[i]\backslash \{(0,0)\}\big|& \gcd(\kappa, \tau)=1, \\ & ~  |\kappa|^2,|\tau|^2\leq N\big\},
\end{align}
\vspace{-0.25in}
\begin{align}\label{set:Qt2}
\nonumber  \mathcal{N}_N\triangleq\big\{(\kappa, \tau)\in & \mathbb{Z}^2[i]\backslash \{(0,0)\}\big| \\
 & (\frac{ \kappa}{\gcd(\kappa,\tau)}, \frac{ \tau}{\gcd(\kappa,\tau)}) \in \mathcal{N}^{char}_N\big\}.
\end{align}

\begin{remark}
  Note that both $\mathcal{N}^{char}_N$ and $\mathcal{N}_N$  have the symmetry property, i.e. given a $(\kappa,\tau)$ in $\mathcal{N}^{char}_N$ and $\mathcal{N}_N$, we can find the other three elements as $\{-(\kappa, \tau), i(\kappa, \tau),-i(\kappa, \tau)\}$  symmetric to $(\kappa,\tau)$ also in $\mathcal{N}^{char}_N$ and $\mathcal{N}_N$, due to $|\kappa|^2,|\tau|^2\leq N$. However, depending on $q$,  $\mathcal{Q}^{char}_q$ and $\mathcal{Q}_q$  may not retain this symmetry. To see this, we know that the elements in $\mathcal{Q}^{char}_q$ and $\mathcal{Q}_q$ are induced from any two distinct elements in $\mathbb{Z}[i]/q$ with a Gaussian prime $q$. The elements in $\mathbb{Z}[i]/q$ may not be symmetric, i.e., given $(w_A, w_B)\in \mathbb{Z}[i]/q$, we  have $\upsilon(w_A, w_B)\notin \mathbb{Z}[i]/q$ for some $\upsilon\in \{\pm 1, \pm i\}$. Therefore, we need to specify  $\upsilon$ that yields $\upsilon(\frac{\kappa}{\gcd(\kappa,\tau)}, \frac{\tau}{\gcd(\kappa,\tau)})\in\mathcal{Q}^{char}_q $ for $(\kappa,\tau)\in \mathcal{Q}_q$ in  \eqref{set:Qt}. In \eqref{set:Qt2}, either all four ways of reducing $(\kappa, \tau)$    give rise to an element in  $\mathcal{N}^{char}_N$, or none of the four reductions does. Hence, we do not distinguish the four ways of reduction in the definition of $\mathcal{N}_N$  in \eqref{set:Qt2}.
\end{remark}\rightline{$\blacksquare$}

Given an arbitrary  pair $(\kappa,\tau)\in\mathcal{N}^{char}_N$, \cite{mah} gives a selection criterion to find distinct elements in $\mathcal{N}^{char}_N$-set that are adjacent to $(\kappa,\tau)$. We refer to it as $\mathcal{N}$-criterion.

\emph{$\mathcal{N}$-criterion} \cite[\emph{Theorem IV}]{mah}: Consider two distinct pairs $(\kappa,\tau)$ and $(\gamma, \delta)\in \mathcal{N}^{char}_N$, $(\kappa,\tau)\neq(\nu\gamma, \nu\delta)$ where $\nu=\pm 1$ or $\pm i$.
A necessary and sufficient condition for $(\kappa,\tau)$ and $(\gamma, \delta)$ to be adjacent is that simultaneously

{\noindent (i)  $|\kappa\delta-\tau\gamma|=1$ or $\sqrt{2}$,}

{\noindent (ii) $(\kappa+\epsilon\gamma, \tau+\epsilon\delta)\notin \mathcal{N}_N$ for some choice of  $\epsilon=\pm 1$ or $\pm i$.}

\rightline{$\blacksquare$}

\subsection{Properties and Identification of Adjacent Characteristic Differences}\label{subsec:qcri}


In this part, we put forth three criteria, referred to as the  \emph{$\mathcal{Q}$-criteria 1-3},   to identify the adjacency relationships among elements in the $\mathcal{Q}^{char}_q$-set.  These criteria are analogous to, but not exactly the same as,   the  $\mathcal{N}$-criterion in  \cite[\emph{Theorem IV}]{mah}.

\emph{$\mathcal{Q}$-criterion 1:} Consider two distinct  pairs $(\kappa,\tau) ,(\gamma, \delta)\in \mathcal{Q}^{char}_q$, where $|\Xi|\triangleq|\kappa\delta-\tau\gamma|=1$ or $\sqrt{2}$. The two pairs are adjacent if and only if $(\kappa+\epsilon\gamma, \tau+\epsilon \delta)\notin {\mathcal{Q}_q}$ for some $\epsilon\in\{\pm 1, \pm i\}$.
 \begin{IEEEproof}[Proof of $\mathcal{Q}$-criterion 1]
 For $|\Xi|=1$ or $\sqrt{2}$, we simply go through the proof of \emph{$\mathcal{N}$-criterion} in \cite[\emph{Theorem IV}]{mah} and verify that after substituting the $\mathcal{Q}^{char}_q$-set for the $\mathcal{N}^{char}_N$-set, the proof remains valid.

\end{IEEEproof}

The following criterion goes beyond the \emph{$\mathcal{N}$-criterion} in \cite{mah} because for the $\mathcal{Q}^{char}_q$-set where two distinct pairs $(\kappa,\tau),(\gamma, \delta)\in\mathcal{Q}^{char}_q$ can also be adjacent if $|\kappa\delta-\gamma\tau|=\sqrt{5}$.

\emph{$\mathcal{Q}$-criterion 2:} Consider two  distinct pairs $(\kappa,\tau),(\gamma, \delta)\in\mathcal{Q}^{char}_q$, where  $|\Xi|\triangleq|\kappa\delta-\tau\gamma|=\sqrt{5}$.
\begin{enumerate}
  \item[(i)]  Let $\epsilon \in \{\pm 1, \pm i\}$. Among the four possible values for $\epsilon$, there exists one and only one value such that $\kappa+\epsilon\gamma=0 ~({\rm mod}~ \Xi)$ and $\tau+\epsilon \delta=0 ~({\rm mod}~ \Xi)$;
  \item[(ii)]  $(\kappa,\tau)$ and $(\gamma, \delta)$ are adjacent if and only if for the $\epsilon\in\{\pm 1, \pm i\}$ that satisfies $\kappa+\epsilon\gamma=0 ~({\rm mod}~ \Xi)$ and $\tau+\epsilon \delta=0 ~({\rm mod}~ \Xi)$, $(\phi,\psi)\triangleq(\kappa+\epsilon\gamma,\tau+\epsilon \delta)\notin \mathcal{Q}_q$.
\end{enumerate}
\begin{IEEEproof}[Proof of $\mathcal{Q}$-criterion 2]
For $|\Xi|=\sqrt{5}$, $\Xi=\upsilon(2+i)$ or $\Xi=\upsilon(2-i)$ where $\upsilon$ is a unit. We focus on $\Xi=\upsilon(2+i)$ (the proof for $\Xi=\upsilon(2-i)$ is similar). We first prove (i). Note that $2+i$ is a Gaussian prime. Thus, $\mathbb{Z}[i]/(2+i)=\{0, \pm 1, \pm i\}$ is a field. We write
\begin{subequations}\label{eqn:qr}
\begin{align}
\label{eqn:qr11}\kappa=\Xi q_{\kappa}+r_{\kappa},\\
\label{eqn:qr21}\tau=\Xi q_{\tau}+r_{\tau},\\
\label{eqn:qr31}\gamma=\Xi q_{\gamma}+r_{\gamma},\\
\label{eqn:qr41}\delta=\Xi q_{\delta}+r_{\delta},
\end{align}
\end{subequations}
where $q_x\in\mathbb{Z}[i]$ and $r_x \in\mathbb{Z}[i]/(2+i)=\{0,\pm 1,\pm i\}$ denote the quotients and remainders, respectively, when $x$ is divided by $\Xi=2+i$. Given $|\Xi|=|\kappa\delta-\tau\gamma|$, we must have $r_{\kappa}r_{\delta}=r_{\tau}r_{\gamma}~({\rm mod}~\Xi)$. Further, by \emph{Lemma \ref{lem:re}} (presented  later), we have three possibilities:
\begin{enumerate}
  \item[(p1)] $r_{\kappa}=r_{\gamma}=0$, $r_{\tau},r_{\delta}\neq 0$;
  \item[(p2)] $r_{\tau}=r_{\delta}=0$, $r_{\kappa},r_{\gamma}\neq 0$;
  \item[(p3)]  $r_{\kappa}, r_{\tau},r_{\gamma},r_{\delta}\neq0$, $r_{\kappa}r_{\gamma}^{-1}=r_{\tau}r_{\delta}^{-1}~({\rm mod}~2+i).$
\end{enumerate}

We further write
\begin{subequations}
\begin{align}
\kappa+\epsilon\gamma=\Xi(q_{\kappa}+\epsilon q_{\gamma})+r_{\kappa}+\epsilon r_{\gamma},\\
\tau+\epsilon\delta=\Xi(q_{\tau}+\epsilon q_{\delta})+r_{\tau}+\epsilon r_{\delta}.
\end{align}
\end{subequations}

In order that $\kappa+\epsilon\gamma=0 ~({\rm mod}~2+i)$ and $\tau+\epsilon\delta=0 ~({\rm mod}~2+i)$, we must have $r_{\kappa}+\epsilon r_{\gamma}=0 ~({\rm mod}~2+i)$ and $r_{\tau}+\epsilon r_{\delta}=0 ~({\rm mod}~2+i)$. If (p1) above applies, we let $\epsilon=-r_{\tau}r_{\delta}^{-1}~({\rm mod}~2+i)$; if (p2) above applies, we let $\epsilon=-r_{\kappa}r_{\gamma}^{-1}~({\rm mod}~2+i)$; if (p3) above applies, we let $\epsilon=-r_{\kappa}r_{\gamma}^{-1}=-r_{\tau}r_{\delta}^{-1}~({\rm mod}~2+i)$. Note that for all three cases,  $\epsilon\in \{\pm 1, \pm i\}$ and there is only one such $\epsilon$ that serves the purpose. This proves (i). We next prove (ii).

``\emph{If}'' part: $(\kappa,\tau)$ and $(\gamma, \delta)$ are adjacent if for the   $\epsilon\in\{\pm 1, \pm i\}$ that satisfies $\kappa+\epsilon\gamma=0 ~({\rm mod}~\Xi)$ and $\tau+\epsilon\delta=0 ~({\rm mod}~\Xi)$, $(\phi,\psi)\triangleq(\kappa+\epsilon\gamma,\tau+\epsilon\delta)\notin {\mathcal{Q}_q}$.

First, we show that both $\phi \neq 0$ and $\psi \neq 0$. Suppose that $\phi=\kappa+\epsilon\gamma=0$. Then,  $\kappa=-\epsilon\gamma$. Substituting this into $\kappa\delta-\tau\gamma$, we have  $\kappa\delta-\tau\gamma=\gamma(-\epsilon\gamma-\tau)$, but this contradicts the fact that $\kappa\delta-\tau\gamma=2+i$  is a Gaussian prime that cannot be factorized. Thus, $\phi\neq 0$. Similarly, $\psi\neq 0$. Given that $\phi= 0 ({\rm mod}~\Xi)$ and $\psi= 0 ({\rm mod}~\Xi)$, and that $\phi\neq 0$  and  $\psi\neq 0$, let us define $(\phi', \psi'
)\triangleq(\frac{\phi}{\Xi},\frac{\psi}{\Xi})$, where $\phi'\neq 0$  and  $\psi'\neq0$. Further, define $z\triangleq\frac{\phi'}{\psi'}$.

If $(\phi,\psi)\notin \mathcal{Q}_q$, we have $(\phi',\psi')\notin \mathcal{Q}_q$ and therefore $\eta=z=\frac{\phi'}{\psi'}$ is not a zero-$l_{\min}$ channel gain  (i.e., $(\phi', \psi')$ is not a distance-valid difference pair). In the following, we show that there is no other distance-valid difference pair that is closer to $\eta=z$ than are  $(\kappa,\tau)$ and $(\gamma, \delta)$  in terms of weighted distance, and that $(\kappa,\tau)$ and $(\gamma, \delta)$  are equidistant to  $\eta=z$. In other words, $\eta=z$ lies on the edge of the Voronoi regions of $(\kappa,\tau)$ and $(\gamma, \delta)$. Thus, $(\kappa,\tau)$ and $(\gamma, \delta)$  must be adjacent.

The weighted distances from $\frac{\kappa}{\tau}$ to $z$ and $\frac{\gamma}{\delta}$ to $z$ are
\begin{align}\label{eqn:wdx}
\nonumber |\tau z-\kappa|&=|\delta z-\gamma|=\frac{|\delta \phi'-\gamma\psi'|}{|\psi'|}\\
&=\frac{|\delta \phi-\gamma\psi|/|\Xi|}{|\psi'|}=\frac{1}{|\psi'|}.
\end{align}
where we can verify that $|\delta \phi-\gamma\psi|=|\Xi|$.

Now, let us consider any arbitrary $(a,b)\in \mathcal{Q}^{char}_q$ where $(a,b)\neq (\kappa,\tau), (\gamma, \delta)$. Since the normalized distance from $\frac{a}{b}$ to $z$ is at least $1$, we have
\begin{align}\label{eqn:wd2}
  |b \phi'-a\psi'|\geq 1 &\Rightarrow |b \frac{\phi'}{\psi'}-a|\geq \frac{1}{|\psi'|}.
\end{align}

From \eqref{eqn:wdx} and \eqref{eqn:wd2}, the weighted distance from $\frac{a}{b}$ to $z$ is not less than $|\tau z-\kappa|=|\delta z-\gamma|=\frac{1}{|\psi'|}$. Therefore, $(\kappa,\tau)$  and $(\gamma, \delta)$ are adjacent in the $\mathcal{Q}^{char}_q$-set.

``\emph{Only if}'' part:  $(\kappa,\tau)$ and $(\gamma, \delta)$ are adjacent only if for the  $\epsilon\in\{\pm 1, \pm i\}$ that satisfies $\kappa+\epsilon\gamma=0 ~({\rm mod}~\Xi)$ and $\tau+\epsilon\delta=0 ~({\rm mod}~\Xi)$, $(\phi,\psi)=(\kappa+\epsilon\gamma,\tau+\epsilon\delta)\notin {\mathcal{Q}_q}$.

Suppose that $(\phi,\psi)=(\kappa+\epsilon\gamma,\tau+\epsilon\delta)\in {\mathcal{Q}_q}$, we want to show that $(\kappa,\tau)$ and $(\gamma, \delta)$ are not adjacent.
Again, as in the proof of the ``\emph{if}'' part, we let
 $(\phi',\psi')=(\frac{\phi}{\Xi}, \frac{\psi}{\Xi})$. Note that    $(\phi',\psi')\in   \mathcal{Q}_q$ given that   $(\phi,\psi)\in {\mathcal{Q}_q}$.
 Consider the boundary (edge) $z$ between $(\kappa,\tau)$ and $(\gamma, \delta)$ defined by
\begin{align}\label{eqnl1:e}
|\tau z-\kappa|=|\delta z-\gamma|.
\end{align}

In \eqref{eqnl1:e1} below, we prove that for any point $z$ that lies on the boundary as specified in  \eqref{eqnl1:e},  $|\psi'z-\phi'|<|\tau z-\kappa|=|\delta z-\gamma|$. In other words, $(\phi',\psi')$  is closer to $z$ than are $(\kappa,\tau)$ and $(\gamma, \delta)$, and thus  $(\kappa,\tau)$ and $(\gamma, \delta)$ are not adjacent.
\begin{align}\label{eqnl1:e1}
\nonumber &|\psi'z-\phi'|=\frac{|\psi z-\phi|}{|\Xi|}=\frac{|\tau z-\kappa+\epsilon(\delta z-\gamma)|}{|\Xi|}\\
\nonumber&=|\tau z-\kappa|\frac{|1+\epsilon(\delta z-\gamma)/(\tau z-\kappa)|}{|\Xi|}
\\\nonumber & \leq |\tau z-\kappa|\frac{1+|\epsilon(\delta z-\gamma)/(\tau z-\kappa)|}{|\Xi|}\\&=|\tau z-\kappa|\frac{1+1}{|\Xi|}
<|\tau z-\kappa|.
\end{align}
where the last inequality holds because $|\Xi|=\sqrt{5}$.

\end{IEEEproof}

 An example showing that it is possible for two characteristic differences to be adjacent if their normalized distance is $\sqrt{5}$  is as follows.  Consider the case of $q=11$. Let $\kappa=10+9i, \tau=1-10i, \gamma=9+8i$,  $\delta=1-9i$, and $\epsilon=1$. In this case, $|\Xi|=|\kappa\delta-\tau\gamma|=|2+i|=\sqrt{5}$. We can verify that (i) in \emph{$\mathcal{Q}$-criterion 2} is satisfied only when $\epsilon=1$, since
\begin{align}\label{eqn:exa}
\nonumber \Xi\phi=\kappa+\epsilon\gamma & \Rightarrow(2+i)(11+3i)=19+17i,\\
\nonumber & \Rightarrow \phi=11+3i,\\
\nonumber \Xi\psi=\tau+\epsilon\delta & \Rightarrow(2+i)(-3-8i)=2-19i,\\
& \Rightarrow \psi=-3-8i.
\end{align}

Given \eqref{eqn:exa}, we can further verify that  $(\kappa,\tau)$ and $(\gamma, \delta)$  are adjacent, since $(\phi,\psi)=(11+3i,-3-8i)  \notin {\mathcal{Q}_{11}}$. This can be easily seen as follows. For  $q=11$, any valid symbol $w=w^R+iw^I$  has its real and imaginary parts bounded as  $-5\leq w^R,w^I\leq 5$. Thus, the difference between two valid symbols is bounded as  $-10\leq \delta^R,\delta^I\leq 10$. Clearly,  the real part of $\phi=11+3i$ does not satisfy this bound.

\begin{lemma}\label{lem:re}
With respect to the statement of \emph{${\mathcal{Q}}$-criterion 2} and the equations as written in \eqref{eqn:qr}, given that $r_\kappa r_\delta=r_\tau r_\gamma ~({\rm mod}~\Xi)$ and $\gcd(\kappa, \tau)=\gcd(\gamma,\beta)=1$, (i) it is not possible for $r_\kappa=r_\tau=0$ or $r_\gamma=r_\delta=0$; (ii) for a Gaussian prime $\Xi$, we have $r_\kappa=0\Leftrightarrow r_\gamma=0$ and $r_\tau=0\Leftrightarrow r_\delta=0$.
\end{lemma}
\begin{IEEEproof}[Proof of Lemma \ref{lem:re}]
(i) is obvious because if $r_\kappa=r_\tau=0$, then from \eqref{eqn:qr}, $\Xi$  is a common factor of $\kappa$ and $\tau$, but this contradicts the fact that $\gcd(\kappa, \tau)=1$ given that $(\kappa, \tau) \in Q_q^{char}$. Similarly,  it is not possible that   $r_\gamma=r_\delta=0$. Therefore, it is not possible for $r_\kappa=r_\tau=0$ or $r_\gamma=r_\delta=0$.

For (ii),  w.l.o.g., suppose that $r_\kappa=0$, then $r_\kappa r_\delta = r_\tau r_\gamma ~({\rm mod}~\Xi)$ implies  either $r_\gamma=0$ or $r_\tau r_\gamma$ is a non-zero multiple of $\Xi$ (this later case, however, is not possible because when $\Xi$ is prime, finite-field arithmetic applies to the remainders. The multiplication of any two nonzero elements of a finite field does not give $0$---i.e., it is not congruent to $\Xi$). Thus, we have $r_\kappa=0\Rightarrow r_\gamma=0$. By symmetry argument, we thus have $r_\kappa=0\Leftarrow r_\gamma=0$. Similarly, we have $r_\tau=0\Leftrightarrow r_\delta=0$. 	

\end{IEEEproof}

\emph{$\mathcal{Q}$-criterion 3:}  Consider two distinct pairs $(\kappa,\tau),(\gamma, \delta)\in\mathcal{Q}^{char}_q$, $(\kappa,\tau)\neq(\nu\gamma, \nu\delta)$ where $\nu=\pm 1$ or $\pm i$, and that $|\Xi|\triangleq|\kappa\delta-\tau\gamma|\neq 1, \sqrt{2}$ or $\sqrt{5}$. The two pairs are not adjacent.
\begin{IEEEproof}[Proof of $\mathcal{Q}$-criterion 3]
The proof of \emph{$\mathcal{Q}$-criterion 3} is given by a series of lemmas. As we will show by the following lemmas, $(\kappa,\tau),(\gamma, \delta)$ are not adjacent if

\begin{itemize}
  \item  $|\Xi|\geq40$  by \emph{Lemmas \ref{lem:81}} and \emph{\ref{lem:82}};
  \item  $|\Xi|=\sqrt{13}, \sqrt{17}, \sqrt{2}\sqrt{13}, \sqrt{29}, \sqrt{2}\sqrt{17},$ or $\sqrt{37}$ by \emph{Lemmas \ref{lem:81}} and  \emph{\ref{lem:83}};
  \item $|\Xi|=\sqrt{10}$ and  $\sqrt{2}\sqrt{10}$ by \emph{Lemmas \ref{lem:81}} and  \emph{\ref{lem:84}};
 \item $|\Xi|=5$ by \emph{Lemmas \ref{lem:81}} and  \emph{\ref{lem:86}};
   \item   $|\Xi|=2, 2\sqrt{2}, 3, 2\cdot2, 3\sqrt{2}, 2\sqrt{8}$, or $2\cdot3$ by  \emph{Lemma  \ref{lem:87}}.
\end{itemize}

\end{IEEEproof}

\begin{lemma}[A modified version of  Lemma 4 in \cite{mah}]\label{lem:81}
Consider two distinct $(\kappa,\tau),(\gamma, \delta)\in \mathcal{Q}^{char}_q$, $(\kappa,\tau)\neq(\nu\gamma, \nu \delta)$ where $\nu=\pm 1$ or $\pm i$, and that $\Xi\triangleq\kappa\delta-\tau\gamma$ contains a factor $\tilde{\Xi}$. The two pairs are not adjacent if there exist $(\zeta, \vartheta),(\phi, \psi)\in \mathbb{Z}^2[i]\backslash \{(0,0)\}$ such that
 \begin{subequations}\label{eqnl81}
\begin{align}
\label{eqnl81:1}\kappa\zeta+\gamma\vartheta=\tilde{\Xi}\phi,\\
\label{eqnl81:2}\tau\zeta+\delta\vartheta=\tilde{\Xi}\psi,\\
\label{eqnl81:3} 0< |\zeta|+|\vartheta|\leq\frac{|\tilde{\Xi}|}{\sqrt{2}}.
\end{align}
\end{subequations}
 \end{lemma}

\begin{IEEEproof}[Proof of Lemma \ref{lem:81}]
Suppose that  $|\Xi|\neq1, \sqrt{2}$ or $\sqrt{5}$, but $(\kappa,\tau)$ and $(\gamma, \delta)$ are adjacent. According  to \eqref{eqnl81:1},  we have
\begin{align}\label{eqn:qc32}
\nonumber |\phi|&=|\frac{\kappa\zeta+\gamma\vartheta}{\tilde{\Xi}}|\leq \frac{|\zeta|+|\vartheta|}{|\tilde{\Xi}|}\max\{|\kappa|,|\gamma|\}\\
&\leq\frac{1}{\sqrt{2}}\max\{|\kappa|,|\gamma|\}\leq\frac{\sqrt{2|q|^2-4q^R+2}}{\sqrt{2}}
<|q|.
\end{align}
 where the second inequality holds due to \eqref{eqnl81:2}   and the third equality is due to \emph{Lemma \ref{lem:suf}}. Similarly, we have $|\psi|<|q|$. Therefore, by \emph{Lemma \ref{lem:nec}}, we have $(\phi,\psi)\in\mathcal{Q}_q$.

Now, if $(\kappa,\tau)$ and $(\gamma, \delta)$ are adjacent, then from \eqref{eqn:com}, there exists a common point $z'$ equidistant to $(\kappa,\tau)$ and $(\gamma, \delta)$ such that no other generators are closer to $z'$ than are $(\kappa,\tau)$ and $(\gamma, \delta)$. Let us compute the weighted distance from $(\phi, \psi)$ to $z'$:
\begin{align}\label{eqn:conex}
\nonumber|\psi z'-\phi|&=\frac{|\zeta(\tau z'-\kappa)+\vartheta(\delta z'-\gamma)|}{|\tilde{\Xi}|}\\ \nonumber & \leq \frac{|\zeta(\tau z'-\kappa)|+|\vartheta(\delta z'-\gamma)|}{|\tilde{\Xi}|}\\
\nonumber & \leq \frac{|\zeta|+|\vartheta|}{|\tilde{\Xi}|}\max\{|\tau z'-\kappa|,|\delta z'-\gamma|\}\\
& \leq\frac{|\tau z'-\kappa|}{\sqrt{2}}<|\tau z'-\kappa|.
\end{align}
  Obviously,   \eqref{eqn:conex} contradicts our assumption that $(\kappa,\tau)$ and $(\gamma, \delta)$ are adjacent.

\end{IEEEproof}

\begin{sublemma}\label{lem:82}
Consider two distinct $(\kappa,\tau),(\gamma, \delta)\in \mathcal{Q}^{char}_q$, $(\kappa,\tau)\neq(\nu\gamma, \nu \delta)$ where $\nu=\pm 1$ or $\pm i$, and that $|\Xi|\triangleq|\kappa\delta-\tau\gamma|\geq 40$. There exist $(\zeta, \vartheta),(\phi, \psi)\in \mathbb{Z}^2[i]\backslash \{(0,0)\}$ such that
 \begin{subequations}\label{eqnl82}
\begin{align}
\label{eqnl82:1} &\kappa\zeta+\gamma\vartheta=\Xi\phi,\\
\label{eqnl82:2}  & \tau\zeta+\delta\vartheta=\Xi\psi,\\
\label{eqnl82:3} & 0< |\zeta|+|\vartheta|\leq\frac{|\Xi|}{\sqrt{2}}.
\end{align}
\end{subequations}
 \end{sublemma}

\begin{IEEEproof}[Proof of Lemma \ref{lem:82}]
If we write $\zeta=\zeta^R+i\zeta^I$ and  $\vartheta=\vartheta^R+i\vartheta^I$ where $\zeta^R,\zeta^I,\vartheta^R,\vartheta^I$ are real, then $(\zeta^R,\zeta^I,\vartheta^R,\vartheta^I)^T$ that satisfy \eqref{eqnl82:1} and \eqref{eqnl82:2} form a four-dimensional lattice $\Lambda$ of determinant $d(\Lambda)=|\Xi|^2$, as explained below.

Let us represent Gaussian integers in \eqref{eqnl82:1} by $2\times 2$ matrices and $2\times 1$ vectors. Specifically, we rewrite \eqref{eqnl82:1} as
 \begin{align}\label{eqn:lat}
\left(\begin{array}{cc}
\bm{\kappa} & \bm{\gamma} \\
\bm{\tau} &  \bm{\delta}\\
\end{array}
\right)           \left(\begin{array}{c}
\bm{\zeta} \\
\bm{\vartheta} \\
                   \end{array}\right)  =\left(\begin{array}{cc}
\bm{\Xi} & \mathbf{0} \\
\mathbf{0} &  \bm{\Xi}\\
\end{array}
\right)           \left(\begin{array}{c}
\bm{\phi} \\
\bm{\psi} \\
                   \end{array}\right),
\end{align}
where
\begin{align}\label{eqn:lat1}
\nonumber \bm{\kappa}=\left(\begin{array}{cc}
 \kappa^R  & -\kappa^I \\
\kappa^I &  \kappa^R
\end{array}
\right),\bm{\gamma}=\left(\begin{array}{cc}
 \gamma^R  & -\gamma^I \\
\gamma^I &  \gamma^R
\end{array}
\right),\\
\nonumber \bm{\tau}=\left(\begin{array}{cc}
 \tau^R  & -\tau^I \\
\tau^I &  \tau^R
\end{array}
\right), \bm{\delta}=\left(\begin{array}{cc}
 \delta^R  & -\delta^I \\
\delta^I &  \delta^R
\end{array}
\right),\\
\nonumber \bm{\Xi}=\left(\begin{array}{cc}
 \Xi^R  & -\Xi^I \\
\Xi^I &  \Xi^R
\end{array}
\right)=\bm{\kappa}\bm{\delta}-\bm{\tau}\bm{\gamma},
\bm{\zeta}=\left(\begin{array}{c}
 \zeta^R   \\
\zeta^I
\end{array}
\right),\\\bm{\vartheta}=\left(\begin{array}{c}
 \vartheta^R   \\
\vartheta^I
\end{array}
\right),\bm{\phi}=\left(\begin{array}{c}
 \phi^R   \\
\phi^I
\end{array}
\right),
\bm{\psi}=\left(\begin{array}{c}
 \psi^R   \\
\psi^I
\end{array}
\right).
\end{align}

Multiplying both sides of \eqref{eqn:lat} by $\left(\begin{array}{cc}
\bm{\Xi}^{-1} & \mathbf{0} \\
\mathbf{0} &  \bm{\Xi}^{-1}\\
\end{array}
\right)  \left(\begin{array}{cc}
\bm{\delta} & \bm{-\gamma} \\
\bm{-\tau} &  \bm{\kappa}
\end{array}
\right)$, we get
 \begin{align}
 \!   \left( \! \begin{array}{c}
\bm{\zeta} \\
\bm{\vartheta} \\
                   \end{array} \! \right) \!  = \!  \left(\begin{array}{cc}
\bm{\Xi}^{-1} & \mathbf{0} \\
\mathbf{0} &  \bm{\Xi}^{-1}\\
\end{array}
\right)  \!   \left(\begin{array}{cc}
\bm{\delta} & \bm{-\gamma} \\
\bm{-\tau} &  \bm{\kappa}\\
\end{array}
\right)   \!  \left(\begin{array}{cc}
\bm{\Xi} & \mathbf{0} \\
\mathbf{0} &  \bm{\Xi}\\
\end{array}
\right)      \!      \left(\!  \begin{array}{c}
\bm{\phi} \\
\bm{\psi}
                   \end{array}\!  \right).
\end{align}

Thus, we see that $\left(\begin{array}{c}
\bm{\zeta} \\
\bm{\vartheta} \\
                   \end{array}\right)$  form a four-dimensional lattice $\Lambda$ with determinant
      \begin{align}
   \nonumber  & \det\bigg(\left(\begin{array}{cc}
\bm{\Xi}^{-1} & \mathbf{0} \\
\mathbf{0} &  \bm{\Xi}^{-1}\\
\end{array}
\right)    \left(\begin{array}{cc}
\bm{\delta} & \bm{-\gamma} \\
\bm{-\tau} &  \bm{\kappa}\\
\end{array}
\right)   \left(\begin{array}{cc}
\bm{\Xi} & \mathbf{0} \\
\mathbf{0} &  \bm{\Xi}\\
\end{array}
\right)   \bigg)\\
\nonumber   & =\det \left(\begin{array}{cc}
\bm{\Xi}^{-1} & \mathbf{0} \\
\mathbf{0} &  \bm{\Xi}^{-1}\\
\end{array}
\right)   \det \left(\begin{array}{cc}
\bm{\delta} & \bm{-\gamma} \\
\bm{-\tau} &  \bm{\kappa}\\
\end{array}
\right)   \det \left(\begin{array}{cc}
\bm{\Xi} & \mathbf{0} \\
\mathbf{0} &  \bm{\Xi}\\
\end{array}
\right)   \\
&=\det \left( \begin{array}{cc}
\bm{\delta} & \bm{-\gamma} \\
\bm{-\tau} &  \bm{\kappa}\\
\end{array}
\right)    = \det(\bm{\kappa}\bm{\delta}-\bm{\tau}\bm{\gamma}) = \det(\bm{\Xi})= |\Xi|^2.
      \end{align}

Next, to prove the validity of \eqref{eqnl82:2}, we have to show the existence of a lattice point other than $(0,0,0,0)$ in the convex region $\mathcal{G}$ as defined below:
\begin{align}
\nonumber \mathcal{G}\triangleq&\big\{(\zeta^R,\zeta^I,\vartheta^R,\vartheta^I)\in \mathbb{R}^4 \big|  \\ & \sqrt{(\zeta^R)^2+(\zeta^I)^2}+\sqrt{(\vartheta^R)^2+(\vartheta^I)^2}\leq \frac{|\Xi|}{\sqrt{2}}\big\}.
\end{align}

Note that the zero lattice point $(\zeta^R,\zeta^I,\vartheta^R,\vartheta^I)=(0,0,0,0)$  is not acceptable because of the statement of the lemma that $|\zeta|+|\vartheta|>0$; on the other hand, a non-zero lattice point of $(\zeta^R,\zeta^I,\vartheta^R,\vartheta^I)$  automatically yields a non-zero solution for $(\phi^R,\phi^I,\psi^R,\psi^I)$  according to \eqref{eqn:lat}.

The volume of $\mathcal{G}$ is given by
\begin{align}\label{eqn:conr}
\mathcal{V}(\mathcal{G})=\int_{(\zeta^R,\zeta^I,\vartheta^R,\vartheta^I)\in \mathcal{G}} d\zeta^R d\zeta^I d\vartheta^R d\vartheta^I.
\end{align}

By change of  rectangular coordinate systems of $(\zeta^R,\zeta^I)$  and $(\vartheta^R,\vartheta^I)$  to polar coordinate systems, where $r=\sqrt{(\zeta^R)^2+(\zeta^I)^2}$  and  $r'=\sqrt{(\vartheta^R)^2+(\vartheta^I)^2}$, we can rewrite \eqref{eqn:conr}   as
\begin{align}
\nonumber \mathcal{V}(\mathcal{G})&=\int^{\frac{|\Xi|}{\sqrt{2}}}_{r=0} \int^{\frac{|\Xi|}{\sqrt{2}}-r}_{r'=0} (2\pi r)(2\pi r') dr' dr
\\
&=\int^{\frac{|\Xi|}{\sqrt{2}}}_{r=0} (2\pi r)\big(\pi(\frac{|\Xi|}{\sqrt{2}}-r)^2\big) dr=\frac{\pi^2 |\Xi|^4}{24}.
\end{align}

By \emph{Minkowski's Convex Body Theorem} \cite{cas}, there is a lattice point other than $(0,0,0,0)$ in   $\mathcal{G}$ if
\begin{align}
\mathcal{V}(\mathcal{G})>2^4 d(\Lambda).
\end{align}
That is, $\frac{\pi^2|\Xi|^4}{24}>2^4|\Xi|^2$ or $|\Xi|^2>\frac{384}{\pi^2}\approx38.9$. Therefore, we have proved the lemma for $|\Xi|^2\geq 40$.

\end{IEEEproof}

\emph{Lemmas \ref{lem:81}} and \emph{\ref{lem:82}} cover the cases with normalized distances  $|\Xi|=|\kappa\delta-\tau\gamma|\geq \sqrt{40}$. There are $16$ remaining cases of $\Xi$  when $|\Xi|^2\neq1, 2$, or $5$ and $|\Xi|^2<40$ as follows:
\begin{align*}
 &|\Xi|^2=4,8,9,10,13,16,17,18,20,25,26,29,32,34,36,37.\\
 &{ \ \rm i.e., \ }|\Xi|=2,2\sqrt{2},3,\sqrt{10},\sqrt{13},2\cdot2,\sqrt{17},3\sqrt{2},\sqrt{2}\sqrt{10},
 5,\\
 &\sqrt{2}\sqrt{13},\sqrt{29},2\sqrt{8},\sqrt{2}\sqrt{17},2\cdot3,\sqrt{37}.
\end{align*}

 \begin{sublemma}\label{lem:83}
Consider two distinct $(\kappa,\tau),(\gamma, \delta)\in \mathcal{Q}^{char}_q$, $(\kappa,\tau)\neq(\nu\gamma, \nu \delta)$ where $\nu=\pm 1$ or $\pm i$, and that $\Xi\triangleq\kappa\delta-\tau\gamma$ contains a factor $\tilde{\Xi}$ with magnitude $|\tilde{\Xi}|=\sqrt{13}, \sqrt{17}, \sqrt{29}, \sqrt{37}$. There exist $(\zeta, \vartheta),(\phi, \psi)\in \mathbb{Z}^2[i]\backslash \{(0,0)\}$ such that
  \begin{subequations} \label{eqnl83}
\begin{align}
\label{eqnl83:1} &\kappa\zeta+\gamma\vartheta=\tilde{\Xi}\phi,\\
 \label{eqnl83:2} & \tau\zeta+\delta\vartheta=\tilde{\Xi}\psi,\\
\label{eqnl83:3}  0 & < |\zeta|+|\vartheta|\leq\frac{|\tilde{\Xi}|}{\sqrt{2}}.
\end{align}
\end{subequations}
 \end{sublemma}

 {\emph{Remark:}} All these $\tilde{\Xi}$ are complex Gaussian-integer primes.

\begin{IEEEproof}[Proof of Lemma \ref{lem:83}]
The proof is given in   {Appendix IV}.

\end{IEEEproof}

\emph{Lemmas \ref{lem:81}} and  \emph{\ref{lem:83}}  cover the cases with   normalized distances  $|\Xi|=\sqrt{13}, \sqrt{17}, \sqrt{2}\sqrt{13}, \sqrt{29}, \sqrt{2}\sqrt{17}, \sqrt{37}$. The remaining cases are
\begin{align}
|\Xi|=2,2\sqrt{2},3,\sqrt{10},2\cdot2,3\sqrt{2},\sqrt{2}\sqrt{10},
 5,2\sqrt{8},2\cdot3.
\end{align}

 \begin{sublemma}\label{lem:84}
Consider two distinct $(\kappa,\tau),(\gamma, \delta)\in \mathcal{Q}^{char}_q$, $(\kappa,\tau)\neq(\nu\gamma, \nu \delta)$ where $\nu=\pm 1$ or $\pm i$, and that $\Xi\triangleq\kappa\delta-\tau\gamma$ contains a factor $\tilde{\Xi}$ with magnitude $|\tilde{\Xi}|=\sqrt{10}$. There exist $(\zeta, \vartheta),(\phi, \psi)\in \mathbb{Z}^2[i]\backslash \{(0,0)\}$ such that
  \begin{subequations} \label{eqnl84}
\begin{align}
\label{eqnl84:1} &\kappa\zeta+\gamma\vartheta=\tilde{\Xi}\phi,\\
 \label{eqnl84:2} & \tau\zeta+\delta\vartheta=\tilde{\Xi}\psi,\\
\label{eqnl84:3}  0 & < |\zeta|+|\vartheta|\leq\frac{|\tilde{\Xi}|}{\sqrt{2}}.
\end{align}
\end{subequations}
 \end{sublemma}

\begin{IEEEproof}[Proof of Lemma \ref{lem:84}]
The proof is given in  {Appendix V}.

\end{IEEEproof}

%
%
%
%
%
%
%
%

\emph{Lemmas \ref{lem:81}} and  \emph{\ref{lem:84}}  cover the cases with normalized distances   $|\Xi|=\sqrt{10}, \sqrt{2}\sqrt{10}$. The remaining cases  are
\begin{align*}
|\Xi|=2,2\sqrt{2},3, 2\cdot2,3\sqrt{2}, 5,2\sqrt{8},2\cdot3.
\end{align*}

\begin{sublemma}\label{lem:86}
Consider two distinct $(\kappa,\tau),(\gamma, \delta)\in \mathcal{Q}^{char}_q$, $(\kappa,\tau)\neq(\nu\gamma, \nu \delta)$ where $\nu=\pm 1$ or $\pm i$, and that $\Xi\triangleq\kappa\delta-\tau\gamma=5$. There exist $(\zeta, \vartheta), (\phi, \psi)\in \mathbb{Z}^2[i]\backslash \{(0,0)\}$ such that
 \begin{subequations} \label{eqnl86}
\begin{align}
\label{eqnl86:1} &\kappa\zeta+\gamma\vartheta={\Xi}\phi,\\
 \label{eqnl86:2} & \tau\zeta+\delta\vartheta={\Xi}\psi,\\
\label{eqnl86:3}  0 & < |\zeta|+|\vartheta|\leq\frac{|{\Xi}|}{\sqrt{2}}.
\end{align}
\end{subequations}
 \end{sublemma}

\begin{IEEEproof}[Proof of Lemma \ref{lem:86}]
The proof is given in  {Appendix VI}.

\end{IEEEproof}

\emph{Lemmas \ref{lem:81}} and  \emph{\ref{lem:86}}  cover the case with  normalized distance   $|\Xi|=5$. The remaining cases are $|\Xi|=2, 2\sqrt{2}, 3, 2\cdot2, 3\sqrt{2}, 2\sqrt{8}, 2\cdot3$.

In the following, we introduce the   concept of convex region for the set of valid differences.

\begin{figure}[t]
  \centering
        \includegraphics[height=0.5\columnwidth]{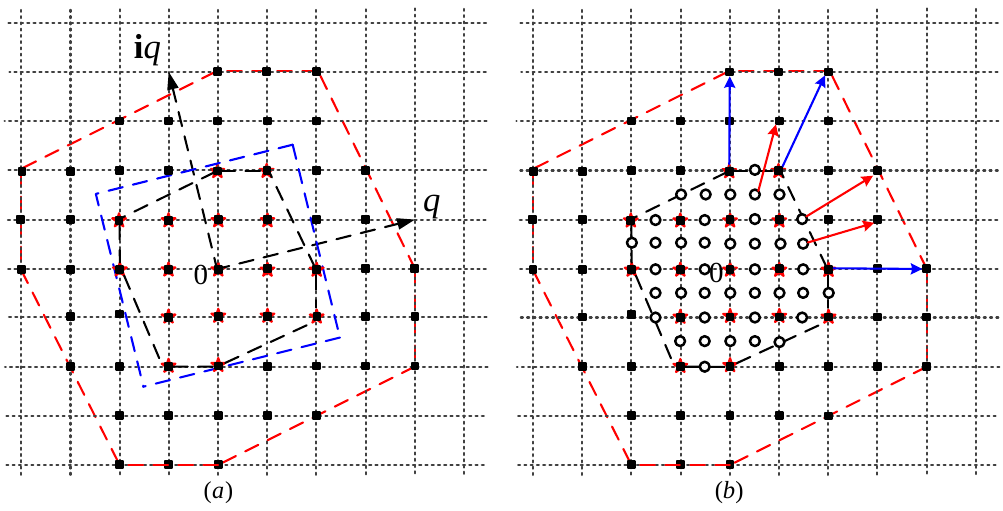}
       \caption{Valid differences of $q=4+i$ and convex region formed by the differences, where red stars are $\mathbb{Z}[i]/q$ and the black squares are distance-valid differences. (a) Gaussian integers within the blue square are elements of $\mathbb{Z}[i]/q$, and the inner octagon in black dashed line denotes the convex region formed by elements in $\mathbb{Z}[i]/q$; (b)  circles  and red stars  within the convex region of $\mathbb{Z}[i]/q$ (inner octagon) are valid differences scaled by half (i.e., the lattice points in the outer octagon scaled by half). }
        \label{fig:conv}
\end{figure}

\begin{definition}\label{def:con}
 Given a Gaussian prime $q$, we have defined $\Lambda$ (see \emph{Definition \ref{def:dvds}}) as a set of Gaussian integers that collects all valid differences (see Fig. \ref{fig:conv}). Given this  $\Lambda$, we can form a \emph{closed convex region} $\mathcal{G}_q$ on the complex plane, defined by
 \begin{align}\label{eqn:conv}
  \nonumber \mathcal{G}_q \triangleq&\big\{g \in \mathbb{C}  \big| g = \sum_{\delta_i \in \Lambda}  a_i  \delta_i, \\&~{\rm where}~  a_i \in \mathbb{R},     0 \leq \forall a_i\leq 1, {\rm and}~\sum_i a_i=1 \big\}.
\end{align}
\end{definition}\rightline{$\blacksquare$}

Given \emph{Definition \ref{def:con}}, we have a lemma as follows:

\begin{lemma}\label{conj:1}
Any Gaussian integer within the convex region $\mathcal{G}_q$ is  a valid difference.
\end{lemma}

\begin{IEEEproof}[Proof of Lemma \ref{conj:1}]

 A sketch of the proof is as follows. With reference to the example with $q=4+i$ in  Fig. \ref{fig:conv}, we can define the convex region formed by valid symbols in $\mathbb{Z}[i]/q$ (see the inner octagon in black dashed line; note that for the case of a real $q$, the convex region will be a square rather than an octagon) as
 \begin{align}
 \nonumber &\big\{x\in \mathbb{C}| x=\sum^{|q|^2}_{i=1}a_i w_i, {\rm~where~}\\   & a_i\in\mathbb{R}, 0\leq \forall a_i \leq 1, \sum^{|q|^2}_{i=1}a_i=1,  {\rm and~} \forall w_i \in \mathbb{Z}[i]/q\big\}.
 \end{align}
 Note that the convex region formed by the valid differences, $\mathcal{G}_q$, (see the outer octagon in red dashed line in Fig. \ref{fig:conv}) is a scaled-up version of this convex region. The scaled-up factor is 2. We need to prove that every Gaussian integer (black squares in Fig. \ref{fig:conv}(a)) within $\mathcal{G}_q$ is a scaled-up point induced by two Gaussian integers (red stars in Fig. \ref{fig:conv}(a)) in $\mathbb{Z}[i]/q$.  In other words, we can express any lattice point  $\delta$ in $\mathcal{G}_q$ as   $\delta=w-w'$, where   $w,w'\in\mathbb{Z}[i]/q$.

 We introduce the concept of \emph{scaled-by-half} lattice as follows: a scaled-by-half lattice is $\frac{1}{2}\mathbb{Z}[i]$, where $z'\in \frac{1}{2}\mathbb{Z}[i]$ if and only if $z'=\frac{1}{2}z$, for some $z\in\mathbb{Z}[i]$. The set of valid symbols is $\mathcal{W}=\mathbb{Z}[i]/q$. We define  $\tilde{\mathcal{W}}=\frac{1}{2}\mathcal{W}=\frac{1}{2}\mathbb{Z}[i]/q$. With respect to Fig. \ref{fig:conv}, the lattice points within the inner octagon in Fig. \ref{fig:conv}(a) (red stars) are  $\mathcal{W}$, and the lattice points within the inner octagon in Fig. \ref{fig:conv}(b) (white circles  and red stars) are $\tilde{\mathcal{W}}$. Note that $\mathcal{W}\subset \tilde{\mathcal{W}}$.

 Denote a Gaussian integer in the outer octagon whose real and imaginary parts are both even by $\delta_e$. We note that each $\delta_e$ is a scaled-up-by-2 version of a $w\in\mathcal{W}$ (see the blue solid lines with arrow in Fig. \ref{fig:conv}(b)). There is a one-to-one mapping between the points in $\mathcal{W}$ and the set of Gaussian integers $\delta_e$. For such a Gaussian integer, we can write
 \begin{align}
   \delta_e=2w=w+w, ~{\rm for ~some~}  w\in\mathcal{W}.
 \end{align}

 We note that since $\mathcal{W}$ is a field, and therefore each element in $\mathcal{W}$ has an additive inverse, and each element is an additive inverse of some other element. Specifically,  $w\in\mathcal{W}$  is the inverse of some $w'\in\mathcal{W}$. We can thus write
  \begin{align}
   \delta_e=2w=w-w', ~{\rm where~both~}  w,w'\in\mathcal{W}.
 \end{align}

 Denote a Gaussian integer in the outer octagon whose real and imaginary parts are not both even by $\delta_o$. We note that each $\delta_o$ is a scaled-up-by-2 version of a  $\tilde{w}\in\tilde{\mathcal{W}}\backslash{\mathcal{W}}$ (see the red solid lines with arrow in Fig. 10 (b)). We further note that for any $\tilde{w}\in\tilde{\mathcal{W}}\backslash{\mathcal{W}}$, we can write $\tilde{w}=\frac{1}{2}w+\frac{1}{2}w'$, for some $w,w'\in\mathcal{W}$ (i.e., $\tilde{w}$ is an equal-weight linear combination of two valid symbols in $\mathcal{W}$). Thus, $\delta_o$  can be expressed as
  \begin{align}
   \delta_o=w+w', ~{\rm where~both~}  w,w'\in\mathcal{W}.
 \end{align}

Again, $w'$ is the inverse of some $w''\in \mathcal{W}$ and vice versa, giving $\delta_o=w-w''$.

 Thus, for any Gaussian integer $\delta$ within the convex region $\mathcal{G}_q$, we can find two $w,w'\in \mathbb{Z}[i]/q$ such that $\delta=w-w'$. This completes the proof.

 \end{IEEEproof}

\begin{lemma}\label{lem:87}
Consider two distinct $(\kappa,\tau),(\gamma, \delta)\in \mathcal{Q}^{char}_q$, $(\kappa,\tau)\neq(\nu\gamma, \nu \delta)$ where $\nu=\pm 1$ or $\pm i$, and that $\Xi\triangleq\kappa\delta-\tau\gamma$ contains a factor $\tilde{\Xi}$ with magnitude $|\tilde{\Xi}|=2$ or $3$. The two pairs $(\kappa,\tau),(\gamma, \delta)$ are non-adjacent.
 \end{lemma}

\begin{IEEEproof}[Proof of Lemma \ref{lem:87}]
The proof is given in  {Appendix VII}.

\end{IEEEproof}

\emph{Lemma  \ref{lem:87}}  covers the cases with  normalized distances   $|\Xi|=2, 2\sqrt{2}, 3, 2\cdot2, 3\sqrt{2}, 2\sqrt{8}, 2\cdot3$.

\subsection{$d^{(\alpha_{opt},\beta_{opt})}_{\min}$ Analysis Within the Voronoi Region}

Given an arbitrary characteristic difference $(\delta^{char}_A,\delta^{char}_B)$, Part E has given a set of criteria for the identification of its adjacent characteristic differences. In this part, we prove (T4-1) and (T4-2) stated at the end of Part C. First, we use these criteria to prove (T4-1), restated as \emph{Lemma \ref{lem:nocluadj}} below.

%
%

\begin{lemma}[T4-1]\label{lem:nocluadj}
Two distinct characteristic differences  $(\delta^{char}_A,\delta^{char}_B)$ and $({\delta^{char}_A}', {\delta^{char}_B}')$  (i.e., $(\delta^{char}_A,\delta^{char}_B)\neq \upsilon({\delta^{char}_A}', {\delta^{char}_B}'),  \forall  \upsilon\in \{\pm 1, \pm i\}$) that   are adjacent cannot be clustered by the same NC mapping.
\end{lemma}
\begin{IEEEproof}[Proof of Lemma \ref{lem:nocluadj}]
Given $(\delta^{char}_A,\delta^{char}_B)$, $({\delta^{char}_A}', {\delta^{char}_B}') \in \mathcal{Q}^{char}_q$, if $(\delta^{char}_A,\delta^{char}_B)$ and $({\delta^{char}_A}', {\delta^{char}_B}')$ are adjacent, the adjacent pair yields $|\Xi|=|\delta^{char}_B{\delta^{char}_A}'-\delta^{char}_A{\delta^{char}_B}'|=1, \sqrt{2},$ or $\sqrt{5}$ by \emph{$\mathcal{Q}$-criteria  1-3}.  Furthermore, $(\delta^{char}_A,\delta^{char}_B)$ and $({\delta^{char}_A}', {\delta^{char}_B}')$  are clustered by the same NC mapping if and only if $\delta^{char (q)}_B{\delta^{char(q)}_A}'-\delta^{char(q)}_A{\delta^{char(q)}_B}'=0 ({\rm mod~} q)$.

{\noindent \textbf{Case 1}: $|q|=\sqrt{2}$}

Consider  a Gaussian prime $q$ with $|q|=\sqrt{2}$ (e.g., $q=1+i$). In this case, we have four zero-$l_{\min}$ gains in the complex plane of $\eta$: two non-trivial zero-$l_{\min}$ channel gains, i.e., $\eta^o=1$ and $-1$, associated with $(\delta^{char}_A,\delta^{char}_B)=(1,-1)$  and $(1,1)$ respectively, and two trivial zero-$l_{\min}$ channel gains $\eta^o=0$ and $\eta^o=\infty$, associated with $(\delta^{char}_A,\delta^{char}_B)=(1,0)$ and $(0,1)$ respectively. For $|q|=\sqrt{2}$, we can prove \emph{Lemma \ref{lem:nocluadj}} by considering the adjacent Voronoi regions of each characteristic difference. For example, at $\eta^o=-1$ associated with $(\delta^{char}_A,\delta^{char}_B)=(1, 1)$, the adjacent characteristic difference is $({\delta^{char}_A}', {\delta^{char}_B}')=(1,0)$   by \emph{$\mathcal{Q}$-criterion 1} (note that the Voronoi regions of $\eta^o=1$ and $-1$ are not adjacent), and $({\delta^{char}_A}', {\delta^{char}_B}')$ cannot be clustered by any NC mapping.

{\noindent \textbf{Case 2}: $|q|\geq\sqrt{5}$}

Consider  a Gaussian prime $q$ with $|q|\geq\sqrt{5}$. In the following, we verify that it is not possible to have
\begin{align}\label{eqn:giclu}
\Xi\triangleq\delta^{char}_B{\delta^{char}_A}'-\delta^{char}_A{\delta^{char}_B}'=mq, \ m\in \mathbb{Z}[i],
 \end{align}
(i.e., not possible to have  $\delta^{char (q)}_B{\delta^{char(q)}_A}'-\delta^{char(q)}_A{\delta^{char(q)}_B}'=0 ({\rm mod~} q)$), if $(\delta^{char}_A,\delta^{char}_B)$ and $({\delta^{char}_A}', {\delta^{char}_B}')$   satisfy \emph{Q-criteria 1-3}.

For $|\Xi|=1$ (i.e., $\Xi\in\{\pm 1, \pm i\}$), \eqref{eqn:giclu} is not possible because it is not possible to have  $1=m q, \forall m\in \mathbb{Z}[i]$.

For $|\Xi|=\sqrt{2}$ (i.e., $\Xi\in \upsilon (1+i)$ and $\upsilon\in\{\pm 1, \pm i\}$), w.l.o.g., let us consider $\Xi=1+i$. It is not possible to satisfy \eqref{eqn:giclu} either because $1+i\neq mq, \forall m\in \mathbb{Z}[i]$, since $1+i$ is prime and cannot be factorized.


For $|\Xi|=\sqrt{5}$ (i.e., $\Xi\in \{\upsilon (1+i), \upsilon (1+2i)\}$ and $\upsilon\in\{\pm 1, \pm i\}$), w.l.o.g., let us consider $\Xi=2+i$. First, let us consider  $|q|>\sqrt{5}$. In this case, \eqref{eqn:giclu} cannot be satisfied because $\Xi=2+i\neq mq, \forall m\in \mathbb{Z}[i]$, since $2+i$ is prime and cannot be factorized.

 Next, consider  $|q|=\sqrt{5}$. In the following, we show that even though \eqref{eqn:giclu} is satisfied, $(\delta^{char}_A,\delta^{char}_B)$ and $({\delta^{char}_A}', {\delta^{char}_B}')$
are not adjacent if $|\Xi|=\sqrt{5}$ (i.e., for $|q|=\sqrt{5}$, two difference pairs with a normalized distance of $\sqrt{5}$ can never be adjacent;  it is only when $|q|\neq\sqrt{5}$ that it is possible for two difference pairs separated by a normalized distance of $\sqrt{5}$ to be adjacent). W.l.o.g., consider $q=2+i$ (the representative elements of $\mathbb{Z}[i]/(2+i)$ and $\mathbb{Z}[i]/(2-i)$ are the same). In this case, the valid symbols are $\{0, \pm 1, \pm i\}$, and $\delta^{char}_A,\delta^{char}_B,{\delta^{char}_A}', {\delta^{char}_B}'\in \{0, \pm 1, \pm i, \pm 2, \pm 2i, \pm (1+i), \pm (1-i)\}$. Let us consider $\Xi=2+i$ (similar proof applies for $\Xi=2-i$).

 By (i) in \emph{$\mathcal{Q}$-criterion 2}, there exists one and only one $\epsilon \in \{\pm 1, \pm i\}$ such that
  \begin{align}\label{eqn:qc2i}
  \nonumber \delta^{char}_A+\epsilon{\delta^{char}_A}'=0 \ ({\rm mod~} \Xi), \\\delta^{char}_B+\epsilon{\delta^{char}_B}'=0\ ({\rm mod~} \Xi).
\end{align}

To satisfy \eqref{eqn:qc2i}, given $\delta^{char}_A,\delta^{char}_B,{\delta^{char}_A}', {\delta^{char}_B}'\in \{0, \pm 1, \pm i, \pm 2, \pm 2i, \pm (1+i), \pm (1-i)\}$, we can verify that the values that can be adopted by $\delta^{char}_A+\epsilon{\delta^{char}_A}'$ and $\delta^{char}_B+\epsilon{\delta^{char}_B}'$  while satisfying \eqref{eqn:qc2i} must be from the set  $\{0, \upsilon(2+i), \upsilon(1+i)(2+i)\}$, where $\upsilon$ is a unit. Furthermore, it is not possible to have $\delta^{char}_A+\epsilon{\delta^{char}_A}'= a(2+i)$ or $\delta^{char}_B+\epsilon{\delta^{char}_B}' = a(2+i)$ where $|a|>\sqrt{2}$. In the following, we list all the possible solutions of \eqref{eqn:qc2i}:

(s1) $(\delta^{char}_A+\epsilon{\delta^{char}_A}', \delta^{char}_B+\epsilon{\delta^{char}_B}')=(0, 0)$

It is not possible to have (s1), since for $(\delta^{char}_A,\delta^{char}_B)\neq \upsilon ({\delta^{char}_A}', {\delta^{char}_B}')$,   $\forall \upsilon\in\{\pm 1, \pm i\}$, according to the statement of lemma (i.e., the two characteristic differences are distinct).

(s2) $(\delta^{char}_A+\epsilon{\delta^{char}_A}', \delta^{char}_B+\epsilon{\delta^{char}_B}')\neq (0, 0)$

\begin{enumerate}
  \item [(s2-i)]  $(\delta^{char}_A+\epsilon{\delta^{char}_A}', \delta^{char}_B+\epsilon{\delta^{char}_B}')=(0, \upsilon(2+i))$ or $(\upsilon(2+i), 0)$
  \item [(s2-ii)]  $(\delta^{char}_A+\epsilon{\delta^{char}_A}', \delta^{char}_B+\epsilon{\delta^{char}_B}')=(0, \upsilon(1+i)(2+i))=(0, \upsilon(1+3i))$ or $(\upsilon(1+3i), 0)$
  \item [(s2-iii)] $(\delta^{char}_A+\epsilon{\delta^{char}_A}', \delta^{char}_B+\epsilon{\delta^{char}_B}')=(\upsilon(2+i), \upsilon(2+i))$
  \item [(s2-iv)] $(\delta^{char}_A+\epsilon{\delta^{char}_A}', \delta^{char}_B+\epsilon{\delta^{char}_B}')=(\upsilon(2+i), \upsilon(1+i)(2+i))$ or $(\upsilon(1+i)(2+i), \upsilon(2+i))$
  \item [(s2-v)] $(\delta^{char}_A+\epsilon{\delta^{char}_A}', \delta^{char}_B+\epsilon{\delta^{char}_B}')=(\upsilon(1+i)(2+i), \upsilon(1+i)(2+i))$
\end{enumerate}

According to the definition of  the $\mathcal{Q}_q$-set in \eqref{set:Qt}, $(\delta^{char}_A+\epsilon{\delta^{char}_A}', \delta^{char}_B+\epsilon{\delta^{char}_B}')\in \mathcal{Q}_q$ for all  subcases in (s2) above, since each belongs to  $\mathcal{Q}^{char}_q$ after factoring of the $\gcd$ (e.g., for  $(\delta^{char}_A+\epsilon{\delta^{char}_A}', \delta^{char}_B+\epsilon{\delta^{char}_B}')=(0, \upsilon(2+i))$, after factoring out the $\gcd$ of $2+i$ we have $\frac{1}{2+i}(\delta^{char}_A+\epsilon{\delta^{char}_A}', \delta^{char}_B+\epsilon{\delta^{char}_B}')=(0, \upsilon) \in \mathcal{Q}^{char}_q$). According to (ii) in \emph{$\mathcal{Q}$-criterion 2}, therefore,  for  $|\Xi|=\sqrt{5}$,  $(\delta^{char}_A,\delta^{char}_B)$ and $({\delta^{char}_A}', {\delta^{char}_B}')$  cannot be adjacent.

\end{IEEEproof}

In the following, we prove (T4-2) by \emph{Lemmas \ref{lem:a1}-\ref{lem:a3}} and \emph{Corollary \ref{cor:2}}.

\begin{lemma}\label{lem:a1}
Consider two non-adjacent pairs $(\kappa, \tau), (\gamma, \delta)\in \mathcal{Q}^{char}_q$  with normalized distance $|\Xi|\triangleq|\kappa\delta-\tau\gamma|>\sqrt{5}$. There exists another pair $(\gamma', \delta')\in \mathcal{Q}^{char}_q$  such that  $|\delta'\eta-\gamma'|<|\delta\eta-\gamma|, \forall \eta\in \mathcal{V}(\frac{\kappa}{\tau})$   and that  $|\Xi'|\triangleq|\kappa\delta'-\tau\gamma'|<|\Xi|$.
\end{lemma}

Remark: The case of $|\Xi|=1,\sqrt{2},\sqrt{5}$ will be  treated separately in \emph{Lemmas \ref{lem:a2}} and  \emph{\ref{lem:a3}}. Note that $(\gamma', \delta')$ may still be non-adjacent to $(\kappa, \tau)$, but the normalized distance with $(\kappa, \tau)$ is getting smaller.

\begin{IEEEproof}[Proof of Lemma \ref{lem:a1}]
  If $|\Xi|>\sqrt{5}$ and $|\Xi|\neq 2, 3$, by \emph{Lemmas \ref{lem:82}-\ref{lem:86}},  there exist $(\zeta, \vartheta), (\phi, \psi)\in \mathbb{Z}^2[i]\backslash\{(0,0)\}$, such that $\phi=\frac{\zeta\kappa+\vartheta\gamma}{\Xi},
\psi=\frac{\zeta\tau+\vartheta\delta}{\Xi}$
  and   $0<|\zeta|+|\vartheta|\leq \frac{|\Xi|}{\sqrt{2}}$. Using similar argument as  \eqref{eqn:qc32}, we conclude that  $(\phi,\psi)\in \mathcal{Q}_q$. Note that it is possible for $\gcd(\phi,\psi)>1$, in which case we can reduce $(\phi,\psi)$  further to $(\gamma', \delta')\triangleq(\frac{\phi}{\gcd(\phi,\psi)},\frac{\psi}{\gcd(\phi,\psi)})\in \mathcal{Q}^{char}_q$. Consider an arbitrary $\eta\in \mathcal{V}(\frac{\kappa}{\tau})$.  Following \eqref{eqn:conex}, we write (note: in \eqref{eqn:conex}, $z'$  is a point equidistant to $(\kappa, \tau)$ and  $(\gamma, \delta)$; here, $\eta$  is not equidistant to $(\kappa, \tau)$ and  $(\gamma, \delta)$)
\begin{align}\label{eqn:la11}
\nonumber|\delta' \eta-\gamma'|&\leq|\psi \eta-\phi|=\frac{|\zeta(\tau \eta-\kappa)+\vartheta(\delta \eta-\gamma)|}{|\Xi|}\\ \nonumber & \leq \frac{|\zeta(\tau \eta-\kappa)|+|\vartheta(\delta \eta-\gamma)|}{|\Xi|}\\
\nonumber & \leq \frac{|\zeta|+|\vartheta|}{|\Xi|}\max\{|\tau \eta-\kappa|,|\delta \eta-\gamma|\}\\
& \leq\frac{|\delta \eta-\gamma|}{\sqrt{2}}<|\delta \eta-\gamma|,
\end{align}
where the first inequality in the last line holds because $0<|\zeta|+|\vartheta|\leq \frac{|\Xi|}{\sqrt{2}}$ and $|\tau\eta-\kappa|\leq|\delta\eta-\gamma|$ for $\forall \eta\in \mathcal{V}(\frac{\kappa}{\tau})$.


Furthermore, we have
\begin{align}\label{eqn:la12}
\nonumber |\Xi'|&\triangleq|\kappa\delta'- \tau\gamma'|\leq |\kappa\psi- \tau\phi|\\ \nonumber & = \frac{|\kappa(\tau\zeta+\delta\vartheta)-\tau(\kappa \zeta+\gamma\vartheta)|}{|\Xi|} \\
 & = \frac{|\vartheta(\kappa \delta-\tau\gamma)|}{|\Xi|}=|\vartheta|  \leq \frac{|\Xi|}{\sqrt{2}} <|\Xi| .
\end{align}

If $|\Xi|= 2$ or  $3$, by \emph{Lemma \ref{lem:87}}, there exist $(\zeta, \vartheta), (\phi, \psi)\in \mathbb{Z}^2[i]\backslash\{(0,0)\}$, such that $\phi=\frac{\zeta\kappa+\vartheta\gamma}{\Xi}$ and $\psi=\frac{\zeta\tau+\vartheta\delta}{\Xi}$. We consider $|\Xi|= 2$ only, and the proof for $|\Xi|= 3$ follows similarly. From the proof of \emph{Lemma  \ref{lem:87}}, we have
\begin{subequations} \label{eqnl85}
\begin{align}
\label{eqnl85:1} &\kappa\zeta+\gamma\vartheta=2\phi,\\
 \label{eqnl85:2} & \tau\zeta+\delta\vartheta=2\psi,
\end{align}
\end{subequations}
 where both $\zeta, \vartheta$ are units and  both signs of $\zeta$ satisfy \eqref{eqnl85}.

 Then,   we rewrite \eqref{eqn:la11} as
\begin{align}\label{eqn:la13}
\nonumber|\delta' \eta-\gamma'|&\leq|\psi \eta-\phi|=\frac{|\zeta(\tau \eta-\kappa)+\vartheta(\delta \eta-\gamma)|}{2}\\ \nonumber & < \frac{|\zeta(\tau \eta-\kappa)|+|\vartheta(\delta \eta-\gamma)|}{2}\\
\nonumber & < \frac{|\zeta|+|\vartheta|}{2}\max\{|\tau \eta-\kappa|,|\delta \eta-\gamma|\}\\
& <|\delta \eta-\gamma|,
\end{align}
where we can find a proper sign of $\zeta$ to validate the second strict inequality.  Then, we further have
\begin{align}\label{eqn:la14}
\nonumber |\Xi'|&\triangleq|\kappa\delta'- \tau\gamma'|\leq |\kappa\psi- \tau\phi|\\
 & = \frac{|\vartheta(\kappa \delta-\tau\gamma)|}{|\Xi|}=|\vartheta|=1   <2 .
\end{align}
\end{IEEEproof}

\begin{lemma}\label{lem:a2}
Consider two non-adjacent pairs $(\kappa, \tau), (\gamma, \delta)\in \mathcal{Q}^{char}_q$  with normalized distance $|\Xi|\triangleq|\kappa\delta-\tau\gamma|=\sqrt{5}$. There exists another pair $(\gamma', \delta')\in \mathcal{Q}^{char}_q$  such that  $|\delta'\eta-\gamma'|<|\delta\eta-\gamma|, \forall \eta\in \mathcal{V}(\frac{\kappa}{\tau})$   and that  $|\Xi'|\triangleq|\kappa\delta'-\tau\gamma'|=1$.
\end{lemma}
\begin{IEEEproof}[Proof of Lemma \ref{lem:a2}]
We follow the proof of \emph{$\mathcal{Q}$-criterion 2}. According to (i) of \emph{$\mathcal{Q}$-criterion 2}, there exist one and only one unit $\epsilon\in \{\pm 1, \pm i\}$ such that $\kappa+\epsilon \gamma=0({\rm mod}~ \Xi)$ and $\tau+\epsilon \delta=0({\rm mod}~ \Xi)$ for $|\Xi|=\sqrt{5}$. Under this $\epsilon$, define  $\phi\triangleq\frac{\kappa+\epsilon\gamma}{\Xi},
\psi\triangleq\frac{\tau+\epsilon\delta}{\Xi}$. Since $(\kappa, \tau)$ and  $(\gamma, \delta)$ are not adjacent,  $(\phi,\psi)\in \mathcal{Q}_q$ by (ii) of \emph{$\mathcal{Q}$-criterion 2}. Note that it is possible that $\gcd(\phi,\psi)>1$, in which case we can reduce $(\phi,\psi)$  further to $(\gamma', \delta')\triangleq(\frac{\phi}{\gcd(\phi,\psi)},\frac{\psi}{\gcd(\phi,\psi)})\in \mathcal{Q}^{char}_q$.

Consider an arbitrary $\eta\in \mathcal{V}(\frac{\kappa}{\tau})$. We have
\begin{align}\label{eqn:la21}
\nonumber|\delta' \eta-\gamma'|&\leq|\psi \eta-\phi|=\frac{| (\tau \eta-\kappa)+\epsilon(\delta \eta-\gamma)|}{|\Xi|} \\
\nonumber &\leq \frac{|\tau \eta-\kappa|+|\delta \eta-\gamma|}{|\Xi|}\\
  & \leq \frac{2}{|\Xi|}\max\{|\tau \eta-\kappa|,|\delta \eta-\gamma|\}<|\delta \eta-\gamma|.
\end{align}
where the last inequality holds since   $|\tau\eta-\kappa|\leq|\delta\eta-\gamma|$ for $\forall \eta\in \mathcal{V}(\frac{\kappa}{\tau})$ and $|\Xi|=\sqrt{5}>2$.
%

Furthermore, we have
\begin{align}\label{eqn:la22}
\nonumber |\Xi'|&\triangleq|\kappa\delta'- \tau\gamma'|\leq |\kappa\psi- \tau\phi|  \leq \frac{|\kappa(\tau +\epsilon\delta )-\tau(\kappa +\epsilon\gamma )|}{|\Xi|} \\
 & = \frac{ |\kappa \delta-\tau\gamma|}{|\Xi|}=1 <|\Xi| =\sqrt{5}.
\end{align}

\end{IEEEproof}

\begin{lemma}\label{lem:a3}
Consider two non-adjacent pairs $(\kappa, \tau), (\gamma, \delta)\in \mathcal{Q}^{char}_q$  with normalized distance $|\Xi|\triangleq|\kappa\delta-\tau\gamma|=1$ or $\sqrt{2}$. There exists another pair $(\gamma', \delta')\in \mathcal{Q}^{char}_q$  such that  $|\delta'\eta-\gamma'|<|\delta\eta-\gamma|, \forall \eta\in \mathcal{V}(\frac{\kappa}{\tau})\backslash \{\frac{\kappa}{\tau}\}$   and $|\delta'\eta-\gamma'|\leq|\delta\eta-\gamma|$ at $\eta=\frac{\kappa}{\tau}$. Furthermore,
 $|\Xi'|\triangleq|\kappa\delta'-\tau\gamma'|\leq|\Xi|=1$ or $\sqrt{2}$.
\end{lemma}

\begin{IEEEproof}[Proof of Lemma \ref{lem:a3}]
 According to  \emph{$\mathcal{Q}$-criterion 1}, $(\phi,\psi)\triangleq({\kappa+\epsilon\gamma},
 {\tau+\epsilon\delta})\in \mathcal{Q}_q$ for $\forall \epsilon \in \{\pm 1, \pm i\}$, since $(\kappa, \tau)$ and $(\gamma, \delta)$ are not adjacent.  Note that it is possible that $\gcd(\phi,\psi)>1$, in which case we can reduce $(\phi,\psi)$  further to $(\gamma', \delta')\triangleq\frac{1}{\gcd(\phi,\psi)}(\phi,\psi)\in \mathcal{Q}^{char}_q$.

Consider an arbitrary $\eta\in \mathcal{V}(\frac{\kappa}{\tau})$. Given $|\tau\eta-\kappa|\leq |\delta\eta-\gamma|$, we can choose a proper $\epsilon\in \{\pm 1, \pm i\}$ such that (we can think of $\tau\eta-\kappa$ and $\delta \eta-\gamma$ in \eqref{eqn:la31} below as two 2-dimensional vectors on the complex plane and that $\epsilon=i,-1$, and $-i$  rotate   $\delta\eta-\gamma$ by $\frac{\pi}{2}, \pi$, and $\frac{3\pi}{2}$ respectively):
\begin{align}\label{eqn:la31}
 \nonumber |\delta' \eta-\gamma'|& \leq|\psi \eta-\phi|=|(\tau\eta-\kappa)+\epsilon(\delta\eta-\gamma)|\\
 & \leq|\delta \eta-\gamma|,
\end{align}
where the  last inequality in \eqref{eqn:la31} is satisfied with equality only at $\eta=\frac{\kappa}{\tau}$. Thus,
\begin{align}\label{eqn:la32}
 &|\delta' \eta-\gamma'|<|\delta \eta-\gamma|, \forall \eta\in \mathcal{V}(\frac{\kappa}{\tau})\backslash \{\frac{\kappa}{\tau}\},\\
\label{eqn:la33} {\rm and~} &|\delta' \eta-\gamma'|\leq|\delta \eta-\gamma|,  ~\eta=\frac{\kappa}{\tau}.
\end{align}

Furthermore, similar to \eqref{eqn:la22}, we can also verify that
\begin{align}\label{eqn:la34}
|\Xi'| \triangleq|\kappa\delta'- \tau\gamma'|\leq |\Xi|=1 \ {\rm~or} \ \sqrt{2}.
\end{align}

\end{IEEEproof}

\begin{remark}\label{rem:only}
 The inequality in \eqref{eqn:la33} may be satisfied with equality at $\eta= \frac{\kappa}{\tau}$. Thus, potentially, a non-adjacent element can still be a  $d^{(\alpha_{opt},\beta_{opt})}_{\min}$-determining difference at the ``singular point'' $\eta= \frac{\kappa}{\tau}$. Note, however, this is the only point at which a non-adjacent element can potentially be a  $d^{(\alpha_{opt},\beta_{opt})}_{\min}$-determining difference. Furthermore, we will argue that there is always an adjacent element that is a  $d^{(\alpha_{opt},\beta_{opt})}_{\min}$-determining difference at  $\eta= \frac{\kappa}{\tau}$ (see \emph{Corollary \ref{cor:2}} below). Thus, we can eliminate non-adjacent elements from consideration when we try to identify the $d^{(\alpha_{opt},\beta_{opt})}_{\min}$ within the Voronoi region $\mathcal{V}(\frac{\kappa}{\tau})$.
\end{remark}\rightline{$\blacksquare$}

\begin{corollary}[T4-2]\label{cor:2}
Consider two non-adjacent pairs $(\kappa, \tau)\in \mathcal{Q}^{char}_q$ and $(\gamma, \delta)\in \mathcal{Q}^{char}_q$ . There exists another pair $(\gamma', \delta')\in \mathcal{Q}^{char}_q$  adjacent to $(\kappa, \tau)$ such that $|\delta'\eta-\gamma'|<|\delta\eta-\gamma|, \forall \eta\in \mathcal{V}(\frac{\kappa}{\tau})\backslash \{\frac{\kappa}{\tau}\}$   and $|\delta'\eta-\gamma'|\leq|\delta\eta-\gamma|$ at $\eta=\frac{\kappa}{\tau}$.
\end{corollary}
\begin{IEEEproof}[Proof of Corollary \ref{cor:2}]
Starting with  $(\gamma, \delta)$ that is non-adjacent to  $(\kappa, \tau)$, we can apply the results of \emph{Lemmas \ref{lem:a1}, \ref{lem:a2}}, and \emph{\ref{lem:a3}} iteratively to eventually find a pair $(\gamma', \delta')\in \mathcal{Q}^{char}_q$   that satisfies one of the following:
\begin{itemize}
  \item [(i)]   $|\Xi|\triangleq|\kappa\delta'-\tau\gamma'|=1, \sqrt{2}$ or $\sqrt{5}$ and that $(\gamma', \delta')$ is adjacent to $(\kappa, \tau)$. We further have that $|\delta'\eta-\gamma'|<|\delta\eta-\gamma|$ for $\forall \eta\in \mathcal{V}(\frac{\kappa}{\tau})\backslash \{\frac{\kappa}{\tau}\}$, and that $|\delta'\eta-\gamma'|\leq|\delta\eta-\gamma|$ at $\eta=\frac{\kappa}{\tau}$.
  \item [(ii)] $|\Xi|\triangleq|\kappa\delta-\tau\gamma|=1$ or $\sqrt{2}$ but $(\gamma', \delta')$ is not adjacent to $(\kappa, \tau)$. That is, we stop at \emph{Lemma \ref{lem:a3}} with $(\gamma', \delta')$  still not adjacent to $(\kappa, \tau)$. Note, however, that for \emph{Lemma \ref{lem:a3}}, we also have  $|\delta'\eta-\gamma'|<|\delta\eta-\gamma|$ for $\forall \eta\in \mathcal{V}(\frac{\kappa}{\tau})\backslash \{\frac{\kappa}{\tau}\}$, and that  $|\delta'\eta-\gamma'|\leq|\delta\eta-\gamma|$ at $\eta=\frac{\kappa}{\tau}$. In this case, we apply \emph{Lemma \ref{lem:a3}} again to find another pair $(\gamma'', \delta'')$  with $|\delta''\eta-\gamma''|<|\delta'\eta-\gamma'|$,  $\forall \eta\in \mathcal{V}(\frac{\kappa}{\tau})\backslash \{\frac{\kappa}{\tau}\}$, and that $|\delta''\eta-\gamma''|\leq|\delta'\eta-\gamma'|$  for $\eta=\frac{\kappa}{\tau}$. If this  $(\gamma'', \delta'')$ is still not adjacent to $(\kappa, \tau)$, we apply  \emph{Lemma \ref{lem:a3}} yet again to find another pair $(\gamma''', \delta''')$. Note that with each successively pair, the weighted distance of the new pair to $\eta\in \mathcal{V}(\frac{\kappa}{\tau})\backslash \{\frac{\kappa}{\tau}\}$ becomes strictly smaller than that of the previous pair. Therefore, the successive pairs are distinct and we will never repeat the same pair in the above iterative argument. Since all these pairs are elements of $\mathcal{Q}^{char}_q$, and $\mathcal{Q}^{char}_q$ is a finite set, we must eventually reach a pair that is adjacent to $(\kappa, \tau)$. Otherwise, we would be able to enumerate an infinite number of non-adjacent pairs within $\mathcal{Q}^{char}_q$.
\end{itemize}
\end{IEEEproof}


Now that we have proved (T4-1) and (T4-2), we can narrow our interest to adjacent elements when we try to identify the $d^{(\alpha_{opt},\beta_{opt})}_{\min}$-determining differences within the weighted Voronoi region.

As an illustration, we revisit the Voronoi diagram for $q=3$ in Fig. \ref{fig:rocd1} to explore the implications of \emph{$\mathcal{Q}$-criteria 1-3} and \emph{Theorem \ref{thm:gdmin}}. We consider a zero-$l_{\min}$ channel gain $\eta^o=\frac{1+i}{2}$ associated with the characteristic difference $(\delta^{char}_A,\delta^{char}_B)=(1+i,-i)$. From Fig. \ref{fig:rocd1}, the $d^{(\alpha_{opt},\beta_{opt})}_{\min}$-determining differences for $\eta$ in $\mathcal{V}(\eta^o)$ are adjacent to $(\delta^{char}_A,\delta^{char}_B)$. Some examples of $d^{(\alpha_{opt},\beta_{opt})}_{\min}$-determining differences are  $({\delta^{char}_A}', {\delta^{char}_B}')=\{(2+2i, -i), (2+2i, -1-2i)\}$ with $|\delta^{char }_B{\delta^{char}_A}'-\delta^{char}_A{\delta^{char}_B}'|=\sqrt{2}$ and $({\delta^{char}_A}', {\delta^{char}_B}')\in \{(1+2i, -i), (2+2i, -1-i), (1+2i, -2i), (2+i, -1-2i)\}$ with $|\delta^{char }_B{\delta^{char}_A}'-\delta^{char}_A{\delta^{char}_B}'|=1$. We can check that these $d^{(\alpha_{opt},\beta_{opt})}_{\min}$-determining differences are consistent with our \emph{Q-criteria 1-3}.

Next,   we show that a non-adjacent characteristic difference can be a $d^{(\alpha_{opt},\beta_{opt})}_{\min}$-determining difference  at a particular zero-$l_{\min}$ channel gain, as stated in \emph{Remark \ref{rem:only}}. For $q=3$, we consider $\eta^o=1$  associated with $(\delta^{char}_A,\delta^{char}_B)=(1,-1)$.
Further consider $({\delta^{char}_A}', {\delta^{char}_B}')=(1,-1-i)$ associated with ${\eta^o}'=1+i$. We can verify that $|\delta^{char }_B{\delta^{char}_A}'-\delta^{char}_A{\delta^{char}_B}'|=1$. However, we can also verify that $({\delta^{char}_A}', {\delta^{char}_B}')$  is not adjacent to $(\delta^{char}_A,\delta^{char}_B)$ according to \emph{$\mathcal{Q}$-criterion 1}, since the four medians between $(\delta^{char}_A,\delta^{char}_B)$ and $({\delta^{char}_A}', {\delta^{char}_B}')$ are all in the $\mathcal{Q}^{char}_q$-set. Therefore, by \emph{Corollary \ref{cor:2}}, $({\delta^{char}_A}', {\delta^{char}_B}')$ is not the $d^{(\alpha_{opt},\beta_{opt})}_{\min}$-determining difference within  $\mathcal{V}(-\frac{\delta^{char}_B}{\delta^{char}_A})$ except at $\eta^o=-\frac{\delta^{char}_B}{\delta^{char}_A}$. As implied by the proof of \emph{Theorem \ref{thm:eta0}}, any characteristic difference whose normalized distance with $(\delta^{char}_A,\delta^{char}_B)$ is $1$ is a $d^{(\alpha_{opt},\beta_{opt})}_{\min}$-determining difference at $\eta^o=-\frac{\delta^{char}_B}{\delta^{char}_A}$. This characteristic difference has weighted distance of $\frac{1}{|\delta^{char}_A|}$ (i.e., $d^{(\alpha_{opt},\beta_{opt})}_{\min}=\frac{1}{|\delta^{char}_A|}$ at $\eta^o=-\frac{\delta^{char}_B}{\delta^{char}_A}$). Since the normalized distance of $({\delta^{char}_A}', {\delta^{char}_B}')$ and $(\delta^{char}_A,\delta^{char}_B)$ is $1$, $({\delta^{char}_A}', {\delta^{char}_B}')$ is $d^{(\alpha_{opt},\beta_{opt})}_{\min}$-determining difference at $\eta^o=-\frac{\delta^{char}_B}{\delta^{char}_A}$. However, the adjacent difference $({\delta^{char}_A}'', {\delta^{char}_B}'')=(2, -2-i)$ is also the $d^{(\alpha_{opt},\beta_{opt})}_{\min}$-determining difference at $\eta^o=-\frac{\delta^{char}_B}{\delta^{char}_A}$, since its normalized distance with $(\delta^{char}_A,\delta^{char}_B)$ is also $1$. Thus, when deriving the $d^{(\alpha_{opt},\beta_{opt})}_{\min}$-determining difference within $\mathcal{V}(-\frac{\delta^{char}_B}{\delta^{char}_A})$, we need not consider $({\delta^{char}_A}', {\delta^{char}_B}')$.


\subsection{Overview of $l_{\min}$ and $d^{(\alpha_{opt},\beta_{opt})}_{\min}$}

With reference to Fig. \ref{fig:rocd1}, we now illustrate how  $l_{\min}$ and $d^{(\alpha_{opt},\beta_{opt})}_{\min}$ change as the channel gain varies. Consider two characteristic differences $(\delta^{char}_A,\delta^{char}_B)$ and $({\delta^{char}_A}', {\delta^{char}_B}')$    associated with $\eta^o$ and ${\eta^o}'$ that are adjacent to each other. For simplicity, we denote the ``new'' weighted Voronoi region of an element $({\delta^{char}_A}', {\delta^{char}_B}')$ after we remove  $(\delta^{char}_A,\delta^{char}_B)$  as $\mathcal{V}_{\backslash\{-\frac{\delta^{char}_B}{\delta^{char}_A}\}}(-\frac{{\delta^{char}_B}'}{{\delta^{char}_A}'})$ by ROCD.
\begin{figure}[t]
  \centering
        \includegraphics[height=0.42\columnwidth]{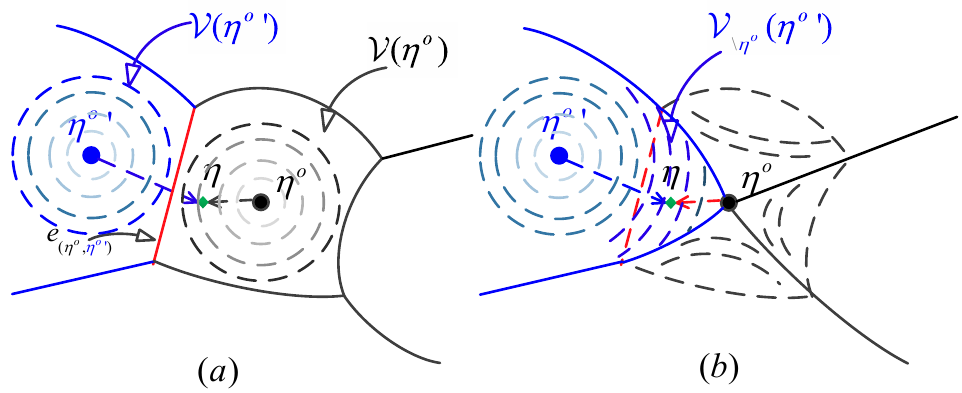}
       \caption{Weighted Voronoi regions of ${\eta^o}'$ (a) with consideration of $\eta^o$ and (b) without consideration of $\eta^o$ (ROCD). The arc denotes the contour line of a generator.}
        \label{fig:dmine}
\end{figure}

Consider $l_{\min}$ shown in Fig. \ref{fig:dmine}(a). In this region, $l_{\min}$   reaches the minimum at $\eta^o$ (the vertex of the cone). Then,  $l_{\min}$ increases as $\eta$  approaches the edges of $\mathcal{V}(-\frac{\delta^{char}_B}{\delta^{char}_A})$ and reaches a local maximum  at an edge of this Voronoi region (the intersections of two cones). When $\eta$ crosses the edges and falls into $\mathcal{V}(-\frac{{\delta^{char}_B}'}{{\delta^{char}_A}'})$ , $({\delta^{char}_A}', {\delta^{char}_B}')$   will yield $l_{\min}$, but the variation of $l_{\min}$ in this Voronoi region still follows the same pattern as above. As long as $\eta$ is within the same weighted Voronoi region, $l_{\min}$ varies in a continuous fashion following the contour as expressed in \eqref{eqn:ld}.

Consider $d^{(\alpha_{opt},\beta_{opt})}_{\min}$ in $\mathcal{V}(-\frac{\delta^{char}_B}{\delta^{char}_A})$ shown in Fig. \ref{fig:dmine}(b). We show   how a subset of the region $\mathcal{V}(-\frac{\delta^{char}_B}{\delta^{char}_A})$ becomes part of $\mathcal{V}_{\backslash\{-\frac{\delta^{char}_B}{\delta^{char}_A}\}}(-\frac{{\delta^{char}_B}'}{{\delta^{char}_A}'})$  after ROCD (i.e., $({\delta^{char}_A}', {\delta^{char}_B}')$   yields $d^{(\alpha_{opt},\beta_{opt})}_{\min}$ within a subset of the region of $\mathcal{V}(-\frac{\delta^{char}_B}{\delta^{char}_A})$). First, $d^{(\alpha_{opt},\beta_{opt})}_{\min}$ reaches a local maximum at $\eta^o=-\frac{\delta^{char}_B}{\delta^{char}_A}$, since we observe that all adjacent elements of $\eta^o$, separated with $({\delta^{char}_A}, {\delta^{char}_B})$ by a normalized distance of $1$, meet at $\eta^o$ after ROCD. Then, within $\mathcal{V}_{\backslash\{-\frac{\delta^{char}_B}{\delta^{char}_A}\}}(-\frac{{\delta^{char}_B}'}{{\delta^{char}_A}'})$  , we observe that $d^{(\alpha_{opt},\beta_{opt})}_{\min}$ decreases as $\eta$ approaches the edge of $\eta^o$ and ${\eta^o}'$ (see contour lines of ${\eta^o}'$ after ROCD in Fig. \ref{fig:dmine}(b)), and reaches a local minimum at the edge, i.e., $d^{(\alpha_{opt},\beta_{opt})}_{\min}=l_{\min}$. Furthermore, when $\eta$ crosses the edge and falls into $\mathcal{V}(-\frac{{\delta^{char}_B}'}{{\delta^{char}_A}'})$, $d^{(\alpha_{opt},\beta_{opt})}_{\min}$  will be determined by a different characteristic difference, but still follows the same pattern as above. In general, the local minima of  $l_{\min}$ correspond to local maxima of  $d^{(\alpha_{opt},\beta_{opt})}_{\min}$ and the local maxima of $l_{\min}$  correspond to the local minima of  $d^{(\alpha_{opt},\beta_{opt})}_{\min}$  in the overall complex plane of $\eta$.

\begin{figure}[t]
  \centering
        \includegraphics[height=0.43\columnwidth]{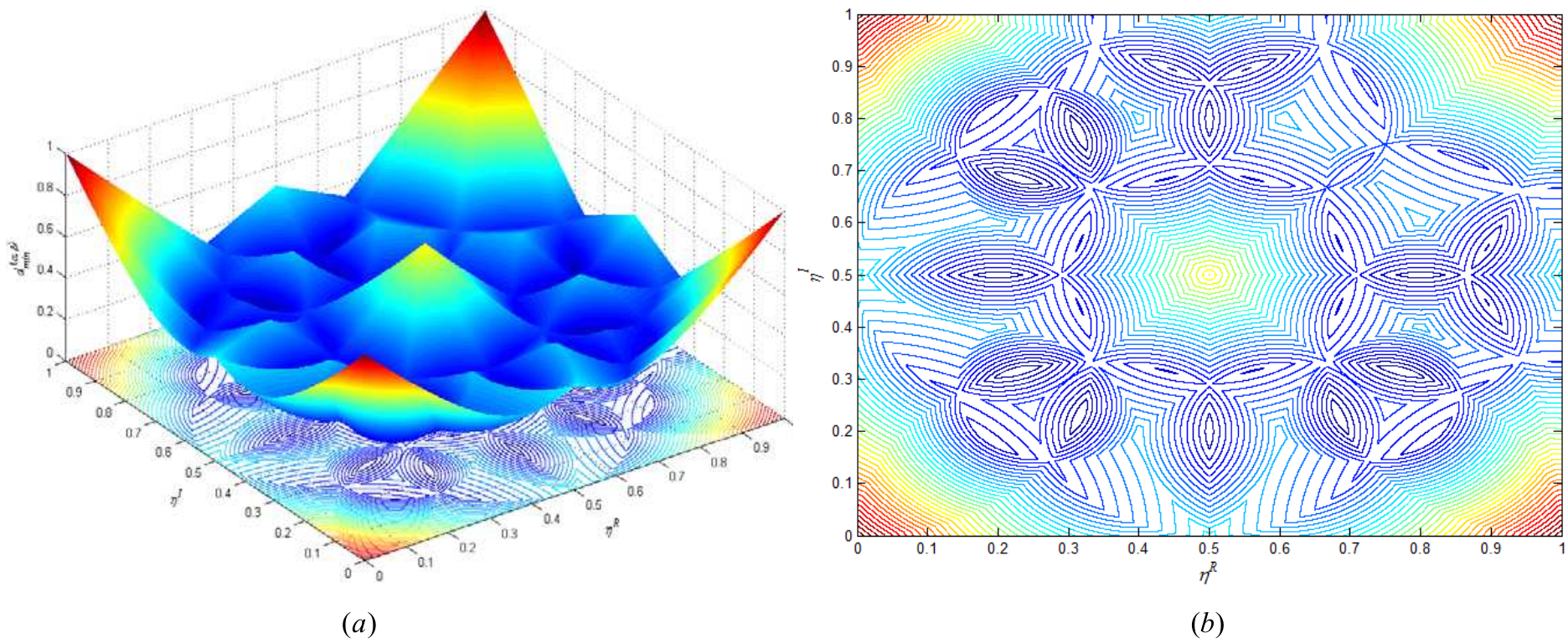}
       \caption{(a) $d^{(\alpha_{opt},\beta_{opt})}_{\min}(\eta)$ surface for $q=3$ and $|\eta|\leq 2$; (b) the corresponding contour graphs of $d^{(\alpha_{opt},\beta_{opt})}_{\min}(\eta)$.}
        \label{fig:dmin}
\end{figure}

As an illustrating example, we plot the $d^{(\alpha_{opt},\beta_{opt})}_{\min}$ versus $\eta$ for $q=3$ in Fig. \ref{fig:dmin}. With respect to the $l_{\min}$  versus $\eta$ plot in Fig. \ref{fig:lmin}, we can observe that the changes of $l_{\min}$ and $d^{(\alpha_{opt},\beta_{opt})}_{\min}$ are consistent   with our analysis above.

\section{Conclusion}

We have investigated a general framework of complex linear PNC for TWRC, where the signals of the two end nodes simultaneously received at the relay incur imbalanced powers and a relative phase offset.  Specifically, we put forth a Gaussian-integer formulation for the complex linear PNC mapping in $\mathbb{Z}[i]/q$. Our Gaussian-integer formulation provides more flexibility for signal constellation designs than the vector formulation in prior work. We further recast the linear PNC mapping based on the coset theory to uncover the isomorphism among PNC mappings. The isomorphism allows us to reduce the search space for the optimal PNC mapping by selecting one representative PNC mapping from each isomorphic group.

For each channel gain ratio $\eta=h_A/h_B$, there is a corresponding optimal PNC mapping  $(\alpha_{opt},\beta_{opt})$. To identify $(\alpha_{opt},\beta_{opt})$  for a given $\eta$, we focused on the characterization of two minimum-distance metrics in the received constellation. The first minimum-distance metric is the minimum \emph{symbol} distance $l_{\min}$, which is the minimum distance among all distances between two constellation points. The second minimum-distance metric is the minimum \emph{NC-symbol} distance $d^{(\alpha,\beta)}_{\min}$ under PNC mapping $(\alpha,\beta)$, which is the minimum distance among all distances between two constellation points mapped to different NC symbols by  $(\alpha,\beta)$. It is $d^{(\alpha,\beta)}_{\min}$ that determines the SER of NC symbols in the high SNR regime. The optimal PNC mapping    is given by   $(\alpha_{opt}, \beta_{opt})=\mathop{\arg\max}\limits_{(\alpha, \beta)} d^{(\alpha,\beta)}_{\min}$.

An important concept put forth in this paper is the \emph{characteristic difference} $(\delta^{char}_A,\delta^{char}_B)=(w_A, w_B)-(w'_A, w'_B)$: the difference between any two distinct joint symbols  $(w_A, w_B)$ and $(w'_A, w'_B)$  for which there is no common factor between $\delta^{char}_A$ and $\delta^{char}_B$  (i.e., $\gcd(\delta^{char}_A,\delta^{char}_B)={\rm unit}$). Given a set of joint symbols $\mathcal{W}_{(A,B)}=\{(w_A,w_B)|w_A,w_B\in \mathbb{Z}[i]/q\}$, there is a corresponding set of characteristic differences. For a given  $\eta$,   $l_{\min}$  is given by the  particular characteristic difference that yields the minimum  $|\eta\delta^{char}_A+\delta^{char}_B|$. The optimal PNC mapping $(\alpha_{opt},\beta_{opt})$ for that $\eta$  is the mapping that maps two pairs of symbols $(w_A, w_B)$  and $(w'_A, w'_B)$  separated by  $(\delta^{char}_A,\delta^{char}_B)$ to the same NC symbol, hence there is no need to distinguish between   $(w_A, w_B)$  and $(w'_A, w'_B)$    although the distance between them in the received constellation is  $l_{\min}$.

For a global understanding of how $l_{\min}$ and $d^{(\alpha,\beta)}_{\min}$ vary with  $\eta$, we investigated the partitioning of the complex plane of $\eta$  into different Voronoi regions. The $\eta$ within a Voronoi region are associated with the same characteristic difference $(\delta^{char}_A,\delta^{char}_B)$  and the same optimal PNC mapping $(\alpha_{opt},\beta_{opt})$  (i.e., $(\delta^{char}_A,\delta^{char}_B)$  induces the $l_{\min}=|\eta\delta^{char}_A+\delta^{char}_B|$, and $(\alpha_{opt},\beta_{opt})$  maps joint symbols separated by  $(\delta^{char}_A,\delta^{char}_B)$ to the same NC symbol). We developed a systematic approach to identify the $d^{(\alpha,\beta)}_{\min}$ for all $\eta$ within a Voronoi region by considering the characteristic differences associated with Voronoi regions adjacent to it.


As a final remark, we believe that our framework of complex linear PNC mapping in the field of Gaussian integer---including the concept of characteristic difference, isomorphism via coset theory, Voronoi-region characterization of optimal PNC mapping, and determination of  $d^{(\alpha_{opt},\beta_{opt})}_{\min}$---is also applicable to complex linear PNC mappings in other fields (e.g., the finite field of Eisenstein integer), since the underlying mathematical concepts are similar.

\section*{Appendix I: Algebraic Construction of Valid NC Mapping}

Consider an NC mapping under $(\alpha,\beta)$ in \eqref{eqn:wn}. \emph{Propositions \ref{pro:5}}  and \emph{\ref{pro:6}}  below specify how the set of joint symbols are partitioned by this NC mapping and show the isomorphism in NC mappings in terms of cosets in group theory.

%

With respect to  $\Delta_{(\alpha,\beta)}$ in \eqref{eqn:cds}, we define a corresponding set within the finite field of $\mathbb{Z}[i]/q$ as follows:
\begin{align}\label{eqn:cdsq}
\nonumber \Delta^{(q)}_{(\alpha,\beta)}=& \big\{(\delta^{(q)}_A,\delta^{(q)}_B)
=(\delta_A ({\rm mod} ~q),\delta_B ({\rm mod} ~q)) \big| \\&(\delta_A,\delta_B)\in\Delta_{(\alpha,\beta)}\big\}.
\end{align}
Given a $(\delta_A,\delta_B)\in \Delta_{(\alpha,\beta)}$ and its corresponding $(\delta^{(q)}_A,\delta^{(q)}_B)= (\delta_A ({\rm mod} ~q),\delta_B ({\rm mod} ~q))\in\Delta^{(q)}_{(\alpha,\beta)}$, we use the terms ``the NC mapping $(\alpha,\beta)$ clusters $(\delta_A,\delta_B)$'' and ``the NC mapping $(\alpha,\beta)$ clusters $(\delta^{(q)}_A,\delta^{(q)}_B)$'' interchangeably in this appendix.

It is easy to show that $\Delta_{(\alpha,\beta)}^{(q)}$  is a group under element-wise addition operation of $\mathbb{Z}^2[i]/q$, where $\Delta_{(\alpha,\beta)}^{(q)}=\Delta_{(\alpha,\beta)} ~({\rm mod} ~q)$. Closure and associativity are obvious. The identity element of the group is simply $e\triangleq(0,0)~({\rm mod} ~q)$,  and the inverse of an element $\delta\triangleq(\delta_A^{(q)},\delta_B^{(q)})\in\mathbb{Z}^2[i]/q$ is simply $-\delta\triangleq(-\delta_A^{(q)},-\delta_B^{(q)}) ~({\rm mod} ~q)$.

Let us see how to enumerate the elements in $\Delta_{(\alpha,\beta)}^{(q)}$. We first note that $({\delta_A^{(q)}}',{\delta_B^{(q)}}')=(-\beta,\alpha)$ is a solution to \eqref{eqn:cds}. We next note that \eqref{eqn:cds} can be satisfied by $(\delta_A^{(q)},\delta_B^{(q)})=v({\delta_A^{(q)}}',{\delta_B^{(q)}}')~({\rm mod} ~q), \forall v\in \mathbb{Z}[i]/q$. Thus, there are altogether $|q|^2 (\delta_A^{(q)},\delta_B^{(q)})$ that can satisfy \eqref{eqn:cds}. Therefore, $\Delta_{(\alpha,\beta)}^{(q)}$ can be rewritten as
\begin{align}\label{eqn:cds1}
\nonumber \Delta_{(\alpha,\beta)}^{(q)}=&\big\{(\delta_A^{(q)},\delta_B^{(q)})\in(\mathbb{Z}[i]/q)^2|\\
&(\delta_A^{(q)},\delta_B^{(q)})=v(-\beta,\alpha)~({\rm mod} ~q)\big\}.
\end{align}

Thus, under $(\alpha,\beta)$, two joint symbols $(w_A,w_B)$ and $(w'_A,w'_B)$ will be mapped to the same NC symbol if and only if
\begin{align}\label{eqn:snc}
(w_A,w_B)=(w'_A,w'_B)+v(-\beta,\alpha)~({\rm mod} ~q)
\end{align}
for some $v\in \mathbb{Z}[i]/q$. Note that, for simplicity, we include the trivial case $v=0$ where $(w_A,w_B)$ and  $(w'_A,w'_B)$ are the same symbol.

\begin{definition}[{Coset}]\label{def:coset}
In algebra, if $G$ is a group with operation $\circ$, $H$ is a subgroup of $G$ and $g\in G$, then

$g \circ H=\{g \circ h|h\in H\}$ is a left coset of $H$ in $G$, and

$H\circ g=\{h \circ g|h\in H\}$ is a right coset of $H$ in $G$.

For abelian groups, the left and right cosets are the same\cite{JBF}.
\end{definition}\rightline{$\blacksquare$}

For us, $G$ is the additive group of $\mathbb{Z}^2[i]/q$ (i.e., the group is $(\mathbb{Z}^2[i]/q, +)$ where $+$ is the element-wise addition). The collection of all joint symbols is the set $\mathcal{W}_{(A,B)}=\mathbb{Z}^2[i]/q$. For a given NC mapping $(\alpha,\beta)$, a subgroup of $G$ is $H\triangleq  \Delta_{(\alpha,\beta)}^{(q)}$.

Consider a joint symbol $(w_A,w_B )\in\mathbb{Z}^2[i]/q$. The coset $(w_A,w_B)+\Delta_{(\alpha,\beta)}^{(q)}~({\rm mod} ~q)$ is the set of all joint symbols mapped to the same NC symbol as $(w_A,w_B)$. That is, a coset consists of all joint symbols mapped to the same NC symbol. Given any element of a coset, i.e., $(w_A,w_B)$, we can find all elements of the coset by $(w_A,w_B)+\Delta_{(\alpha,\beta)}^{(q)}~({\rm mod} ~q)$ if we know $\Delta_{(\alpha,\beta)}^{(q)}$. The elements in $\Delta_{(\alpha,\beta)}^{(q)}$ can be found from \eqref{eqn:cds1}.

The following proposition summarizes the discussion above.

\begin{proposition}\label{pro:5}

A linear NC mapping under $(\alpha,\beta)$ partitions the set of joint symbols $\mathcal{W}_{(A,B)}=\mathbb{Z}^2[i]/q$ into $|q|^2$ subsets, each mapped to a unique NC symbol. Consider the additive group of $\mathcal{W}_{(A,B)}$. Each of the $|q|^2$ subset is a coset generated by the subgroup $\Delta_{(\alpha,\beta)}^{(q)}$ of $\mathcal{W}_{(A,B)}$, described as follows:
\begin{align}\label{eqn:cos}
\Delta_{(\alpha,\beta)}^{(q)}=\big\{(\delta_A^{(q)},\delta_B^{(q)})\in\mathbb{Z}^2[i]/q\big|\alpha\delta_A^{(q) }+\beta\delta_B^{(q)}=0~({\rm mod} ~q)\big\}.
\end{align}

The subgroup $\Delta_{(\alpha,\beta)}^{(q)}$ contains $|q|^2$ elements and they can be found as follows:
\begin{align}\label{eqn:cos1}
\nonumber\Delta_{(\alpha,\beta)}^{(q)}&=\big\{(\delta_A^{(q)},\delta_B^{(q)})\in\mathbb{Z}^2[i]/q\big|\\
&(\delta_A^{(q)},\delta_B^{(q)})=v(-\beta,\alpha)~({\rm mod} ~q),  v\in \mathbb{Z}[i]/q\big\}.
\end{align}

For  a joint symbol $(w_A,w_B)\in\mathbb{Z}^2[i]/q$,
\begin{align}\label{eqn:cos2}
\nonumber \mathcal{C}_w=& \big\{(w'_A,w'_B)|(w'_A,w'_B)=(w_A,w_B)+(\delta_A^{(q)},\delta_B^{(q)})~({\rm mod} ~q),\\&(\delta_A^{(q)},\delta_B^{(q)})\in \Delta_{(\alpha,\beta)}^{(q)}\big\}.
\end{align}
is a coset of $\Delta_{(\alpha,\beta)}^{(q)}$ in $\mathbb{Z}^2[i]/q$. Each coset contains $|q|^2$ distinct joint symbols.
\end{proposition}\rightline{$\blacksquare$}

From \emph{Proposition \ref{pro:5}}, the complex NC mapping $f_N^{(\alpha, \beta)}:\mathcal{W}_{(A,B)}\rightarrow \mathcal{W}_N^{(\alpha,\beta)}$ is a $|q|^2$-to-$1$ mapping. This NC mapping partitions $\mathcal{W}_{(A,B)}$ into $|q|^2$ subsets (i.e., $|q|^2$ cosets). We say that these $|q|^2$ cosets are \emph{generated} by $(\alpha,\beta)$. Each of these cosets is \emph{labeled} by an NC symbol to which the elements within the coset is mapped. To find the NC symbol that serves as the label, we simply take an element $(w_A,w_B)$ from the coset, and then compute $\alpha w_A+\beta w_B ~({\rm mod} ~q)$.

\begin{proposition}\label{pro:6}
The cosets generated by $\Delta_{(\alpha,\beta)}^{(q)}$ are the same as the cosets generated by $\Delta_{(\gamma\alpha,\gamma\beta)~({\rm mod} ~q)}^{(q)}$ where $\gamma\in\mathbb{Z}[i]/q\backslash\{0\}$).
\end{proposition}
\begin{IEEEproof}[Proof of Proposition \ref{pro:6}]
Consider an arbitrary element $(\delta_A^{(q)},\delta_B^{(q)})\in \Delta_{(\alpha,\beta)}^{(q)}$. Then, we have $\alpha\delta_A^{(q)}+\beta\delta_B^{(q)}=0 ~({\rm mod} ~q)$. For $\gamma\in \mathbb{Z}[i]/q\backslash\{0\}$, we further have
\begin{eqnarray}\label{eqn:61}
\gamma(\alpha \delta_A^{(q)} +\beta\delta_B^{(q)})=(\gamma\alpha)\delta_A^{(q)}+(\gamma\beta)\delta_B^{(q)}=0 ({\rm mod} \ q).
\end{eqnarray}
Thus, $(\delta_A^{(q)},\delta_B^{(q)})\in \Delta_{(\gamma\alpha,\gamma\beta) ~({\rm mod} ~q)}^{(q)}$.

Similarly, consider an arbitrary element $(\delta_A^{(q)},\delta_B^{(q)})\in \Delta_{(\gamma\alpha,\gamma\beta) ~({\rm mod} ~q)}^{(q)}$. Then, $(\gamma\alpha)\delta_A^{(q)}+(\gamma\beta)\delta_B^{(q)}=0 ~({\rm mod} ~q)$. Since $\gamma\in\mathbb{Z}[i]/q\backslash\{0\}$, we further have $\gamma^{-1} \big((\gamma\alpha) \delta_ A^{(q)}+(\gamma\beta)\delta_B^{(q)} \big)=\alpha\delta_A^{(q)}+\beta\delta_B^{(q)}=0 ~({\rm mod} ~q)$, where $\gamma^{-1}$ is the multiplicative inverse of $\gamma$. Thus, $(\delta_A^{(q)},\delta_B^{(q)})\in \Delta_{(\alpha,\beta)}^{(q)}$.

Therefore, $\Delta_{(\alpha,\beta)}^{(q)}=\Delta_{(\gamma\alpha,\gamma\beta)~({\rm mod} ~q)}^{(q)}$ and the cosets generated by $\Delta_{(\alpha,\beta)}^{(q)}$ and $\Delta_{(\gamma\alpha,\gamma\beta)~({\rm mod} ~q)}^{(q)}$ are the same.

\end{IEEEproof}

\begin{remark}\label{rem:iso}
 The above results mean that the NC partitioning of $\mathcal{W}_{(A,B)}$ into $|q|^2$ cosets are the same under $(\alpha,\beta)$ and $(\gamma\alpha,\gamma\beta)~({\rm mod} ~q), \forall\gamma\in \mathbb{Z}[i]/q\backslash \{0\}$. However, the NC symbols used to label the same coset are different under $(\alpha,\beta)$ and $(\gamma\alpha,\gamma\beta)$ when $\gamma\neq 1$. Specifically, the NC mappings induced by $(\alpha,\beta)$ and $(\gamma\alpha,\gamma\beta)~({\rm mod} ~q)$ are isomorphic as per \emph{Definition \ref{def:iso}}. Note that given $(\delta_A^{(q)}, \delta_B^{(q) })$, the NC mapping $(\alpha,\beta)$ that clusters $(\delta_A^{(q)},\delta_B^{(q)})$ is unique except for $(\gamma\alpha,\gamma\beta)~({\rm mod} ~q)$. In this paper, we say that $(\alpha,\beta)$ clusters $(\delta_A^{(q)}, \delta_B^{(q) })$ uniquely if any other $(\alpha',\beta')$ that can also cluster $(\delta_A^{(q)}, \delta_B^{(q) })$ does not alter the partitioning of $\mathcal{W}_{(A,B)}$ into $|q|^2$ cosets. Therefore, the NC mappings under $(\alpha,\beta)$ and $(\gamma\alpha,\gamma\beta)~({\rm mod} ~q)$ yield the same NC partitioning. Note that uniqueness in this sense can be assured for Gaussian prime $q$ thanks to finite-field arithmetic.
\end{remark}

Below is a corollary of \emph{Proposition \ref{pro:6}}, which rephrases \emph{Proposition \ref{pro:3}} from the coset perspective.
\begin{corollary}\label{cor:1}
 For a specific set of cosets generated by $(\alpha,\beta)$, there exists a corresponding $(\alpha',\beta')=(\beta^{-1}\alpha,1)~({\rm mod} ~q)$ generating the same set of cosets.
\end{corollary}

\begin{IEEEproof}[Proof of Proposition \ref{cor:1}]
Consider a set of cosets generated by $\Delta_{(\alpha,\beta)}^{(q)}$. We choose $\gamma$ to be the multiplicative inverse of $\beta$, i.e., $\beta^{-1}$. Here, $\gamma$ exists since $\beta\in \mathbb{Z}[i]/q\backslash\{0\}$. By \emph{Proposition \ref{pro:6}}, $\Delta_{(\alpha,\beta)}^{(q)}=\Delta_{(\beta^{-1} \alpha,1) ~({\rm mod} ~q)}^{(q)}$. Therefore, the corresponding $(\alpha',\beta')=(\beta^{-1} \alpha,1)~({\rm mod} ~q)$.

\end{IEEEproof}

\section*{Appendix II - Proof of Lemma \ref{lem:suf}}

When $|q|<\sqrt{5}$ (i.e., $|q|=\sqrt{2}$), we have $\mathbb{Z}[i]/q=\{0,1\}$ by \emph{Remark \ref{rem:sqrt2}}. Then, $\delta_A,\delta_B\in\{0,\pm 1\}$ and $\delta_A$ and $\delta_B$ cannot be both zero if the $(\delta_A,\delta_B)$ is a distance-valid difference pair. Therefore, $|\delta_A|,|\delta_B|\leq 1 \leq \sqrt{2|q|^2-4q^R+2}=\sqrt{2{\sqrt{2}}^2-4+2}=\sqrt{2}$ .

In the following, we consider $|q|\geq\sqrt{5}$. We consider $\delta_A$ only and the proof for $\delta_B$ is similar. The difference between any two distinct representative elements $w_A$ and $w'_A$ in $\mathbb{Z}[i]/q$ is upper-bounded as follows:
\begin{align}
|\delta_A|=|w_A-w'_A|\leq|w_A|+|w'_A|\leq2\max_{w_A\in\mathbb{Z}[i]/q}|w_A|.
\end{align}

To derive  $\max|w_A|$ in $\mathbb{Z}[i]/q$, let us consider the center of the square formed by $q$ and $iq$, i.e., the point A (see the cross in Fig. \ref{fig:suf}) and
\begin{align}
\nonumber A&=\frac{1}{2}(q^R+iq^I)+\frac{1}{2}(-q^I+iq^R)\\
&=\frac{1}{2}(q^R-q^I)+i\frac{1}{2}(q^R+q^I).
\end{align}
 Note that point A is a vertex of the square within which all valid symbols lie (i.e., the blue square in Fig. \ref{fig:suf}; according to \emph{Definition \ref{def:res}}, representative elements lie within the zero-centered square of side length $|q|$ with orientation aligned with the directions as indicated by the basis $(x,y)$---see the red stars within the blue squares in Fig. \ref{fig:suf}).

 Point A is the point with the largest magnitude within the square. However, A is not a valid symbol (i.e., $A\notin\mathbb{Z}[i]/q$), since it is not a Gaussian integer. Thus, $|A| > \max_{w_A\in\mathbb{Z}[i]/q}|w_A|$. We claim that the valid symbol that is closest to A is a symbol with the maximum magnitude.

 Specifically, we claim that $w^*_A\triangleq A-(\frac{1}{2}+i\frac{1}{2})=\max_{w_A\in\mathbb{Z}[i]/q}|w_A|$. First, for $w^*_A=A-(\frac{1}{2}+i\frac{1}{2})$,  we can easily verify that $w^{*x}_A,w^{*y}_A<\frac{|q|}{2}$ by \emph{Definition \ref{def:res}}, and therefore $w^{*}_A$ is a valid symbol. Furthermore,
 \begin{align}\label{eqn:aws}
\nonumber |w^*_A|&=|\frac{1}{2}(q^R-q^I-1)+i\frac{1}{2}(q^R+q^I-1)|\\
&=\sqrt{\frac{|q|^2}{2}-q^R+\frac{1}{2}}.
\end{align}

In the following, we prove that $w^*_A=\arg\max_{w_A\in\mathbb{Z}[i]/q}|w_A|$   by showing that $|w^*_A|\geq |w_A|, \forall w_A\in \mathbb{Z}[i]/q$. Given an arbitrary valid symbol $w_A\in\mathbb{Z}[i]/q$, we can verify that three other Gaussian integers, i.e., $\{-w_A, iw_A, -iw_A\}$, are also valid symbols in $\mathbb{Z}[i]/q$, symmetric to $w_A$ with respect to four quadrants in the complex plane. By this symmetry property, we focus on the complex quadrant (angle from $0$ to $\pi/2$) in which A lies.

Consider a real Gaussian prime $q$ (i.e., $q^R\neq 0$ and $q^I= 0$). W.l.o.g., suppose that $q>0$. The representative elements of $\mathbb{Z}[i]/q$ by \emph{Definition \ref{def:res}} are shown in Fig. \ref{fig:suf}(b). In this case, the point A is $A=\frac{1}{2}(q+iq)$, and we can see that $w^*_A=A-(\frac{1}{2}+i\frac{1}{2})$ is a valid symbol in $\mathbb{Z}[i]/q$ and $w^*_A=\arg\max_{w_A\in\mathbb{Z}[i]/q}|w_A|$. Then, we have $|\delta_A|\leq 2|A-(\frac{1}{2}+i\frac{1}{2})|=2|\frac{1}{2}(q-1)+i\frac{1}{2}(q-1)|=\sqrt{2}(q-1)$. Therefore, we have proved \emph{Lemma \ref{lem:nec}} when $q=q^R$.

Consider a complex Gaussian prime $q$, where $|q|\geq\sqrt{5}$ and $q^R\neq0$, $q^I\neq 0$. W.l.o.g., we assume $q^R>q^I\geq 1$ (note that $q^R\neq q^I$ because $q$ is prime). We want to prove that $w^*_A=A-(\frac{1}{2}+i\frac{1}{2})$ is the largest valid symbol within this quadrant. Consider an arbitrary valid symbol $w$ in $\mathbb{Z}[i]/q$ within this quadrant
 \begin{eqnarray}
\label{eqn:su1}w=A-(a+ib)=(\frac{q^R-q^I}{2}-a)+i(\frac{q^R+q^I}{2}-b),
\end{eqnarray}
 \begin{align}
\label{eqn:su11}\nonumber |w|^2&=(\frac{q^R-q^I}{2}-a)^2+(\frac{q^R+q^I}{2}-b)^2\\
&=\frac{|q|^2}{2}+a[a-(q^R-q^I)]+b[b-(q^R+q^I)].
\end{align}
where
 \begin{subequations}
 \begin{align}
\label{eqn:su2}&\frac{q^R-q^I}{2}-a \in \mathbb{Z}, \frac{q^R+q^I}{2}-b \in \mathbb{Z},\\
\label{eqn:su3} &\frac{q^R-q^I}{2}\geq a, \frac{q^R+q^I}{2}\geq b.
\end{align}
\end{subequations}

 From \eqref{eqn:su2}, $|a|\geq\frac{1}{2}$ if $w$ in \eqref{eqn:su1} is a Gaussian integer. Since $w\in \mathbb{Z}[i]/q$, we have $|w^x|,|w^y|<\frac{|q|}{2}$ by \emph{Definition \ref{def:res}}, where
  \begin{align}\label{eqn:su4}
 \nonumber \left(\begin{array}{c}
     w^x \\
     w^y \\
   \end{array}\right)&=\frac{1}{|q|}\left(
                        \begin{array}{cc}
                          q^R & q^I \\
                          -q^I & q^R
                        \end{array}
                      \right)\left(
                               \begin{array}{c}
                                 \frac{q^R-q^I}{2}-a \\
                                 \frac{q^R+q^I}{2}-b
                               \end{array}
                             \right)\\&=\left(
                                       \begin{array}{c}
                                        \frac{|q|^2-2aq^R-2bq^I}{2|q|} \\
                                         \frac{|q|^2-2bq^R+2aq^I}{2|q|} \\
                                       \end{array}
                                     \right).
 \end{align}

 From \eqref{eqn:su4}, we further have
    \begin{subequations}
  \nonumber \begin{align}
   0<aq^R+bq^I<|q|^2,\\
   \label{eqn:suf52}0<bq^R-aq^I<|q|^2.
     \end{align}
    \end{subequations}

Given  $|a|\geq 1/2$, we next prove that $|w|\leq |w^*_A|, \forall w\in \mathbb{Z}[i]/q$ by considering the following cases.

\textbf{Case 1}: $a\geq 1/2$ and $b\geq 1/2$

Obviously, from \eqref{eqn:su11}, $|w|\leq|w^*_A|$.

\textbf{Case 2}: $a\geq 1/2$ and $b<1/2$

This case is not possible if $w$ is to be valid in $\mathbb{Z}[i]/q$  (see the blue dashed squares in Fig. \ref{fig:suf}.).

\textbf{Case 3}: $a\leq -1/2$ and $b\geq1/2$

From \eqref{eqn:su2}, we let $a=-\frac{1+2m}{2}$ and $b=\frac{1+2n}{2}$, where $m\geq0$ and $n\geq0$. Therefore, \eqref{eqn:su11} can be rewritten as
 \begin{align}
\nonumber|w|^2=&\frac{|q|^2}{2}+a[a-(q^R-q^I)]+b[b-(q^R+q^I)]\\
\nonumber=&\frac{|q|^2}{2}+m^2+n^2+m+n+\frac{1}{2}\\
& -(n-m)q^R-(m+n+1)q^I.
\end{align}

From \eqref{eqn:su3} and \eqref{eqn:suf52}, we have
   \begin{subequations}\label{eqn:suf6}
   \begin{align}
   \label{eqn:suf61} q^R+q^I&\geq1+2n,\\
   \label{eqn:suf62} \frac{1+2m}{1+2n}&<\frac{q^I}{q^R}.
     \end{align}
    \end{subequations}

From \eqref{eqn:suf62}, we have
\begin{align} \label{eqn:mn}
  n>m,
\end{align}
since $q^R>q^I$. Furthermore, from \eqref{eqn:suf6}, we have
\begin{eqnarray}\label{eqn:suf63}
q^I>\frac{1+2m}{1+2n}q^R\geq\frac{1+2m}{q^R+q^I}q^R
>\frac{1+2m}{2q^R}q^R=\frac{1}{2}+m.
\end{eqnarray}

To find a valid $w\in\mathbb{Z}[i]/q$ such that $|w|>|w^*_A|$, we need
\begin{align}\label{eqn:suf7}
\nonumber &\frac{|q|^2}{2}+m^2+n^2+m+n+\frac{1}{2}-(n-m)q^R-(m+n+1)q^I\\
 &>\frac{|q|^2}{2}-q^R+\frac{1}{2},\\
\label{eqn:suf71}\nonumber & \Rightarrow m^2+n^2+m+n\\
&-(n-m-1)q^R-(m+n+1)q^I>0.
\end{align}

To check whether $w$ yielding \eqref{eqn:suf71} exists, we consider two subcases as follows:

\begin{enumerate}

\item [(3-1)] $2q^I>1+2n$

In this subcase, we have $q^R>q^I>\frac{1}{2}+n$. Then, LHS of \eqref{eqn:suf71} can be upper bounded as
\begin{align}\label{eqn:suf8}
\nonumber &m^2+n^2+m+n-(n-m-1)q^R-(n+m+1)q^I\\
\nonumber &<m^2+n^2+m+n-(n-m-1)(\frac{1}{2}+n)\\
\nonumber & -(n+m+1)(\frac{1}{2}+n)\\
\nonumber & =m^2+n^2+m+n-2n(\frac{1}{2}+n)\\
&= (m+n)(m-n)+m<0,
\end{align}
where the first inequality holds since $q^R>q^I>\frac{1}{2}+n$, and the last inequality holds since $n>m$ because of \eqref{eqn:mn}. This contradicts with \eqref{eqn:suf7}. Therefore, in this subcase, $w^*$ has the largest magnitude among the valid symbols in $\mathbb{Z}[i]/q$.

  \item [(3-2)] $2q^I<1+2n$

Since $q^I<\frac{1}{2}+n$ and $q^I$ is an integer, we have $q^I\leq n$. From \eqref{eqn:suf61}, we further have $q^R>1+n$, since $q^R+n\geq q^R+q^I>1+2n$.

Then,  LHS of \eqref{eqn:suf7} can be upper bounded as
\begin{align}
\nonumber &m^2+n^2+m+n-(n-m-1)q^R-(n+m+1)q^I\\
\nonumber & = m^2+n^2-(n-m-1)q^R-(n+m+1)(q^I-1)-1 \\
\nonumber &<m^2+n^2-(n-m-1)(n+1)\\
\nonumber  & -(m+n+1)(q^I-1)-1\\
\nonumber & = m^2+mn+m-(m+n+1)(q^I-1)\\
& =(m+n+1)(m-q^I+1)\leq 0,
\end{align}
where the first inequality holds since $q^R>1+n$, and the last inequality holds since   $m+1\leq q^I$ in \eqref{eqn:suf63}. Therefore, in this subcase, $w^*$ has the largest magnitude among the valid symbols in $\mathbb{Z}[i]/q$.
\end{enumerate}
\textbf{Case 4}: $a\leq -\frac{1}{2}$ and $b<\frac{1}{2}$

This case is not possible if $w$ is to be valid in $\mathbb{Z}[i]/q$.

\section*{Appendix III - Proof of Lemma \ref{lem:nec}}

We need to show that we can find two valid joint symbols $(w_A, w_B)$  and $(w'_A, w'_B)$  (i.e., $w_A, w_B, w'_A, w'_B$) such that $w_A-w'_A=\delta_A$ and  $w_B-w'_B=\delta_B$ in the statement of the lemma. In the following, we show that we can find $w_A$ and  $w'_A$ such that $w_A-w'_A=\delta_A$ (similar proof applies for $\delta_B$).

When $|q|<\sqrt{5}$ (i.e.,$|q|=\sqrt{2}$), we have $\mathbb{Z}[i]/q=\{0,1\}$ by \emph{Remark \ref{rem:sqrt2}}. Therefore, if $|\delta_A|<|q|=\sqrt{2}$ and $|\delta_A|\neq 0$, we must have  $|\delta_A|=1$ (since $\delta_A$ is a Gaussian integer). We can choose $w_A=\delta_A$ and $w'_A= 0$.

\begin{figure}[h]
  \centering
        \includegraphics[height=1\columnwidth]{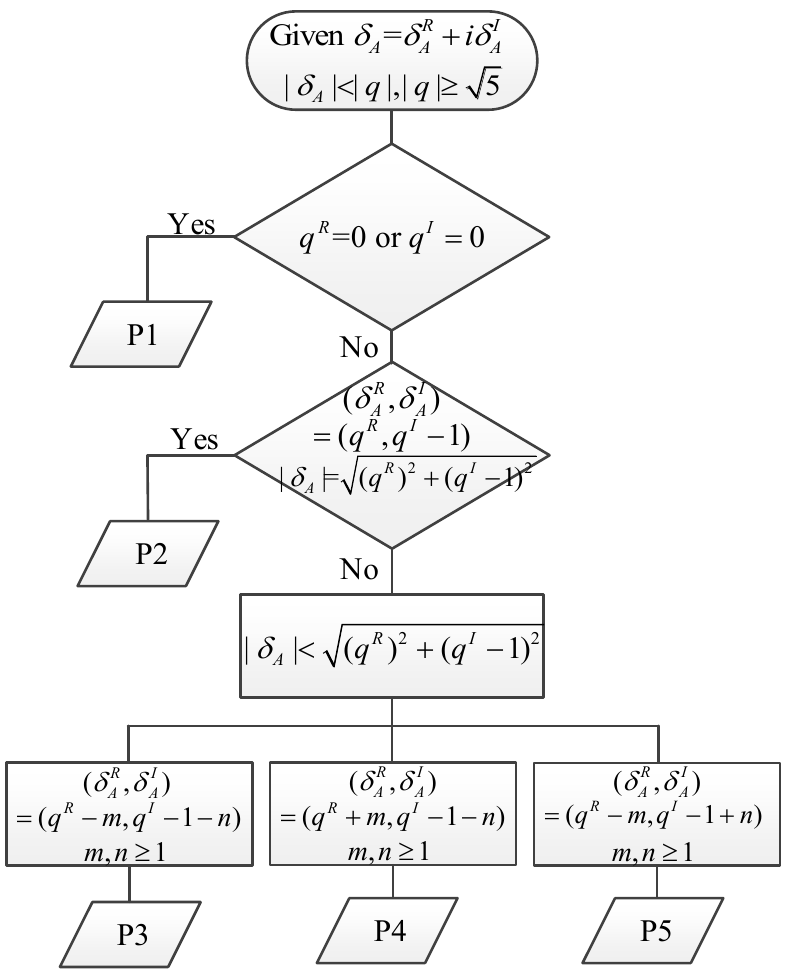}
       \caption{The proof sketch of \emph{Lemma \ref{lem:nec}}.}
        \label{fig:flow}
\end{figure}

We next consider $|q|\geq\sqrt{5}$.   Fig. \ref{fig:flow} gives the roadmap of the lengthy proof. First, P1 below gives the proof for the case of $q^R=0$ or $q^I=0$ (thus, this includes the case of real $q$). Then, P2 proves the case of $q^R\neq0$ or $q^I\neq0$, assuming $q^R>q^I\geq1$, focusing on $\delta_A=q^R+i(q^I-1)$ (this is the $\delta_A$ with the largest magnitude that yields $|\delta_A|<|q|$). After that, P3-P5 prove the cases of $\delta_A$ with  $|\delta_A|<|q^R+i(q^I-1)|$.

\textbf{P1)} We first consider the case when $q$ is a real integer prime (i.e., $q^I=0$). The proof of the case when $q^R=0$ is similar. Since $q$ is real, we have $q=|q|$. Furthermore, we verify that $w^R_A=w^x_A, w^I_A=w^y_A$ by \emph{Definition \ref{def:res}}. If $|\delta_A|=|\delta^R_A+i\delta^I_A|$ and $q$ is real, we have $|\delta^R_A|, |\delta^I_A|\leq q-1$. Then, we consider the following cases:

P1-1) $\delta^R_A$ is even,  $\delta^I_A$ is even

In this case, we choose $w_A=\frac{\delta^R_A}{2}+i\frac{\delta^I_A}{2}$ and $w'_A=-\frac{\delta^R_A}{2}-i\frac{\delta^I_A}{2}$. Therefore, by \emph{Definition \ref{def:res}}, $w_A, w'_A\in\mathbb{Z}[i]/q$ since $w^R_A, w^I_A, {w'^R_A}, {w'^I_A}<q/2$.

P1-2) $\delta^R_A$ is odd,  $\delta^I_A$ is odd

In this case, we further have $\delta^R_A, \delta^I_A\leq q-2$, since the integer prime $q$ is odd and $\delta^R_A, \delta^I_A\leq q-1$. We choose $w_A=\frac{\delta^R_A-1}{2}+i\frac{\delta^I_A-1}{2}$ and $w'_A=-\frac{\delta^R_A+1}{2}-i\frac{\delta^I_A+1}{2}$. Therefore, by \emph{Definition \ref{def:res}}, $w_A, w'_A\in\mathbb{Z}[i]/q$ since $w^R_A, w^I_A, {w'^R_A}, {w'^I_A}<q/2$.

P1-3) $\delta^R_A$ is even,  $\delta^I_A$ is odd

In this case, we further have $\delta^I_A\leq q-2$. We choose $w_A=\frac{\delta^R_A}{2}+i\frac{\delta^I_A-1}{2}$ and $w'_A=-\frac{\delta^R_A}{2}-i\frac{\delta^I_A+1}{2}$. Therefore, by \emph{Definition \ref{def:res}}, $w_A, w'_A\in\mathbb{Z}[i]/q$ since $w^R_A, w^I_A, {w'^R_A}, {w'^I_A}<q/2$.

P1-4) $\delta^R_A$ is odd,  $\delta^I_A$ is even

In this case, we choose $w_A=\frac{\delta^R_A-1}{2}+i\frac{\delta^I_A}{2}$ and $w'_A=-\frac{\delta^R_A+1}{2}-i\frac{\delta^I_A}{2}$. Similar to P1-3), $w_A, w'_A\in\mathbb{Z}[i]/q$.

\textbf{P2)}  We consider a complex $q=q^R+iq^I$ and $q^R\neq0$, $q^I\neq0$. Since $q$ is a complex Gaussian prime, we must have $|q|^2=(q^R)^2+(q^I)^2=4k+1$ where $k$ is an integer. Thus, either $q^R$  is even and $q^I$  is odd, or $q^R$  is odd and $q^I$  is even. Suppose that $q^R> q^I\geq 1$. Thus, the Gaussian integer with the largest magnitude that yields $|\delta_A|<|q|$ is $(\delta^R_A, \delta^I_A)=(q^R,q^I-1)$.

P2-1) $q^R$ is even and $q^I$ is odd

In this case, given $q^R> q^I\geq 1$, we further have $q^R\geq2$, $q^I\geq1$,  and $q^R> q^I$.

When $\delta_A=q^R+i(q^I-1)$, we let $w_A=\frac{q^R}{2}+i\frac{q^I-1}{2}$ and $w'_A=-\frac{q^R}{2}-i\frac{q^I-1}{2}$. Then, $\delta_A=w_A-w'_A$.

Since both $q^R$ and $q^I-1$  are even, by \emph{Lemma \ref{lem:even}} (as below), both $w_A, w'_A$ are valid, i.e., $w_A, w'_A\in\mathbb{Z}[i]/q$.

\begin{lemma}\label{lem:even}
Given a Gaussian integer $\delta\in \mathbb{Z}[i]$ where $\delta=\delta^R+i\delta^I$ and a Gaussian prime $q$ that defines valid symbols in $\mathbb{Z}[i]$ according to \emph{Definition \ref{def:res}}. If $|\delta|<|q|$ and both $\delta^R, \delta^I$ are even integers, there exists at least one a pair of $w, w'\in \mathbb{Z}[i]$ such that $\delta=w-w'$.
\end{lemma}
\begin{IEEEproof}[Proof of Lemma \ref{lem:even}]
Since both $\delta^R, \delta^I$ are even, there exist two Gaussian integers $w=\frac{\delta^R}{2}+i\frac{\delta^I}{2}$ and  such that $\delta=w-w'$. Furthermore, we can verify that $w, w'\in \mathbb{Z}[i]$, since  $|w|=|w'|=\frac{\sqrt{(\delta^R)^2+(\delta^I)^2}}{2}<|q|/2$. By \emph{Proposition \ref{pro:scw}}, both  $w, w'\in \mathbb{Z}[i]$.
\end{IEEEproof}

P2-2) $q^R$ is odd and $q^I$ is even

In this case, given $q^R> q^I\geq 1$, we further have $q^R\geq3$, $q^I\geq2$,  and $q^R> q^I$.

When $\delta_A=q^R+i(q^I-1)$,  we propose to have $w_A=\frac{q^R+1}{2}+i\frac{q^I-2-2k}{2}$  and $w'_A=-\frac{q^R-1}{2}-i\frac{q^I+2k}{2}$  for some non-negative integer $k\geq0$, where we will find a suitable $k$ to ensure both $w_A$ and $w'_A$ are valid symbols. We first note that $\delta_A=w_A-w'_A$. By \emph{Definition \ref{def:res}}, the corresponding coordinates of $w_A$ and $w'_A$ with the basis $(x,y)$ are given by
\begin{subequations}\label{eqn:nec1}
\begin{align}\label{eqn:nec11}
\nonumber \left[\begin{array}{c}
   w^{x}\\
     w^{y}
  \end{array}
  \right] &=\frac{1}{|q|}\left[\begin{array}{cc}
   q^R & q^I\\
    -q^I & q^R
  \end{array}
  \right]\left[\begin{array}{cc}
   \frac{q^R+1}{2} \\
   \frac{q^I-2-2k}{2}  \end{array}
  \right]\\&=\left[\begin{array}{cc}
  \frac{|q|}{2}+\frac{q^R-(2+2k)q^I}{2|q|} \\
  \frac{-q^I-(2+2k)q^R}{2|q|} \end{array}
  \right],\\
  \label{eqn:nec12}\nonumber \left[\begin{array}{c}
   w'^{x}\\
     w'^{y}
  \end{array}
  \right] &=\frac{1}{|q|}\left[\begin{array}{cc}
   q^R & q^I\\
    -q^I & q^R
  \end{array}
  \right]\left[\begin{array}{cc}
   -\frac{q^R-1}{2} \\
   -\frac{q^I+2k}{2}  \end{array}
  \right]\\&=\left[\begin{array}{cc}
  -\frac{|q|}{2}+\frac{q^R-2kq^I}{2|q|} \\
  \frac{-q^I-2k q^R}{2|q|} \end{array}
  \right].
    \end{align}
\end{subequations}

To ensure $|w^x_A|, |w^y_A|, |w'^x_A|, |w'^y_A|<|q|/2$ in $\eqref{eqn:nec1}$, we require
\begin{align}\label{eqn:nec2}
\nonumber &\max\big\{0, \lfloor\frac{q^R}{2q^I}-1\rfloor+1\big\}\leq k\\
&\leq \min\big\{\lceil\frac{q^R}{2q^I}\rceil-1,\lceil\frac{|q|^2-q^I}{2q^R}-1\rceil-1 \big\},
\end{align}
where is $\lfloor m\rfloor$ the largest integer that is smaller than $m$ and $\lceil m\rceil$ is the smallest integer that is larger than $m$. Next, we verify that there exists at least one non-negative integer $k$ in \eqref{eqn:nec2} such that $|w^x_A|, |w^y_A|, |w'^x_A|, |w'^y_A|<q/2$ (i.e., $w_A, w'_A\in \mathbb{Z}[i]/q$). Let $x=\frac{q^R}{2q^I}$ and $y=\frac{|q|^2-q^I}{2q^R}$. In case P2-2) here, $x$ is not an integer since  $q^R$ is odd. Therefore, we can reduce \eqref{eqn:nec2} to
\begin{align}\label{eqn:nec3}
 \max\{0, \lfloor x\rfloor\big\}\leq k\leq \min\{\lfloor x \rfloor,\lceil y-2\rceil \},
\end{align}
where $\lfloor x-1\rfloor+1=\lfloor x\rfloor$ and  $\lceil y-1\rceil-1=\lceil y-2\rceil$. Note also that $\lceil x\rceil-1=\lfloor x\rfloor$  if $x$ is not an integer.
According to \eqref{eqn:nec3}, we consider the following possible ranges of $k$:

P2-2-i) $\max\{0, \lfloor x\rfloor\}=0$ and $\min\{\lfloor x\rfloor, \lceil y-2\rceil\}=\lfloor x\rfloor\Rightarrow 0\leq k\leq \lfloor x\rfloor$

Since $\max\{0, \lfloor x\rfloor\}=0$, we have $\lfloor x\rfloor=0$. Since $x$ is positive, it is not possible for $\lfloor x\rfloor<0$. Thus,  $0\leq k\leq \lfloor x\rfloor$ implies that $k=0$ is the only solution. This applies for the case of $q=5+4i$.

P2-2-ii) $\max\{0, \lfloor x\rfloor\}=0$ and $\min\{\lfloor x\rfloor, \lceil y-2\rceil\}=\lceil y-2\rceil\Rightarrow 0\leq k\leq \lceil y-2\rceil$

Since $\max\{0, \lfloor x\rfloor\}=0$, we have $\lfloor x\rfloor=0$. Furthermore, since $\min\{\lfloor x\rfloor, \lceil y-2\rceil\}=\lceil y-2\rceil$, we have $\lfloor x\rfloor \geq \lceil y-2\rceil$. Given $q^R\geq 3$ and $q^I\geq 2$, we can verify that $\lceil y-2\rceil\geq 0$, since $\lceil y-2\rceil=\lceil \frac{|q|^2-q^I}{2q^R}-2\rceil=\lceil \frac{q^R}{2}+\frac{q^I(q^I-1)}{2q^R}-2\rceil\geq \lceil\frac{3}{2}-2\rceil=0$. Thus, we have  $\lfloor x\rfloor=\lceil y-2\rceil=0$. Therefore,  $0\leq k\leq \lceil y-2\rceil$ implies $k=0$ is the only solution. This applies for the case of $q=3+2i$.

P2-2-iii) $\max\{0, \lfloor x\rfloor\}=\lfloor x\rfloor$ and $\min\{\lfloor x\rfloor, \lceil y-2\rceil\}=\lfloor x\rfloor\Rightarrow \lfloor x\rfloor\leq k\leq \lfloor x\rfloor$

Thus, we have $k=\lfloor x\rfloor$ as the only solution. This applies for the case of $q=7+2i$.

P2-2-iv) $\max\{0, \lfloor x\rfloor\}=\lfloor x\rfloor$ and $\min\{\lfloor x\rfloor, \lceil y-2\rceil\}=\lceil y-2\rceil\Rightarrow \lfloor x\rfloor\leq k\leq \lceil y-2\rceil$

The case where $\lfloor x\rfloor=0$ has been dealt with in P2-2-i). Here, we assume $\lfloor x\rfloor>0$. Therefore, we have $q^R\geq 2q^I$. However, since $q^R$ is odd, we must have $q^R>2q^I$. Since $\min\{\lfloor x\rfloor,\lceil y-2\rceil\}=\lceil y-2\rceil$, we have $\lfloor x\rfloor\geq\lceil y-2\rceil$. Furthermore, given $q^R\geq3$, $q^I\geq2$, and $q^R>2q^I$, we can verify $\lfloor x\rfloor \leq \lceil y-2\rceil$. The proof is given as follows:
\begin{align}\label{eqn:nec4}
\nonumber \lceil y-2\rceil&=\lceil \frac{q^I|q|^2-(q^I)^2-(q^R)^2}{2q^R q^I}-2+\frac{(q^R)^2}{2q^Rq^I}\rceil\\
&=\lceil \frac{(q^I-1)|q|^2-4q^R q^I}{2q^R q^I}+x\rceil.
\end{align}

When $q^I>3$, given $q^R>2q^I$, we have $\frac{(q^I-1)|q|^2-4q^R q^I}{2q^R q^I}>\frac{2|q|^2-4q^R q^I}{2q^R q^I}>0$. Thus, $\lfloor x\rfloor\leq \lceil y-2\rceil$. When $q^I=2$, given $q^R>2q^I$, we further have
\begin{align}\label{eqn:nec5}
 \nonumber \lceil y-2\rceil&=\lceil \frac{|q|^2-q^I}{2q^R}-2\rceil=\lceil \frac{(q^R)^2+2}{2q^R}-2\rceil\\
 &=\lceil \frac{q^R}{2}+\frac{1}{q^R}-2\rceil.
\end{align}
From \eqref{eqn:nec5}, we have $\lfloor x\rfloor\leq \lceil y-2\rceil$  when $q^R\geq8$. When $q^R<8$ and  $q^R$ is odd, the possible $q^R$ are $5$ and $7$, since $q^R>4$. In this case of $q^I=2$, we can verify that $\lfloor x\rfloor= \lceil y-2\rceil$ when $q^R=5$ and $\lfloor x\rfloor< \lceil y-2\rceil$ when $q^R=7$. The proof of $\lfloor x\rfloor\leq \lceil y-2\rceil$ is completed.

Thus, we have $\lfloor x\rfloor= \lceil y-2\rceil$, which implies that $k=\lfloor x\rfloor$ is the only solution. This applies for the case of $q=5+2i$.

As we discussed in P2), assuming $q^R_A>q^I_A\geq 1$, $(\delta^R_A,\delta^I_A)=(q^R, q^I-1)$  has the largest magnitude that yields $|\delta_A|<|q|$. W.l.o.g., the Gaussian integers $\delta_A$ where $\delta^R_A,\delta^I_A\geq 0$ with the smaller magnitude, i.e., $|\delta_A|<|q^R+i(q^I-1)|$, belong to the following subcases:

P3) $\delta_A=(q^R-m)+i(q^I-1-n)$;

P4) $\delta_A=(q^R+m)+i(q^I-1-n)$;

P5) $\delta_A=(q^R-m)+i(q^I-1+n)$;

{\noindent where $m,n\geq 1$ and $n,m\in \mathbb{Z}$  in P3)-P5). The proofs of   P3), P4), and P5) are given as follows:}

 \textbf{P3)} We consider $\delta_A=\delta^R+i\delta^I$ with $\delta^R=q^R-m$ and $\delta^I=q^I-1-n$. Suppose that $\delta^R$ is odd and $\delta^I$ is odd. We choose $w_A=\frac{\delta^R+1}{2}+i\frac{\delta^I-1}{2}$ and $w'_A=-\frac{\delta^R-1}{2}-i\frac{\delta^I+1}{2}$. Furthermore, we can verify that both $w_A$ and $w'_A$  are valid, since $(\delta^R+1)^2+(\delta^I)^2<|q|^2$  in this case. For other subcases in P3) (i.e., even $\delta^R$ and even $\delta^I$, even $\delta^R$ and odd $\delta^I$, odd $\delta^R$ and even $\delta^I$), the proofs follow similarly.

\textbf{P4)}  We consider with $\delta_A=\delta^R+i\delta^I$ with $\delta^R=q^R+m$ and $\delta^I=q^I-1-n$. First, we consider the subcase where  $\delta^R$ is even and  $\delta^I$ is odd. In this subcase, $\delta^R\geq 2$ and $\delta^R\geq 1$. W.l.o.g., we assume $\delta^R, \delta^I\geq 0$. Furthermore, we deduce that $n>m$, since $q^R>q^I$. Given that  $\delta^I$ is odd and nonnegative, we have
\begin{align}\label{eqn:necp41}
q^I\geq n+2\geq m+1,
\end{align}
since $q^I-n-1\geq 1$. Thus, we have
\begin{align}\label{eqn:necp42}
q^R> n+2,
\end{align}
since $q^R>q^I$.

Since $(\delta^R)^2+(\delta^I)^2<(q^R)^2+(q^I-1)^2$, we further have $m^2+2q^Rm+n^2-2n(q^I-1)<0$, i.e., $q^In-q^Rm>\frac{m^2+n^2+2n}{2}>0$.

We let $w=\frac{\delta^R+2k}{2}+i\frac{\delta^I-2j-1}{2}$ and $w'=-\frac{\delta^R-2k}{2}-i\frac{\delta^I+2j+1}{2}$ where $k,j\in\mathbb{Z}$, such that $\delta=w-w'$. Then, we have
\begin{subequations}\label{eqn:nec51}
\begin{align}\label{eqn:nec511}
\nonumber \left[\begin{array}{c}
   w^{x}\\
     w^{y}
  \end{array}
  \right] &=\frac{1}{|q|}\left[\begin{array}{cc}
   q^R & q^I\\
    -q^I & q^R
  \end{array}
  \right]\left[\begin{array}{cc}
   \frac{\delta^R+2k}{2} \\
   \frac{\delta^I-2j-1}{2}  \end{array}
  \right]\\&=\left[\begin{array}{cc}
  \frac{q^R(\delta^R+2k)+q^I(\delta^I-2j-1)}{2|q|} \\
  \frac{-q^I(\delta^R+2k)+q^R(\delta^I-2j-1)}{2|q|} \end{array}
  \right],\\
  \label{eqn:nec512}\nonumber \left[\begin{array}{c}
   w'^{x}\\
     w'^{y}
  \end{array}
  \right] &=\frac{1}{|q|}\left[\begin{array}{cc}
   q^R & q^I\\
    -q^I & q^R
  \end{array}
  \right]\left[\begin{array}{cc}
   -\frac{\delta^R-2k}{2} \\
   -\frac{\delta^I+2j+1}{2}  \end{array}
  \right]\\&=\left[\begin{array}{cc}
  \frac{-q^R(\delta^R-2k)-q^I(\delta^I+2j+1)}{2|q|} \\
  \frac{q^I(\delta^R-2k)-q^R(\delta^I+2j+1)}{2|q|} \end{array}
  \right].
    \end{align}
\end{subequations}

To ensure $|w^x_A|, |w^y_A|, |w'^x_A|, |w'^y_A|<|q|/2$ in $\eqref{eqn:nec51}$, we require
\begin{subequations}\label{eqn:nec6}
\begin{align}\label{eqn:nec61}
  &\frac{q^Rm-q^I(2+n)}{2q^I}<kq^R-jq^I <\frac{q^In-q^Rm}{2q^I},\\
\label{eqn:nec62}&\frac{-|q|^2+q^Rn+q^Im}{2q^I}<jq^R+kq^I <\frac{|q|^2-q^R(n+2)-q^Im}{2}
\end{align}
\end{subequations}
We can verify that  $(k,j)=(0,0)$, i.e.,  $w=\frac{\delta^R}{2}+i\frac{\delta^I-1}{2}$ and $w'=-\frac{\delta^R}{2}-i\frac{\delta^I+1}{2}$, is a solution of \eqref{eqn:nec61} and \eqref{eqn:nec62} at the same time. The proof is as follows: First, LHS of \eqref{eqn:nec61} is less than $0$ and RHS of \eqref{eqn:nec61} is larger than $0$, since $q^In-q^Rm>0$. Second, LHS of \eqref{eqn:nec62} is less than 0 and RHS of \eqref{eqn:nec62} is larger than $0$. The proofs of the other subcases in P4) follow similarly. Due to space limit, we omit the derivations here.

\textbf{P5)} We consider $\delta_A=\delta^R+i\delta^I$ with $\delta^R=q^R-m$ and $\delta^I=q^I-1+n$. We use the same way in P4) to prove P5), by choosing proper pairs of $w$ and $w'$. Suppose that $\delta^R$ is even and $\delta^I$ is odd. We can also verify that $w=\frac{\delta^R}{2}+i\frac{\delta^I-1}{2}$ and $w'=-\frac{\delta^R}{2}-i\frac{\delta^I+1}{2}$ are two valid symbols, such that $\delta_A=w-w'$.

\section*{Appendix IV - Proof of Lemma \ref{lem:83}}

We prove $|\tilde{\Xi}|=\sqrt{13}$  only, and the proofs for $\sqrt{17}, \sqrt{29}, \sqrt{37}$ follow similarly.

W.l.o.g., consider $\tilde{\Xi}=3+2i$ (the treatment of other $\tilde{\Xi}$ with $|\tilde{\Xi}|^2=13$ is similar). First, we express $\kappa,\tau,\gamma,\delta$ in the form of a quotient and  a remainder as in \eqref{eqn:qr}. Then, we have
\begin{align}\label{eqn:c1}
\nonumber &\kappa\delta-\tau\gamma\\
\nonumber =&\tilde{\Xi}^2(q_\kappa q_\delta-q_\tau q_\gamma )+\tilde{\Xi}[(q_\kappa r_\delta+q_\delta r_\kappa)-(q_\tau r_\gamma+q_\gamma r_\tau)]\\
& +(r_\kappa r_\delta-r_\tau r_\gamma)=\tilde{\Xi}.
\end{align}
We see that in order that the last equality in \eqref{eqn:c1} applies, we must have $r_\kappa r_\delta=r_\tau r_\gamma ({\rm mod}~ \tilde{\Xi})$.  Since $\tilde{\Xi}$ is Gaussian-integer prime in this case, finite-field arithmetic applies to the remainders. The elements in the field $\mathbb{Z}[i]/(3+2i)$ are $\{0,\epsilon,\epsilon(1+i), 2\epsilon\}$, where $\epsilon$ is a unit.

Given $r_\kappa r_\delta=r_\tau r_\gamma ({\rm mod}~ \tilde{\Xi})$ as implied by \eqref{eqn:c1}, there are three possibilities as follows:
\begin{eqnarray}\label{eqn:c2}
 \nonumber ({\rm p1}) \ \  \ \ &r_{\kappa}, r_{\tau}, r_{\gamma}, r_{\delta} {\rm ~are ~all ~nonzero};\\
 \nonumber ({\rm p2}) \ \  \ \ &  r_\kappa=r_\gamma=0, r_{\tau},r_{\delta}\neq 0; \\
 \nonumber ({\rm p3}) \ \  \ \ & r_\tau=r_\delta=0, r_{\kappa}, r_{\gamma}\neq 0.
\end{eqnarray}

Note that it is not possible for $r_{\kappa}=r_{\tau}=0$ because that would imply $\gcd(\kappa, \tau)\neq 1$ according to \eqref{eqn:c1}. Similarly, it is not possible for $r_\gamma=r_\delta=0$.

We can write \eqref{eqnl83:1} and \eqref{eqnl83:2}  as
\begin{align}\label{eqn:sc1}
\nonumber \tilde{\Xi}\phi=\zeta\kappa+\vartheta\gamma=\tilde{\Xi}(\zeta q_\kappa+\vartheta q_\gamma)+(\zeta r_\kappa+\vartheta r_\gamma),\\
\tilde{\Xi}\psi=\zeta\tau+\vartheta\delta=\tilde{\Xi}(\zeta q_\tau+\vartheta q_\delta)+(\zeta r_\tau+\vartheta r_\delta).
\end{align}
Then, we let $\zeta=1$ and rewrite \eqref{eqn:sc1} as
\begin{align}\label{eqn:sc1x}
\nonumber \tilde{\Xi}\phi=\kappa+\vartheta\gamma=\tilde{\Xi}(q_\kappa+\vartheta q_\gamma)+(r_\kappa+\vartheta r_\gamma),\\
\tilde{\Xi}\psi=\tau+\vartheta\delta=\tilde{\Xi}(q_\tau+\vartheta q_\delta)+(r_\tau+\vartheta r_\delta).
\end{align}
To satisfy \eqref{eqn:sc1x}, we further let
\begin{align}\label{eqn:vart}
 \nonumber &\vartheta=-r_{\kappa}r_{\gamma}^{-1}=-r_{\tau}r_{\delta}^{-1}~({\rm mod}~\tilde{\Xi}) ~({\rm if ~(p1) ~above ~applies});\\
\nonumber &\vartheta=-r_{\tau}r_{\delta}^{-1}~({\rm mod}~\tilde{\Xi})~ ({\rm if  ~(p2) ~above ~applies});\\
 &\vartheta=-r_{\kappa}r_{\gamma}^{-1} ~({\rm mod}~\tilde{\Xi})~ ({\rm if ~(p3) ~above ~applies}).
 \end{align}
Given $\zeta=1$, if $\vartheta\in \{\epsilon,\epsilon(1+i)\}$ in \eqref{eqn:vart},  \eqref{eqnl83:3} is satisfied, since $0<1+|\vartheta|\leq \sqrt{\frac{13}{2}}$. On the other hand, if $\vartheta\in\{2\epsilon\}$, \eqref{eqnl83:3} cannot be satisfied, since $1+|\vartheta|>\sqrt{\frac{13}{2}}$. Note that if $\vartheta \in \{2\epsilon\}$, then $\vartheta^{-1} \in \{\epsilon(1+i)\}$.
 So, we multiply \eqref{eqn:sc1x} by $\vartheta^{-1}$ so that $\zeta$ becomes $\vartheta^{-1}$ and  $\vartheta$ becomes $1$ in \eqref{eqn:sc1x}. Doing so gives us:
\begin{align}\label{eqn:sc2x}
\nonumber \tilde{\Xi}\phi=\zeta\kappa+\gamma=\tilde{\Xi}(\zeta q_\kappa+ q_\gamma)+(\zeta r_\kappa+ r_\gamma),\\
\tilde{\Xi}\psi=\zeta\tau+\delta=\tilde{\Xi}(\zeta q_\tau+ q_\delta)+(\zeta r_\tau+ r_\delta).
\end{align}
where   $\zeta=-r_{\delta}r_{\tau}^{-1}$ and/or $-r_{\gamma}r_{\kappa}^{-1} ~({\rm mod}~\tilde{\Xi})$. Then, we can verify that \eqref{eqnl83:2} is satisfied.

Therefore, we have proved \emph{Lemma \ref{lem:83}} under $|\tilde{\Xi}|^2=13$.

\section*{Appendix V - Proof of Lemma \ref{lem:84}}

W.l.o.g., we consider $\tilde{\Xi}=3+i=-i(1+i)(1+2i)$. First, we express $\kappa,\tau,\gamma,\delta$ in the form of a quotient and  a remainder as in \eqref{eqn:qr}. Eqn. \eqref{eqn:c1} is still valid, and we also have  $r_\kappa r_\delta=r_\tau r_\gamma ({\rm mod}~ \tilde{\Xi})$.  The remainders $r_\kappa, r_\tau,r_\gamma, r_\delta\in \mathbb{Z}[i]/(3+i)=\{0,\epsilon,\epsilon(1+i),1+2i\}$, where $\epsilon$ is a unit.

Case 1: One of the remainders is $0$

W.l.o.g., suppose that $r_\kappa=0$. By \emph{Lemma 8.3.1}  (presented later), we cannot have $r_\tau=0$ or $\epsilon(1+i)$ or $(1+2i)$. Therefore, $r_\tau=\epsilon$. Then, $r_\kappa r_\delta=r_\tau r_\gamma ({\rm mod}~ \tilde{\Xi})$  implies $r_\gamma=0$. Now, since $r_\gamma=0$, by \emph{Lemma 8.3.1}, $r_\delta=\upsilon$ for some  $\upsilon\in\{\pm 1, \pm i\}$. Overall, we have $r_{\kappa}=r_{\gamma}=0, r_{\tau}=\epsilon,$ and  $r_{\delta}=\upsilon$. We can write \eqref{eqnl84:1} and \eqref{eqnl84:2} as
 \begin{subequations} \label{eqn:sc11}
\begin{align}
\label{eqn:sc111} \tilde{\Xi}\phi=\zeta\kappa+\vartheta\gamma=\tilde{\Xi}(\zeta q_\kappa+\vartheta q_\gamma)+(\zeta r_\kappa+\vartheta r_\gamma),\\
\label{eqn:sc112} \tilde{\Xi}\psi=\zeta\tau+\vartheta\delta=\tilde{\Xi}(\zeta q_\tau+\vartheta q_\delta)+(\zeta r_\tau+\vartheta r_\delta).
\end{align}
\end{subequations}
A way to satisfy \eqref{eqn:sc11} is to let $\zeta=r_\delta=\upsilon, \vartheta=-r_\tau=-\epsilon$. We see that $|\zeta|+|\vartheta|=2<\frac{|\tilde{\Xi}|}{\sqrt{2}}=\frac{\sqrt{10}}{\sqrt{2}}$ satisfying \eqref{eqnl84:3}. Now, we see that the only possible way for $(\phi,\psi)=(0,0)$ is for $(\kappa, \tau)=\frac{\epsilon}{\upsilon}(\gamma, \delta)$. However, this contradicts  our statement that $(\kappa, \tau)\neq (\nu \gamma, \nu \delta)$. Therefore,  $(\phi, \psi)\in \mathbb{Z}^2[i]\backslash \{(0,0)\}$.

Case 2: No remainder is $0$, and one remainder is a unit.

W.l.o.g., suppose that $r_\kappa$ is a unit and $r_\kappa=1$.  Then, from \eqref{eqn:c1}, we have $r_\delta=r_\tau r_\gamma ({\rm mod}~\tilde{\Xi}),$ $ r_\tau,r_\gamma,r_\delta\in\{\epsilon,\epsilon(1+i), 1+2i\}$.
\begin{itemize}
\item[(2-i)] $r_\delta=$ unit

Suppose that $r_\delta=\epsilon, \epsilon\in \{\pm 1, \pm i\}$, giving $\epsilon=r_\tau r_\gamma ({\rm mod}~\tilde{\Xi})$. We can verify that for this case, both $r_\tau$ and $r_\gamma$ must also be units (otherwise, $r_\tau r_\gamma$  is not congruent to a unit $({\rm mod}~\tilde{\Xi})$). Overall, all remainders are units. W.l.o.g., we let $r_\gamma=\omega$ and $r_\tau=\upsilon$, where $\omega, \upsilon$ are units. Then we have $\epsilon=\omega \upsilon$. Again, with the writing of \eqref{eqnl84:1} and \eqref{eqnl84:2} as \eqref{eqn:sc1}, we can let $\zeta=r_\delta=\epsilon$, $\vartheta=-r_\tau=-\upsilon$ to satisfy \eqref{eqn:sc1}. The rest is the same as the last part of case 1.

\item[(2-ii)] $r_\delta=\epsilon(1+i)$, where $\epsilon\in \{\pm 1, \pm i\}$

   In this case, $r_\delta=\epsilon(1+i)=r_\tau r_\gamma ({\rm mod}~\tilde{\Xi})$. By \emph{Lemma 8.3.1}, given $r_\delta=\epsilon(1+i)$, then $r_\gamma\neq\upsilon (1+i), \forall\upsilon\in \{\pm 1, \pm i\}$. We can also rule out the possibility of $r_\gamma=\upsilon(1+2i)$ and $ r_\tau=\upsilon(1+i)$ where  $\upsilon$ denotes any unit, since $(1+i)(1+2i)=0 ({\rm mod}~\tilde{\Xi})$.

Furthermore, we can verify that the way to satisfy  $r_\delta=r_\tau r_\gamma ({\rm mod}~\tilde{\Xi})$  is for $r_\gamma=\omega$, $r_\tau=\upsilon(1+i)$, and $\epsilon=\omega\upsilon$, where $\omega, \upsilon$ are both units. Overall, $r_\kappa=1, r_\tau=\upsilon(1+i),r_\gamma=\omega$, and $r_\delta=\epsilon(1+i)$. To satisfy \eqref{eqn:sc1}, we can let $\zeta=1, \vartheta=-\epsilon^{-1}\upsilon$. The rest is the same as the last part of case 1.

\item[(2-iii)] $r_\delta=1+2i$

Although by the relationship $r_\delta=r_\tau r_\gamma ~({\rm mod}~ \tilde{\Xi})$, it is possible for $r_\gamma=r_\tau=1+2i$  (because $(1+2i)^2=(1+2i)~({\rm mod}~ \tilde{\tilde{\Xi}})$). However, this possibility is ruled out because it will violate $\gcd(\gamma,\delta)=1$. The only possibility is   $r_\tau=1+2i$ and $r_\gamma=1$. Overall, we have $r_\kappa=1, r_\tau= 1+2i,r_\gamma=1$, and $r_\delta=1+2i$.
The rest is the same as the last part of case 1.
\end{itemize}

Case 3: No remainder is $0$ or a unit, one remainder is $\epsilon(1+i)$.

W.l.o.g., suppose that $r_\kappa=1+i$. Then, $r_\tau=1+2i$  by  \emph{Lemma 8.3.1}. Then, $(1+i) r_\delta=r_\tau r_\gamma ({\rm mod}~ \tilde{\Xi})$  means $r_\gamma=1+i$ and $r_\delta=1+2i$. To satisfy \eqref{eqn:sc1}, we can let $\zeta=1$ and $\vartheta=-1$. The rest is the same as the last part of subcase 1.

Subcase 4: No remainder is $0$ or a unit, or $\epsilon(1+i)$, all remainders are $1+2i$.

This case is obviously not possible because of the requirement $\gcd(\kappa,\tau)=\gcd(\gamma,\delta)=1$.

Therefore, we have proved \emph{Lemma \ref{lem:84}} under $|\tilde{\Xi}|^2=10$.

\emph{Lemma 8.3.1}:
 With respect the proof in \emph{Lemma \ref{lem:84}}, given $\gcd(\kappa,\tau)=\gcd(\gamma,\delta)=1$ and  $r_\kappa r_\delta=r_\tau r_\gamma ~({\rm mod}~ \tilde{\Xi})$, for $\Xi=3+i=-i(1+i)(1+2i)$, the following are not possible:

{($r_\kappa=0$ or $\epsilon(1+i)$) and ($r_\tau=0$ or  $\upsilon(1+i)$);}

{($r_\gamma=0$ or $\epsilon(1+i)$) and ($r_\delta=0$ or  $\upsilon(1+i)$);}

{($r_\kappa=0$ or $1+2i$) and ($r_\tau=0$ or  $1+2i$);}

{($r_\gamma=0$ or $1+2i$) and ($r_\delta=0$ or  $1+2i$);}

{\noindent where $\epsilon, \upsilon\in \{\pm 1, \pm i\}$ denote some arbitrary units.}

\begin{IEEEproof}[Proof of Lemma 8.3.1]

Each of the cases in the above is disallowed because it will lead to $\gcd(\kappa,\tau)\neq 1$ or $\gcd(\gamma,\delta)\neq 1$. For example, if $r_\kappa=0$ or $(1+i)$, $r_\tau=-(1+i)$,  then $\gcd(\kappa,\tau)=1+i$.

\end{IEEEproof}

\section*{Appendix VI - Proof of Lemma \ref{lem:86}}

\begin{figure*}[t]
  \centering
        \includegraphics[height=0.5\columnwidth]{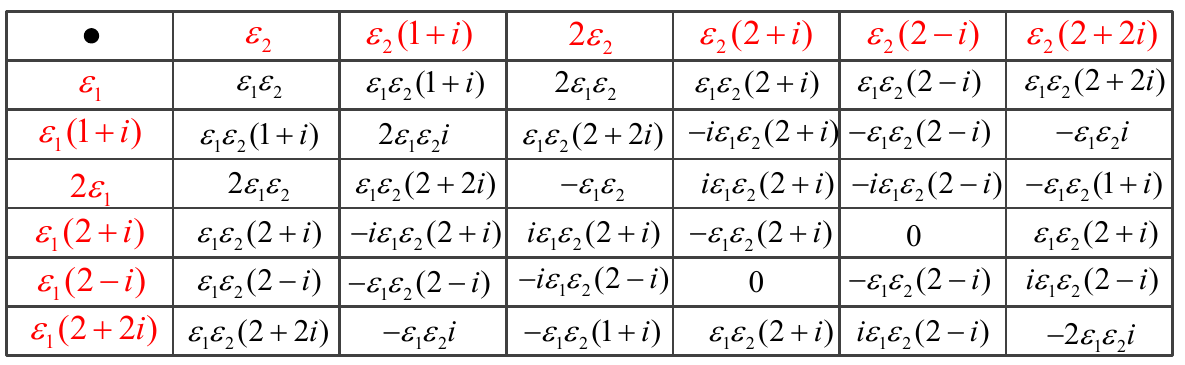}
       \caption{Multiplication for non-zero elements in $\mathbb{Z}[i]/5$, where $\varepsilon_1$ and $\varepsilon_2$ are units.}
        \label{fig:sqrt5}
\end{figure*}

The representative elements of $\mathbb{Z}[i]/5$ are $\{0, \varepsilon, \varepsilon(1+i), 2\varepsilon, \varepsilon(2+i), \varepsilon(2-i), \varepsilon(2+2i)\}$, where $\varepsilon\in \{\pm 1, \pm i\}$. From Fig. \ref{fig:sqrt5}, we see that the non-zero elements other than $\varepsilon(2+i)$ and $\varepsilon(2-i)$ all have inverses.

We express $\kappa,\tau,\gamma,\delta$ in the form of a quotient and  a remainder with division by $\Xi=5$, as in \eqref{eqn:qr}. Eqn. \eqref{eqn:c1} is still valid, and we also have  $r_\kappa r_\delta=r_\tau r_\gamma ({\rm mod}~ \tilde{\Xi})$.  The remainders $r_\kappa, r_\tau,r_\gamma, r_\delta\in \mathbb{Z}[i]/5=\{0, \varepsilon, \varepsilon(1+i), 2\varepsilon, \varepsilon(2+i), \varepsilon(2-i), \varepsilon(2+2i)\}$.

Case 1:  $r_\kappa, r_\tau, r_\gamma, r_\delta\in \{ \varepsilon, \varepsilon(1+i), 2\varepsilon,    \varepsilon(2+2i)\}$

In this case, since inverses of  $r_\kappa, r_\gamma, r_\tau, r_\delta$ exists, we can apply the same argument as in the proof of the case of  $|\Xi|=\sqrt{13}$. Note that  $\vartheta\notin \{\varepsilon(2+i), \varepsilon(2-i)\}$, according to the multiplication in Fig. \ref{fig:sqrt5}. In case  $\vartheta=\varepsilon(2+2i)$, we do a transformation to make  $\zeta=\vartheta^{-1}=\upsilon(1+i)$, $\upsilon$ is unit, and $\vartheta=1$. Thus, we can make sure
\begin{align}
 |\zeta|+|\vartheta|\leq1+\sqrt{2}<\frac{5}{\sqrt{2}}=\frac{|{\Xi}|}{\sqrt{2}}.
\end{align}

Case 2: One of $r_\kappa, r_\tau, r_\gamma, r_\delta$ is $0$

W.l.o.g., suppose that $r_\kappa=0$. Then, $r_\tau\notin\{0, \varepsilon(2+i),  \varepsilon(2-i)\}$ because $\gcd(\kappa, \tau)=1$. Given that $r_\kappa r_\delta=r_\tau r_\gamma ({\rm mod}~ {\Xi})$, we must have $r_\gamma=0$. Given $r_\gamma=0$, $r_\delta\notin\{0, \varepsilon(2+i),  \varepsilon(2-i)\}$. Thus, inverses of $r_\tau, r_\delta$ exist. We can choose $\zeta=1, \vartheta=-r_\tau r_\delta^{-1}$ if $-r_\tau r_\delta^{-1}\neq \varepsilon(2+2i)$; and $\zeta=-r_\tau^{-1} r_\delta, \vartheta=1$ otherwise. The statement of the lemma is thus fulfilled.

Case 3: One of $r_\kappa, r_\tau, r_\gamma, r_\delta$ is $\varepsilon(2+i)$, none of $r_\kappa, r_\tau, r_\gamma, r_\delta$ is 0

W.l.o.g., suppose that  $r_\kappa=\varepsilon(2+i)$. Then, $r_\tau\neq \upsilon(2+i)$  because  $\gcd(\kappa, \tau)=1$. Given that  $r_\kappa r_\delta=r_\tau r_\gamma ({\rm mod}~ {\Xi})$ and   $r_\kappa=\varepsilon(2+i)$,  $r_\tau\notin\{0,\upsilon(2+i)\}$,  $r_\delta\neq 0$, we must have  $r_\gamma=\upsilon(2+i)$ for some unit  $\upsilon$, according to Fig. \ref{fig:sqrt5}. That $r_\gamma=\upsilon(2+i)$ also means  $r_\delta\neq\omega(2+i)$, where $\omega$ is a unit. Overall, we have the following possibilities: $r_\kappa=\varepsilon(2+i), r_\gamma=\upsilon(2+i), r_\tau, r_\delta\in \{\varepsilon, \varepsilon(1+i), \varepsilon(2-i), \varepsilon(2+2i)\}$.

\begin{itemize}
\item[(3-i)] $r_\tau, r_\delta\neq\varepsilon(2-i)$

 In this subcase, both $r_\tau, r_\delta$ have an inverse. We can choose $\zeta=1, \vartheta=-r_\tau r_\delta^{-1}$ if $-r_\tau r_\delta^{-1}\neq \varepsilon(2+2i)$; and $\zeta=-r_\tau^{-1} r_\delta, \vartheta=1$ otherwise. This ensures $\zeta r_\tau+\vartheta r_\delta=0 \ ({\rm mod}~ {5})$ (note that $r_\kappa r_\delta-r_\tau r_\gamma=0 ({\rm mod}~ {\Xi}) \Rightarrow$ $\zeta, \vartheta$ selected above are such that $\zeta r_\tau+\vartheta r_\delta=0 \ ({\rm mod}~ {5})$.) Overall, we also have $|\zeta|+|\vartheta|\leq 1+\sqrt{2}<\frac{5}{\sqrt{2}}=\frac{|\Xi|}{\sqrt{2}}$.

\item[(3-ii)] One of $r_\tau,r_\delta =\varepsilon(2-i)$

W.l.o.g, suppose that $r_\tau=\omega(2-i)$. Then, $r_\kappa r_\delta-r_\tau r_\gamma=0 ({\rm mod}~ {\Xi}) \Rightarrow \varepsilon(2+i)r_\delta-\omega(2-i)\upsilon(2+i)=0 ({\rm mod}~ {\Xi})$. Therefore, we have $r_\delta=\mu(2-i)$, where $\mu$ is a unit.

We want to find $\zeta, \vartheta$ such that
\begin{subequations}  \label{eqnl87}
\begin{align}
 \label{eqnl87:1} &\kappa\zeta+\gamma\vartheta=0 \ ({\rm mod}~ {5}),\\
 \label{eqnl87:2} & \tau\zeta+\delta\vartheta=0 \ ({\rm mod}~ {5}),\\
  \label{eqnl87:3} & 0 < |\zeta|+|\vartheta|\leq\frac{5}{\sqrt{2}}.
\end{align}
\end{subequations}

Substituting the above values of $r_\kappa, r_\gamma, r_\tau, r_\delta$ and setting $\vartheta=1$ in \eqref{eqnl87:1} and  \eqref{eqnl87:2}, we have
\begin{subequations}  \label{eqnl88}
\begin{align}
 \label{eqnl88:1} &\varepsilon[\zeta(2+i)+\varepsilon^{-1}\upsilon(2+i)]=0 \ ({\rm mod}~ {5}),\\
 \label{eqnl88:2} & \omega[\zeta(2-i)+\omega^{-1}\mu(2-i)]=0  \ ({\rm mod}~ {5}).
\end{align}
\end{subequations}

If $\varepsilon^{-1}\upsilon=\omega^{-1}\mu$, then we just set $\zeta=-\varepsilon^{-1}\upsilon$, giving $0 < |\zeta|+|\vartheta|=1+1\leq\frac{5}{\sqrt{2}}$.

If $\varepsilon^{-1}\upsilon=-\omega^{-1}\mu$, we note from Fig. \ref{fig:sqrt5} that multiplying $(2+i)$ by $2$ is the same as the multiplying it by $i$; and multiplying $(2-i)$ by $2$ is the same as multiplying it by $-i$. Accordingly, we set $\zeta=-\varepsilon^{-1}\upsilon\cdot -i \cdot2$. This  fulfills \eqref{eqnl88}. We have $0 < |\zeta|+|\vartheta|=2+1\leq\frac{5}{\sqrt{2}}$.

If $\varepsilon^{-1}\upsilon=i\omega^{-1}\mu$, we note from Fig. \ref{fig:sqrt5} that multiplying $(2+i)$ by $(1+i)$ is the same as the multiplying it by $-i$; and multiplying $(2-i)$ by $(1+i)$ is the same as multiplying it by $-1$. Accordingly, we set $\zeta=-\varepsilon^{-1}\upsilon\cdot i \cdot(1+i)$. This  fulfills \eqref{eqnl88}. We have $0 < |\zeta|+|\vartheta|=\sqrt{2}+1\leq\frac{5}{\sqrt{2}}$.

If $\varepsilon^{-1}\upsilon=-i\omega^{-1}\mu$, we take the complex conjugates on both sides to get $\varepsilon\upsilon^{-1}=i\omega\mu^{-1}$. Instead of setting $\vartheta=1$, we set $\zeta=1$, and instead of \eqref{eqnl88}, we have
\begin{subequations}  \label{eqnl89}
\begin{align}
 \label{eqnl89:1} &\upsilon[\varepsilon\upsilon^{-1} (2+i)+\vartheta(2+i)]=0 \ ({\rm mod}~ {5}),\\
 \label{eqnl89:2} & \mu[\omega\mu^{-1}(2-i)+\vartheta(2-i)]=0 \ ({\rm mod}~ {5}).
\end{align}
\end{subequations}
We then set $\vartheta=-\varepsilon\upsilon^{-1}\cdot i \cdot(1+i)$.   We again have $0 < |\zeta|+|\vartheta|=\sqrt{2}+1\leq\frac{5}{\sqrt{2}}$.

This completes the proof of subcase (3-ii).

\end{itemize}

Case 4: One of $r_\kappa, r_\gamma, r_\tau, r_\delta$ is $\varepsilon(2-i)$, none of $r_\kappa, r_\gamma, r_\tau, r_\delta$ is 0

The proof of case 4 is similar to case 3 by symmetry.

\section*{Appendix VII - Proof of Lemma \ref{lem:87}}

This proof consists two parts. First, Part 1  proves that there exists $(\zeta, \vartheta)$ such that \eqref{eqnl81:1} and \eqref{eqnl81:2} are satisfied. Then, Part 2  proves that   $(\phi, \psi)$ satisfying \eqref{eqnl81:1} and \eqref{eqnl81:2} is a  distance-valid difference pair, i.e.,  $(\phi, \psi)\in\Delta$.

\emph{Remark:} In this proof , we will not follow \emph{Lemmas} \emph{\ref{lem:82}-\ref{lem:86}} to prove \eqref{eqnl81:3}, since $(\zeta, \vartheta)$  yielding \eqref{eqnl81:1} and \eqref{eqnl81:2}  cannot satisfy  \eqref{eqnl81:3}  for the case of $|\tilde{\Xi}|=2$ or $3$.

\noindent{\textbf{Part 1:}}

 \underline{(1) $|\tilde{\Xi}|=2$}

W.l.o.g., we consider $\tilde{\Xi}=2=(1+i)(1-i)$. First, we express $\kappa,\tau,\gamma,\delta$ in the form of a quotient and  a remainder as in \eqref{eqn:qr}. Eqn. \eqref{eqn:c1} is still valid, and we also have  $r_\kappa r_\delta=r_\tau r_\gamma ({\rm mod}~ \tilde{\Xi})$.  The remainders $r_\kappa, r_\tau,r_\gamma, r_\delta\in \mathbb{Z}[i]/2=\{0,1, i, 1+i\}$.

Case 1: One of the remainders is $0$

W.l.o.g., suppose that $r_\kappa=0$. Since $\gcd(\kappa,\tau)=1$, $r_{\tau}\notin\{0 ,1+i\}$. Therefore, $r_{\tau}=1$ or $i$.

\begin{itemize}
\item[(1-i)] $r_\tau=1$

In this subcase, the only possibility for $0=r_\gamma ({\rm mod}~2)$ is $r_\gamma=0$. Given $r_\gamma=0$, we  have $r_\gamma\in\{0, 1, i, 1+i\}$.  However, we can rule out the possibility of $r_\gamma=1+i$, since this contradicts $\gcd(\gamma,\delta)=1$. Therefore, we have $r_\delta=1$ or $i$. Overall, we have $r_\kappa=0, r_\tau= 1,r_\gamma=0$, and $r_\delta\in\{1, i\}$. To satisfy \eqref{eqn:c1}, if $r_\delta=1$, we can choose $\zeta= 1, \vartheta=\pm 1$, where the sign of $\vartheta$ does not matter (i.e., both signs will work); if $r_\delta=i$, we can choose $\zeta=1, \vartheta=\pm i$, where the sign of $\vartheta$ does not matter.

\item[(1-ii)] $r_\tau=i$

    This subcase is similar to subcase (1-i).  We can also either choose $\zeta= 1, \vartheta=\pm1$  or  $\zeta= 1, \vartheta=\pm i$.
\end{itemize}

Case 2: No remainder is $0$, one remainder is $1$.

W.l.o.g., suppose that $r_\kappa=1$. Since $\gcd(\gamma,\delta)=1$, $r_{\tau}\in\{1,i,1+i\}$.

\begin{itemize}
\item[(2-i)] $r_\tau=1$

Since $r_\delta=r_\gamma ({\rm mod}~2)$, we have $(r_\gamma, r_\delta)\in\{(1,1), (i,i), (1+i, 1+i)\}$. However, since $\gcd(\gamma,\delta)=1$, we can rule out the possibility of $(r_\gamma, r_\delta)=(1+i, 1+i)$.  If $(r_\gamma, r_\delta)=(1, 1)$, we choose $\zeta=1, \vartheta=\pm 1$, where the sign of $\vartheta$ does not matter. If $(r_\gamma, r_\delta)=(i, i)$,  we   choose $\zeta=1, \vartheta= \pm i$, where the sign of $\vartheta$ does not matter.

\item[(2-ii)] $r_\tau=i$

    This subcase is similar to subcase (2-i).

 \item[(2-iii)] $r_\tau=1+i$

 Since $r_\delta=(1+i)r_\gamma ({\rm mod}~2)$, we have $(r_\gamma, r_\delta)\in\{(1,1+i), (i, 1+i)\}$.   If $(r_\gamma, r_\delta)=(i, 1+i)$, we choose $\zeta= 1, \vartheta= \pm 1$, where the sign of $\vartheta$ does not matter. If $(r_\gamma, r_\delta)=(i, 1+i)$,  we   choose $\zeta= 1, \vartheta=\pm i$, where the sign of $\vartheta$ does not matter.

\end{itemize}

Case 3: No remainder is $0$ or 1, one remainder is $i$.

W.l.o.g., suppose that $r_\kappa=i$. Since $\gcd(\gamma,\delta)=1$, $r_{\tau}\in\{i,1+i\}$.

\begin{itemize}
\item[(3-i)] $r_\tau=i$

Since $ir_\delta=ir_\gamma ({\rm mod}~2)$, we have $(r_\gamma, r_\delta)\in\{(i,i), (1+i, 1+i)\}$. However, since $\gcd(\gamma,\delta)=1$, we can rule out the possibility of $(r_\gamma, r_\delta)=(1+i, 1+i)$.
Overall, we have $r_\kappa=i, r_\tau= i,r_\gamma=i$, and $r_\delta=i$. To satisfy \eqref{eqn:c1}, we choose $\zeta= 1, \vartheta=\pm  1$, where the sign of $\vartheta$ does not matter.

\item[(3-ii)] $r_\tau=1+i$

   In this subcase, the only possibility for $ir_\delta=(1+i)r_\gamma ({\rm mod}~2)$ is $(r_\gamma, r_\delta)=(i, 1+i)$. Overall, we have $r_\kappa=i, r_\tau= 1+i,r_\gamma=i$, and $r_\delta=1+i$. To satisfy \eqref{eqn:c1}, we choose $\zeta= 1, \vartheta=\pm 1$, where the sign of $\vartheta$ does not matter.

\end{itemize}

Case 4: No remainder is $0$ or 1  or $i$, all remainders are $1+i$.

W.l.o.g., suppose that $r_\kappa=1+i$. However, this subcase is not possible because of the requirement $\gcd(\kappa,\tau)=1$.


For all of cases 1, 2, and 3 above, one of the following two choices must be able to satisfy \eqref{eqnl81:1} and \eqref{eqnl81:2}:

\emph{Choice 1}: we can choose  either  $\zeta=1, \vartheta=1$ or $\zeta=1, \vartheta=-1$  (specifically, the sign of $\vartheta$ does not matter; both $\vartheta= 1$ and $\vartheta = -1$ will work if this choice is the valid choice);

\emph{Choice 2}: we can choose  either  $\zeta=1, \vartheta=i$ or $\zeta=1, \vartheta=-i$  (again, the sign of $\vartheta$ does not matter here);

Since $(\kappa,\tau)\neq(\nu\gamma, \nu \delta)$ where $\nu=\pm 1$ or $\pm i$ from the statement of the lemma, we have $\phi\neq 0$ and $\psi\neq 0$.

\underline{(2) $|\tilde{\Xi}|=3$}

With a proof substantially similar in spirit  to that of $|\tilde{\Xi}|=2$, we can show that one of the following three choices will satisfy  \eqref{eqnl81:1} and \eqref{eqnl81:2} for $|\tilde{\Xi}|=3$:

\emph{Choice 1}: we can choose  either  $\zeta=1, \vartheta=1$ or $\zeta=1, \vartheta=-1$;

\emph{Choice 2}: we can choose  either  $\zeta=1, \vartheta=i$ or $\zeta=1, \vartheta=-i$;

\emph{Choice 3}: we can choose  either  $\zeta=1, \vartheta=1+i$ or $\zeta=1, \vartheta=-1-i$;

Again, for the valid choice, the sign of $\vartheta$ does not matter here.

\noindent{\textbf{Part 2:}}

 We first prove  $|\tilde{\Xi}|=2$. In the following, we  prove the case where  \emph{Choice 1} is the valid choice to satisfy \eqref{eqnl81:1} and \eqref{eqnl81:2}. Similar proof applies if  \emph{Choice 2}  is the valid choice.

Given \emph{Choice 1}, we first show  that there exists  $(\phi, \psi)=(\frac{\kappa+\vartheta \gamma}{\tilde{\Xi}}, \frac{\tau+\vartheta \delta}{\tilde{\Xi}})\in \Delta$ (i.e., $(\phi, \psi)$ is a distance-valid pair). According to the convex regions in \emph{Definition \ref{def:con}}, we introduce a property of convex regions,   as follows:

\emph{Convex Combination \cite{conv}:} Given $\sum_k a_k\leq 1$ where $a_k\geq 0$, and $\forall  c_k\in \mathcal{G}_q$, we have $\sum_k a_k c_k\in \mathcal{G}_q$.

\rightline{$\blacksquare$}

\begin{statement}\label{stm:2}
If $a$ is a  valid difference (i.e., $a\in \Lambda$), so is $i^n a, \forall n\in \{1,2,3\}$. Similarly, if $b\in \mathcal{G}_q$, then $i^n b\in \mathcal{G}_q, \forall n\in \{1,2,3\}$.
\end{statement}
\rightline{$\blacksquare$}

The proof of \emph{Statement \ref{stm:2}} is given as follows. If  $a\in \Lambda$,  we can find $w,w'\in \mathbb{Z}[i]/q$ such that $a=w-w'$. Then, $i^n a=i^n(w-w') \in \Lambda$ for $n\in \{1,2,3\}$, since  $i^n w, i^n w'\in \mathbb{Z}[i]/q$ by \emph{Definition \ref{def:res}}.

 Since $b\in \mathcal{G}_q$, we have $b=\sum^K_{k=1} a_k c_k$, where $\sum_k a_k\leq 1$, $a_k\geq 0$, and $\forall c_k\in \mathcal{G}_q$. Then, $i^n b=\sum^K_{k=1} a_k (i^n c_k)$.  Given $c_k\in \Lambda$,  $i^n c_k$ is also a valid difference as we proved above. Therefore,  $i^n b\in \mathcal{G}_q$, since $i^n b$ is a also convex combination.


\begin{statement}\label{stm:3}
When $\tilde{\Xi}=2$,  for any $\kappa, \gamma \in \Lambda$,   $\phi=\frac{\kappa+\vartheta \gamma}{\tilde{\Xi}}\in \Lambda$, where $\vartheta$ can be $1$ or $-1$.
\end{statement}
\rightline{$\blacksquare$}

Proof of \emph{Statement \ref{stm:3}} is given as follows.
By \emph{Statement 1}, $\epsilon \gamma  \in \mathcal{G}_q$, where $\epsilon$ can be any unit. Since convex region is closed under linear combination, we have $\phi\in \mathcal{G}_q$.
 Therefore, by \emph{Lemma 9}, we have $\phi\in \Lambda$. Similarly, we can also prove that $\psi\in \Lambda$.

Suppose that   $(\kappa,\tau)$ and $(\gamma, \delta)$ are adjacent under $\tilde{\Xi}=\kappa\delta-\tau\gamma=2$. The following part is similar to the proof in \eqref{eqn:conex}.  If $(\kappa,\tau)$ and $(\gamma, \delta)$ are adjacent,  then there exists a common point $z'$ equidistant to $(\kappa,\tau)$ and $(\gamma, \delta)$ such that no other generators are closer to $z'$ than are $(\kappa,\tau)$ and $(\gamma, \delta)$. From \eqref{eqnl81:1} and \eqref{eqnl81:2},  the weighted distance from $(\phi, \psi)$ to $z'$ is
\begin{align}\label{eqnl85:6}
|\psi z'-\phi|&=\frac{| (\tau z'-\kappa)+\vartheta(\delta z'-\gamma)|}{2},
\end{align}
where $\vartheta$ can be $1$ or $-1$ in \emph{Choice 1}.

 Then, we have
\begin{align}\label{eqnl85:7}
\nonumber& |(\tau z'-\kappa)+\vartheta(\delta z'-\gamma)| \\ \nonumber &= \sqrt{|\tau z'-\kappa|^2+|\vartheta(\delta z'-\gamma)|^2-2|\tau z'-\kappa||\vartheta(\delta z'-\gamma)|\cos \theta_\vartheta}\\
  &= \sqrt{2|\tau z'-\kappa|^2(1-\cos \theta_\vartheta)} \leq \sqrt{2}|\tau z'-\kappa|,
\end{align}
where $\theta_\vartheta$ denote the angle between two vectors $\tau z'-\kappa$ and $\vartheta(\delta z'-\gamma)$. Note that $\theta_\vartheta$ depends on  $\vartheta$. Furthermore, the first equality holds because of \emph{cosine rule}, the second equality holds because of $|\tau z'-\kappa|=|\vartheta(\delta z'-\gamma)|$, and the inequality holds since we can either choose $\vartheta=1$ or $-1$ such that $0<\theta_\vartheta\leq90^o$ yielding $|1-\cos \theta|\leq1$.

From \eqref{eqnl85:6} and \eqref{eqnl85:7}, we further have
\begin{align}\label{eqnl85:8}
& |\psi z'-\phi| < \frac{\sqrt{2}}{2}|\tau z'-\kappa|<|\tau z'-\kappa|.
\end{align}
Obviously,   \eqref{eqnl85:8} contradicts our assumption that $(\kappa,\tau)$ and $(\gamma, \delta)$ are adjacent.

The proof of $|\tilde{\Xi}|=3$ follows similarly.   First, by \emph{Statement 1} and \emph{Lemma \ref{conj:1}}, we can also show that $\phi=\frac{\kappa+\vartheta\gamma}{\tilde{\Xi}}\in \Lambda$ where $\vartheta$ is chosen from any choice in Part I. Second,  we    follow  the proof by contradiction in \eqref{eqnl85:6} and \eqref{eqnl85:7}. For example, suppose that   \emph{Choice 3}, where $\vartheta=1+i$ or $-1-i$,  is the valid choice. Then,  $|(\tau z'-\kappa)+\vartheta(\delta z'-\gamma)|=\sqrt{3}|\tau z'-\kappa|\sqrt{1-\frac{2\sqrt{2}}{3}\cos\theta_\vartheta}< \sqrt{3}|\tau z'-\kappa|$, where the inequality holds since we can either choose $\vartheta=1+i$ or $-1-i$ such that $0<\theta_\vartheta\leq90^o$. Furthermore, we  have $ |\psi z'-\phi| < \frac{\sqrt{3}}{3}|\tau z'-\kappa|<|\tau z'-\kappa|$. Therefore, we can prove that $(\kappa,\tau)$ and $(\gamma, \delta)$ are non-adjacent for $|\tilde{\Xi}|=3$ under \emph{Choice 3}.


\end{document}